\documentclass[final,5p,times,twocolumn]{elsarticle}
\usepackage{color}
\usepackage{graphicx}
\usepackage{amssymb}

\biboptions{sort&compress }

\newcommand{\nue}{\ensuremath{\nu_{e}}}

\newcommand{\nuebar}{\ensuremath{\overline{\nu}_{e}}}
\newcommand{\g}{\gamma}
\newcommand{\pspt}{PROSPECT}

\newcommand{\NIST}{National Institute of Standards and Technology (NIST) \renewcommand{\NIST}{NIST}}
\newcommand{\NBSR}{National Bureau of Standards Reactor (NBSR) \renewcommand{\NBSR}{NBSR}}
\newcommand{\INL}{Idaho National Laboratory (INL) \renewcommand{\INL}{INL}}
\newcommand{\ATR}{Advanced Test Reactor (ATR) \renewcommand{\ATR}{ATR}}
\newcommand{\ORNL}{Oak Ridge National Laboratory (ORNL) \renewcommand{\ORNL}{ORNL}}
\newcommand{\HFIR}{High Flux Isotope Reactor (HFIR) \renewcommand{\HFIR}{HFIR}}
\newcommand{\LLNL}{Lawrence Livermore National Laboratory (LLNL) \renewcommand{\LLNL}{LLNL}}

\newcommand{\ptwenty}{PROSPECT-20 \renewcommand{\ptwenty}{P20}}

\journal{Nuclear Instruments and Methods}

\begin{document}

\begin{frontmatter}




\title{The PROSPECT Reactor Antineutrino Experiment}


\author{
J.\,Ashenfelter$^{p}$,
A.\,B.\,Balantekin$^{m}$,
C.\,Baldenegro$^{i}$,
H.\,R.\,Band$^{p}$,
C.\,D.\,Bass$^{f}$,
D.\,E.\,Bergeron$^{g}$,
D.\,Berish$^{j}$,
L.\,J.\,Bignell$^{a}$,
N.\,S.\,Bowden$^{e}$,
J.\,Boyle$^{i}$,
J.\,Bricco$^{n}$,
J.\,P.\,Brodsky$^{e}$,
C.\,D.\,Bryan$^{h}$,
A.\,Bykadorova~Telles$^{p}$,
J.\,J.\,Cherwinka$^{n}$,
T.\,Classen$^{e}$,
K.\,Commeford$^{b}$,
A.\,J.\,Conant$^{c}$,
A.\,A.\,Cox$^{l}$,
D.\,Davee$^{o}$,
D.\,Dean$^{i}$,
G.\,Deichert$^{h}$,
M.\,V.\,Diwan$^{a}$,
M.\,J.\,Dolinski$^{b}$,
A.\,Erickson$^{c}$,
M.\,Febbraro$^{i}$,
B.\,T.\,Foust$^{p}$,
J.\,K.\,Gaison$^{p}$,
A.\,Galindo-Uribarri$^{i,k}$,
C.\,E.\,Gilbert$^{i,k}$,
K.\,E.\,Gilje$^{d}$,
A.\,Glenn$^{e}$,
B.\,W.\,Goddard$^{b}$,
B.\,T.\,Hackett$^{i,k}$,
K.\,Han$^{p}$,
S.\,Hans$^{a}$,
A.\,B.\,Hansell$^{j}$,
K.\,M.\,Heeger$^{p}$,
B.\,Heffron$^{i,k}$,
J.\,Insler$^{b}$,
D.\,E.\,Jaffe$^{a}$,
X.\,Ji$^{a}$,
D.\,C.\,Jones$^{j}$,
K.\,Koehler$^{n}$,
O.\,Kyzylova$^{b}$,
C.\,E.\,Lane$^{b}$,
T.\,J.\,Langford$^{p}$,
J.\,LaRosa$^{g}$,
B.\,R.\,Littlejohn$^{d}$,
F.\,Lopez$^{p}$,
X.\,Lu$^{i,k}$,
D.\,A.\,Martinez~Caicedo$^{d}$,
J.\,T.\,Matta$^{i}$,
R.\,D.\,McKeown$^{o}$,
M.\,P.\,Mendenhall$^{e}$,
H.\,J.\,Miller$^{g}$,
J.\,M.\,Minock$^{b}$,
P.\,E.\,Mueller$^{i}$,
H.\,P.\,Mumm$^{g}$,
J.\,Napolitano$^{j}$,
R.\,Neilson$^{b}$,
J.\,A.\,Nikkel$^{p}$,
D.\,Norcini$^{p}$,
S.\,Nour$^{g}$,
D.\,A.\,Pushin$^{l}$,
X.\,Qian$^{a}$,
E.\,Romero-Romero$^{i,k}$,
R.\,Rosero$^{a}$,
D.\,Sarenac$^{l}$,
B.\,S.\,Seilhan$^{e}$,
R.\,Sharma$^{a}$,
P.\,T.\,Surukuchi$^{d}$,
C.\,Trinh$^{b}$,
M.\,A.\,Tyra$^{g}$,
R.\,L.\,Varner$^{i}$,
B.\,Viren$^{a}$,
J.\,M.\,Wagner$^{b}$,
W.\,Wang$^{o}$,
B.\,White$^{i}$,
C.\,White$^{d}$,
J.\,Wilhelmi$^{j}$,
T.\,Wise$^{p}$,
H.\,Yao$^{o}$,
M.\,Yeh$^{a}$,
Y.-R.\,Yen$^{b}$,
A.\,Zhang$^{a}$,
C.\,Zhang$^{a}$,
X.\,Zhang$^{d}$,
M.\,Zhao$^{a}$}
\address{$^{a}$Brookhaven National Laboratory, Upton, NY, USA}
\address{$^{b}$Department of Physics, Drexel University, Philadelphia, PA, USA}
\address{$^{c}$George W.\,Woodruff School of Mechanical Engineering, Georgia Institute of Technology, Atlanta, GA USA}
\address{$^{d}$Department of Physics, Illinois Institute of Technology, Chicago, IL, USA}
\address{$^{e}$Nuclear and Chemical Sciences Division, Lawrence Livermore National Laboratory, Livermore, CA, USA}
\address{$^{f}$Department of Physics, Le Moyne College, Syracuse, NY, USA}
\address{$^{g}$National Institute of Standards and Technology, Gaithersburg, MD, USA}
\address{$^{h}$High Flux Isotope Reactor, Oak Ridge National Laboratory, Oak Ridge, TN, USA}
\address{$^{i}$Physics Division, Oak Ridge National Laboratory, Oak Ridge, TN, USA}
\address{$^{j}$Department of Physics, Temple University, Philadelphia, PA, USA}
\address{$^{k}$Department of Physics and Astronomy, University of Tennessee, Knoxville, TN, USA}
\address{$^{l}$Institute for Quantum Computing and Department of Physics and Astronomy, University of Waterloo, Waterloo, ON, Canada}
\address{$^{m}$Department of Physics, University of Wisconsin, Madison, Madison, WI, USA}
\address{$^{n}$Physical Sciences Laboratory, University of Wisconsin, Madison, Madison, WI, USA}
\address{$^{o}$Department of Physics, College of William and Mary, Williamsburg, VA, USA}
\address{$^{p}$Wright Laboratory, Department of Physics, Yale University, New Haven, CT, USA}

\begin{abstract}
The Precision Reactor Oscillation and Spectrum Experiment, PROSPECT, is designed to make both a precise measurement of the antineutrino spectrum from a highly-enriched uranium reactor and to probe eV-scale sterile neutrinos by searching for neutrino oscillations over meter-long baselines. 
PROSPECT utilizes a segmented $^6$Li-doped liquid scintillator detector for both efficient detection of reactor antineutrinos through the inverse beta decay reaction and excellent background discrimination. PROSPECT is a movable 4-ton antineutrino detector covering distances of 7\,m to 13\,m from the High Flux Isotope Reactor core. 
It will probe the best-fit point of the $\bar\nu_e$ disappearance experiments at 4\,$\sigma$ in 1 year and the favored regions of the sterile neutrino parameter space at more than $3\,\sigma$ in 3 years. 
PROSPECT will test the origin of spectral deviations observed in recent $\theta_{13}$ experiments, search for sterile neutrinos, and address the hypothesis of sterile neutrinos as an explanation of the reactor anomaly.
This paper describes the design, construction, and commissioning of PROSPECT and reports first data characterizing the performance of the PROSPECT antineutrino detector.
\end{abstract}


\end{frontmatter}
\tableofcontents


\section{Introduction }
\label{introduction}

\begin{figure}
   \centering
    \includegraphics[width=0.45\textwidth]{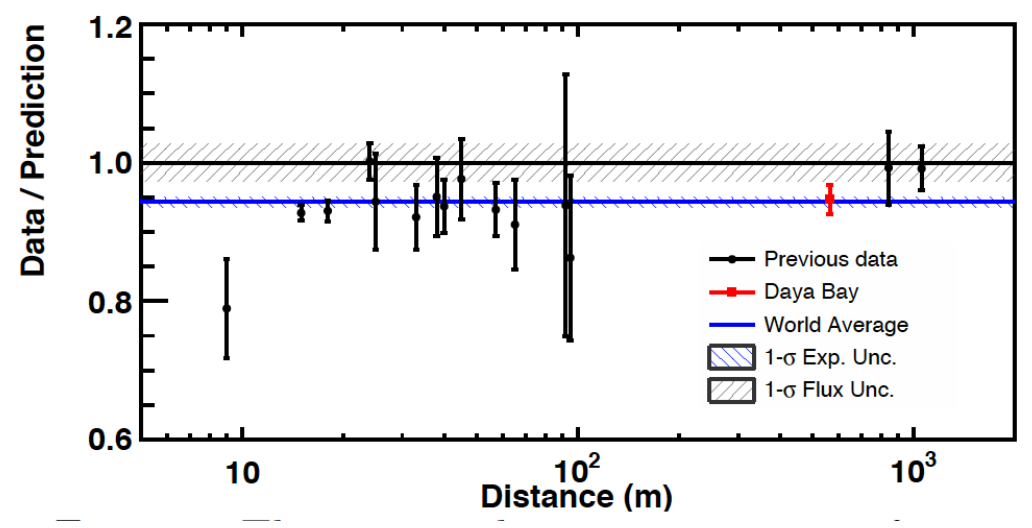}
   \caption{Comparison of previously measured reactor antineutrino fluxes over theoretical predictions with a recent Daya Bay flux measurement (from ~\cite{An:2015nua}). Predictions are
based on models for the emission of reactor antineutrinos from~\cite{Mueller:2011nm,Huber:2011wv}. 
   The measured deficit relative to prediction is known as the ``reactor antineutrino anomaly''~\cite{Mention:2011rk}.}
   \label{fig:reactorAnomaly}
\end{figure}

\begin{figure*}
   \centering
    \includegraphics[clip=true, trim=0mm 20mm 0mm 30mm, width=0.95\textwidth]{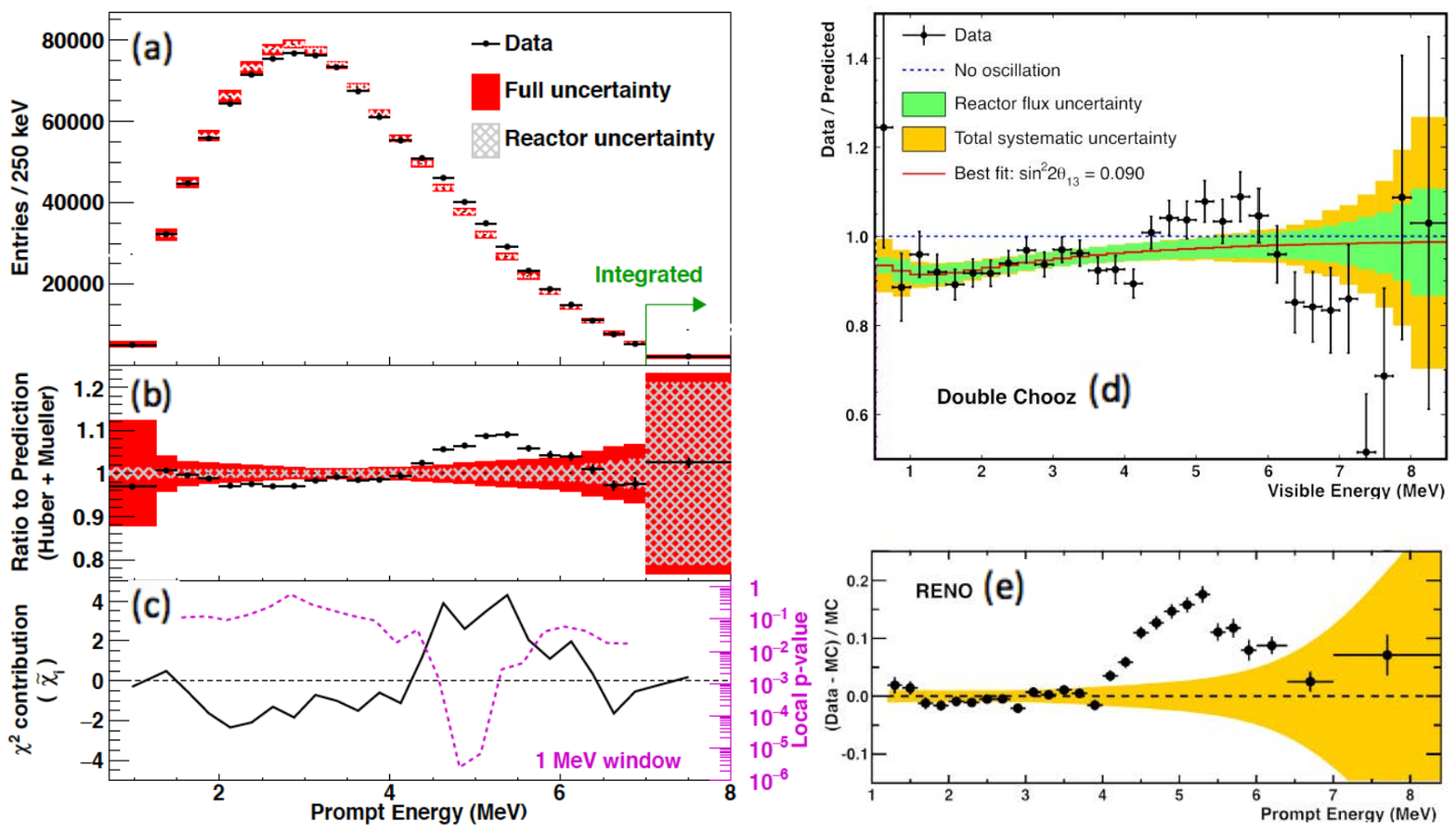}
  \caption{
Measured prompt energy spectra and comparison to model predictions of antineutrino emission from pressurized water reactors (PWR) for kilometer-baseline experiments. (a-c): near detector Daya Bay~\cite{An:2015nua} (The oscillated prediction is normalized to the observed number of events in the entire energy range). (d): far detector Double Chooz~\cite{Abe:2014bwa} (The un-oscillated prediction is normalized to the observed number of events in the entire energy range). (e): near detector RENO~\cite{,Seo:2016uom} The oscillated prediction is normalized to the observed number of events in the energy range E$<$3.6 MeV).}
   \label{fig:spectrumAnomaly}
\end{figure*}

Recent neutrino experiments have provided a coherent picture of neutrino flavor change and mixing, and allowed the precise determination of oscillation parameters in the 3-neutrino model. However, anomalous results in the measurement of the reactor \nuebar{} flux and spectrum have suggested this picture is incomplete and may be interpreted as indicators of new physics.  
Reactor \nuebar{} experiments (Fig.~\ref{fig:reactorAnomaly}) 
observe a $\sim$6\,\% deficit in the absolute flux when compared to predictions~\cite{Mueller:2011nm,Huber:2011wv}. The observed flux deficit, the ``reactor antineutrino anomaly'', has led to the hypothesis of oscillations involving a sterile neutrino state with $\sim$1~eV$^2$ mass splitting~\cite{Mention:2011rk,Abazajian:2012ys,Kopp:2013vaa}. Moreover, measurements of the reactor \nuebar{} spectrum by recent $\theta_{13}$ experiments (Daya Bay, RENO, Double Chooz) observe spectral discrepancies compared to predictions, particularly at \nuebar{} energies of 5-7~MeV~\cite{An:2015nua,Abe:2014bwa,Seo:2016uom}(Fig.~\ref{fig:spectrumAnomaly}), 
possibly indicating deficiencies in current prediction methods and/or the nuclear data underlying them.  The reactor anomaly and the measured spectral discrepancies are open issues in a suite of anomalous results~\cite{Abazajian:2012ys} that may hint at revolutionary new physics in the neutrino sector.
Observation of an eV-scale sterile neutrino would have a profound impact on our understanding of neutrino physics and the Standard Model of particle physics with wide-ranging implications for the physics reach of the planned US long-baseline experiment DUNE~\cite{Gandhi:2015xza}, searches for neutrinoless double beta decay, neutrino mass constraints from cosmology and beyond.

The Precision Reactor Oscillation and Spectrum Experiment, PROSPECT~\cite{Ashenfelter:2015uxt}, is designed to comprehensively address this situation by making a search for \nuebar{} oscillations at short baselines from a compact reactor core while concurrently making the world's most precise \nuebar{} energy spectrum measurement from a highly-enriched uranium (HEU) research reactor. In particular, a first-ever precision measurement of the $^{235}$U spectrum would highly constrain predictions for a  static single fissile isotope system ($> 99$\% $^{235}$U) as compared to commercial power reactors that have evolving fuel mixtures of multiple fissile isotopes ($^{235}$U fission fraction typically changes from $\approx 73$\% to $\approx 45$\% during a reactor cycle).  
Simultaneously measuring the relative \nuebar{} flux and spectrum at multiple distances from the core within the same detector provides a method independent of any reactor model prediction for  PROSPECT to probe for oscillations into additional neutrino states in the parameter space favored by reactor and radioactive source experiments~\cite{Kopp:2013vaa}. 

In addition to directly addressing the sterile neutrino interpretation of the reactor anomaly~\cite{Ashenfelter:2018osc}, PROSPECT can also provide new experimental data to test for deficiencies in reactor \nuebar{} flux predictions. 
By making a high-resolution energy spectrum measurement, PROSPECT will  determine if the observed spectral deviations in Daya Bay and other $\theta_{13}$ experiments at commercial nuclear power plants persist in a HEU fueled research reactor and provide a precision benchmark spectrum to test and constrain the modeling of reactor \nuebar{} production. 
A better understanding of the reactor \nuebar{} spectrum will aid precision medium-baseline reactor experiments such as JUNO \cite{Capozzi:2015bpa} and improve reactor monitoring capabilities for nonproliferation and safeguards.

The goals of the \pspt{} experiment are to:
\begin{itemize}
\item Make an unambiguous discovery of eV-scale sterile neutrinos through the observation of energy and baseline dependent oscillation effects, or exclude the existence of this particle in the allowed parameter region with high significance. Accomplishing this addresses the proposed sterile neutrino explanation of the reactor anomaly using a method that is independent of reactor flux predictions;
\item Directly test reactor antineutrino spectrum predictions using a well-understood reactor  dominated by fission of $^{235}$U, while also providing information that is complementary to nuclear data measurement efforts;
\item Demonstrate techniques for antineutrino detection on the surface with little overburden;
\item Develop technology for use in nonproliferation applications.
\end{itemize}

 PROSPECT is located at the \HFIR{}~\cite{HFIR} at \ORNL\ and consists of a 3760 liter, segmented $^6$Li-doped liquid scintillator antineutrino detector accessing baselines in the range 7~m to 13~m from the reactor core.   
PROSPECT combines competitive exposure, baseline mobility for increased physics reach and systematic checks, good energy and position resolution, and efficient background discrimination.
PROSPECT has already demonstrated  a signal over correlated background ratio of $\gtrsim 1:1$~\cite{Ashenfelter:2018osc} and set new limits on sterile neutrino oscillations 
based on its first 33 days of reactor operation.
Within a single calendar year, \pspt{} can probe the best-fit region for all current global analyses of \nue{} and \nuebar{} disappearance~\cite{Abazajian:2012ys,Kopp:2013vaa} at 4\,$\sigma$ confidence level. Over 3 years of operation, \pspt{} can discover oscillations as a sign of sterile neutrinos with a significance of 5\,$\sigma$ for the best-fit point and $>3$\,$\sigma$ over the majority of the suggested parameter space.

\section{Nuclear reactor antineutrinos}

\subsection{Antineutrino flux and spectrum}

Neutron-rich isotopes produced from fission processes within power reactors undergo a series of decays as shown in equation \ref{eqn:betadecay}, producing approximately six antineutrinos per fission.  

\begin{equation}
^{A}_{Z}X \to _{Z+1}^{A}Y+\beta^{-} + \nuebar
\label{eqn:betadecay}
\end{equation}

\noindent The mixture of isotopes produced is complex, leading to a continuous spectrum of electron flavored antineutrinos with energies primarily between 0~MeV and 8~MeV.  
Given the generally short half-life of the fission by-products, the flux of antineutrinos is proportional to the thermal power of the reactor core.   
A variety of methods have been used over many decades to calculate the $\overline{\nu}_e$ flux and spectrum.  
As early as 1948, statistical modeling of known nuclear physics was used to estimate the expected flux~\cite{Way:1948zz}.  
Over the years, tabulation of careful experimental measurements of isotope yields and isotope decay schemes lead to the summation or 
\begin{itshape} 
ab initio 
\end{itshape} approach~\cite{Avignone:1980qg,Klapdor:1982sf}.   
Incorporating precision studies of the beta spectra from fission by-products (beta conversion method~\cite{Vogel:2007du}) resulted in more precise estimates.  
However, given that thousands of beta-branches contribute to the observed spectrum, these calculations remained challenging.  
In recent years, new techniques and methods~\cite{Mueller:2011nm,Huber:2011wv} have produced tension with previous calculations.  

\subsection{The High Flux Isotope Reactor (HFIR)}

\begin{figure}
   \centering
      \includegraphics[width=0.4\textwidth]{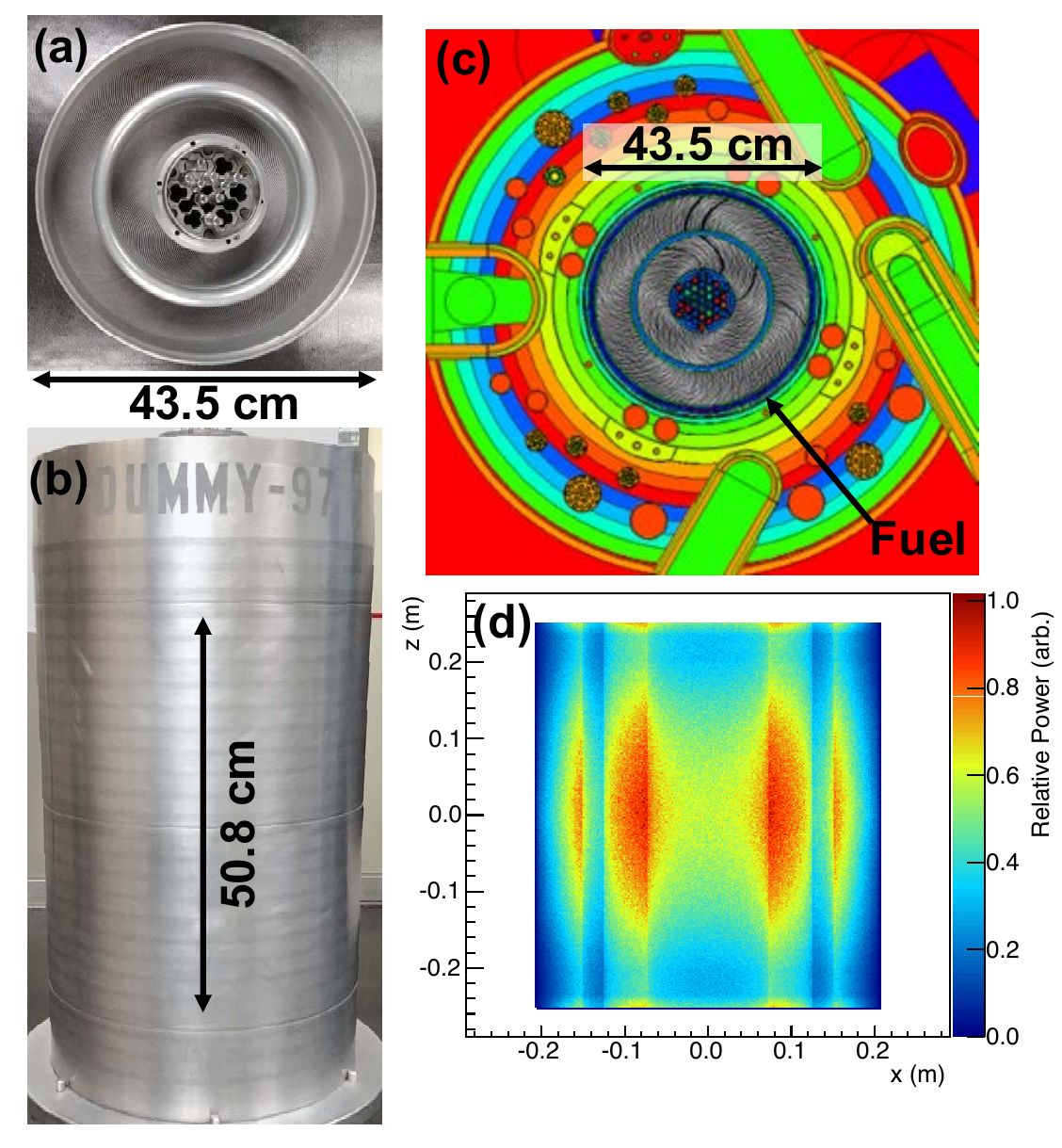}
   \caption{Photographs of a dummy HFIR fuel element with active fuel diameter of $0.435$~m and length of $0.508$~m are shown in (a) \& (b). 
Colors in (c) represent different components of the Monte Carlo N-Particle~\cite{MCNP} (MCNP) model of the HFIR core~\cite{HFIR_model}.
A projection of the cylindrically symmetric core fission power density (i.e. antineutrino production source term) onto the x-z plane is shown in (d).}
   \label{fig:HFIR}
\end{figure}

HFIR is a compact research reactor located at ORNL, and is described in great detail elsewhere~\cite{HFIR_model}.  
It burns highly enriched uranium fuel ($^{235}$U), and was designed primarily to support neutron scattering and radiation damage experiments, trace element detection, and the production of radioactive isotopes for medical and industrial purposes.   
Operating at 85~MW, HFIR is also a steady and reliable source of antineutrinos with minimal fuel evolution ($>99$\,\% of fissions are from $^{235}$U throughout each cycle).  
As seen in  Fig.~\ref{fig:HFIR} the HFIR core has two cylindrical fuel elements with the outer element having a diameter of 0.435~m and a height of 0.508~m. 
The HFIR facility typically operates  seven 24-day  cycles per year for a duty cycle (Reactor On) of $\sim$\,46\,\%.
The entire fuel assembly is replaced after each  cycle.
Reactor Off data can be used to accurately measure backgrounds from coincident cosmogenic sources during Reactor On data.

\subsection{Antineutrino detection}

Antineutrinos with energy $\geq1.8$\,MeV are detected  via the inverse beta-decay (IBD) reaction on protons in the liquid scintillating target:
\begin{equation}
\nuebar + p \to e^+ + n
\label{eqn:IBD}
\end{equation}
\noindent The positron carries most of the antineutrino energy  and rapidly annihilates with an electron 
producing a prompt signal with energy ranging from 1~MeV to 8~MeV.
The neutron, after thermalizing, captures on a $^6$Li or H nucleus, 
with a typical capture time of 40\,$\mu s$. 
The correlation in time and space between the prompt and delayed signals provides a 
distinctive $\bar{\nu}_e$ signature, greatly suppressing backgrounds. 

Liquid scintillators have historically been the standard detection medium for large volume antineutrino detectors. Gadolinium has often been used for the neutron capture signal in large, monolithic detectors~\cite{An:2015nua,Abe:2014bwa,Seo:2016uom}, emitting a robust 8\,MeV signal in $\g$-rays. However, for a smaller (few ton) highly segmented detector such as PROSPECT, the spatial extent of the $\g$-ray signal compromises segmentation. Furthermore, the $\g$-rays  will escape detection near the sides of the detector, leading to a spatial dependence of detection efficiency.
Additionally, since PROSPECT will operate in a high-$\g$-ray background environment, the $\g$-rays from the neutron capture on gadolinium could be  mimicked by random coincidences of the predominant $\g$-ray backgrounds.

In contrast, neutron captures on $^6$Li produce well localized energy depositions\footnote{The very high energy deposition density from low energy nuclear fragments or proton recoils, suppresses the light output in liquid scintillator. For this reason, we refer to energies observed in such reactions in terms of their ``electron equivalent'', or ``ee''.} from the reaction
 $n + ^6$Li$ \rightarrow \alpha + t + 0.55$~MeV$_{ee}$ which are most often contained within a single segment of a divided detector.
Since this capture only produces heavy charged particles, a pulse-shape discriminating LiLS is able to separate neutron captures from 
background $\g$-ray events reducing the likelihood of random coincidences.

Pulse-shape discrimination (PSD) is a long studied property of many liquid scintillators that allows for the isolation of interactions with high $dE/dx$, typically heavy charged particles, from those with low $dE/dx$, such as muons and electrons. 
Previous experiments using LiLS were based on scintillators that are toxic, flammable, and are not suitable for operating inside a reactor facility.
Also many of these scintillators have had insufficient light yields for realizing the 
energy resolution needed by PROSPECT.
A multi-year research and development effort by PROSPECT collaborators 
developed a new low-toxicity and low-flashpoint liquid scintillator utilizing a  commercial scintillator base
(Section~\ref{sec-scintillator}).

\section{PROSPECT goals and design concept }
\label{sec-concept}

\subsection{Goals}
Previous optimization studies of short baseline antineutrino detectors~\cite{Heeger:2012tc}
identified as key parameters:  an energy resolution of $\le\!10\%/\sqrt{E(\rm MeV)} $ (RMS), a position resolution
$\le\!0.20$ m, a signal to background ratio better than 1:1, a mass of a few tons and a baseline coverage of about 3 m. 
A segmented liquid scintillator detector utilizing $^6$Li to identify the neutrons from the IBD interaction and having  good PSD to separate signals from $\g$-rays, electrons and other minimum ionization background signals from hadronic particles can meet these goals.
The modularity improves background suppression by allowing spatial correlation of the prompt and delayed signals while  
naturally dividing the data into bins of known position and size.
The non-scintillator material defining the segments should be minimized  to achieve an acceptable energy response  
for accurate measurement of the antineutrino energy spectrum.

Multiple calibration methods are needed to establish the efficiency as well as the energy and time response of the detector to IBD interactions. 
The PROSPECT detector design  should allow 
the insertion of radioactive sources or optical pulses into the active detector volume as needed. 
Radioactive sources   such as
${}^{137}{\rm Cs}$ or ${}^{60}{\rm Co}$ are needed to establish the overall energy scale. 
Positron annihilation $\g$-rays such as
${}^{68}{\rm Ge}$ or ${}^{22}{\rm Na}$ can establish the detector response and detection efficiency 
to positrons from IBD events.  
A neutron source such as ${}^{\rm 252}{\rm Cf}$ is needed to determine the IBD neutron detection efficiency. 
Signals from background radioactivity in the LiLS 
should also be used to track performance over time.

\subsection{Shielding design studies} 
\label{sec-shielddesign}
PROSPECT operates on the Earth's surface with 
$<1$~m  overburden and is within 7~m of a nuclear reactor core.
Singles rates from $\g$-rays or neutrons from the reactor  or cosmogenic sources exceed those from antineutrino interactions  by $> 10^{7}$.
Background to PROSPECT antineutrino detection by IBD falls into two categories: single energy deposits, 
mainly due to $\g$-rays entering the detector, and coincident energy deposits largely from the recoil and capture of fast neutrons. 
The former needs to be suppressed  to limit the data acquisition rate and minimize IBD backgrounds due to accidental coincidences.
The latter is more pernicious as it closely mimics the IBD signal. 

Neutron and $\g$-ray background measurements 
performed at HFIR~\cite{Ashenfelter:2015tpm} found multiple $\g$-ray background sources  associated with penetrations  in 
the reactor pool shielding wall.  
Backgrounds were much lower over the many-meters-thick solid concrete monolith which supports most of PROSPECT in the shortest baseline position. 
Diffuse background rates rose next to the base of the pool wall at the front of the detector and over the floor at the back of the detector.

Single segment detector prototypes were run at HFIR~\cite{Ashenfelter:2015uxt}  with different shielding configurations to test the layered shielding approach. 
Layers of water, polyethylene, borated polyethylene (BPE), and 0.05~m to 0.1~m of lead suppressed
reactor associated $\g$-ray and neutron  backgrounds  
sufficiently to minimize random IBD-like coincidences, leaving a coincident background that was cosmogenic in origin.
These time correlated backgrounds were attributed to the interactions of energetic cosmic ray neutrons or neutron showers in the shielding close to the active detector.
Extrapolating this single segment data to a full size detector through background simulations revealed two important insights.
Keeping the lead thickness of 0.05\,m to 0.1\,m for a full size detector was untenable
due to weight limitations. 
Using the outermost active detector layer to veto cosmogenic neutron 
interactions in an inner ``fiducial" volume could reduce coincident backgrounds below the rate expected from IBD interactions.

Since most of the $\gamma$-ray  backgrounds originated in the reactor pool wall, the shielding design was split into a fixed lead wall mounted close to the $\gamma$-ray  sources
(local shield wall, Section~\ref{sec-fixed}) and a shielding package that surrounded the detector volume and moved with it during baseline moves (passive shielding, Section~\ref{sec-passive}). 
The local shield wall was less constrained in total weight, allowing thicknesses from 0.05~m to as much as 0.2~m of lead in certain locations. 
The passive shielding design contained a single 0.025~m hermetic lead layer surrounded by  layers of polyethylene, borated polyethylene, and water
to mitigate the cosmogenic backgrounds.

\begin{figure}
\centering
\includegraphics[width=3.4in]{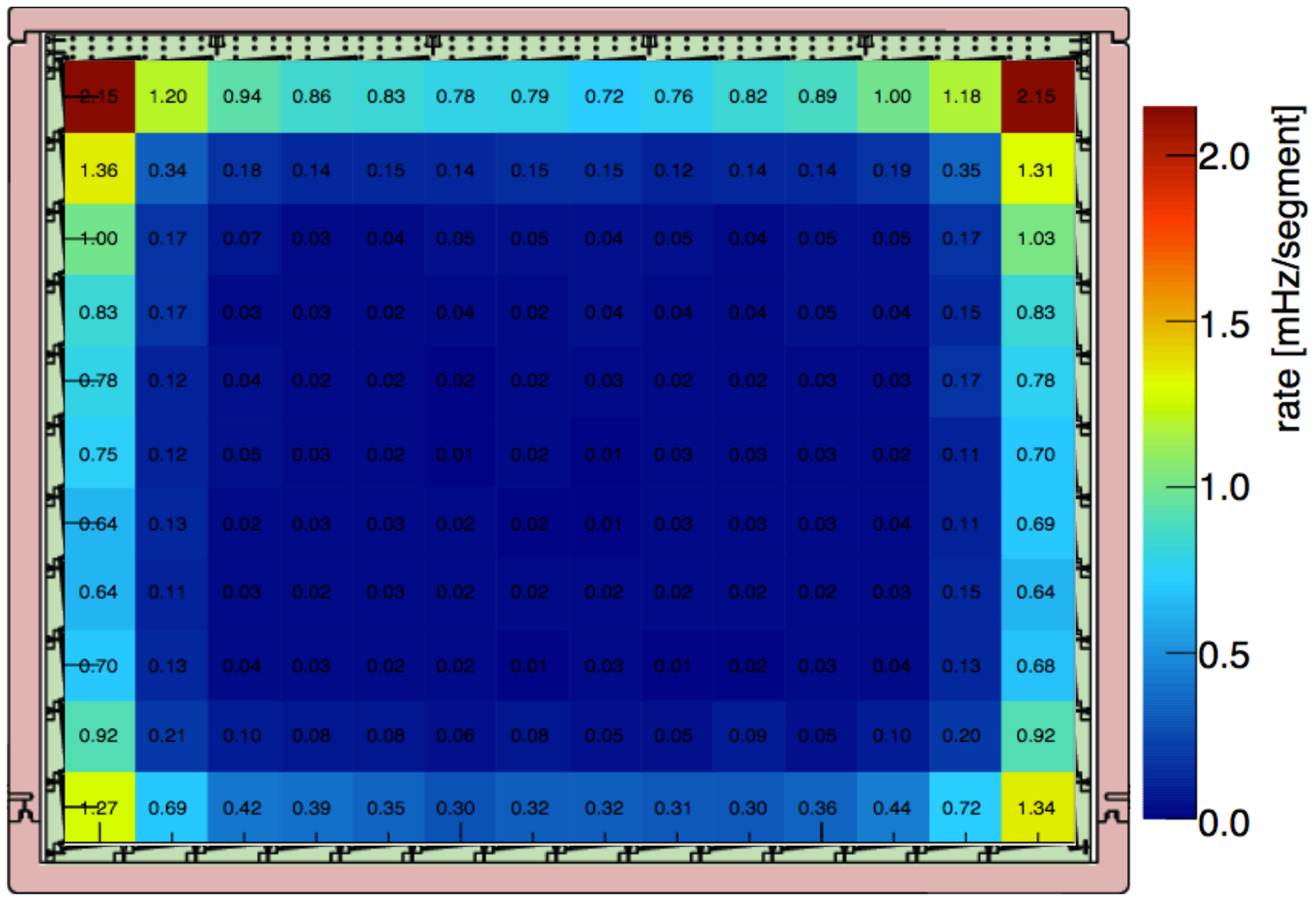}
\caption{
Simulated background rate of cosmogenic neutron interactions that mimic the IBD signal after topology cuts and segment-end fiducialization. The background rate in the outermost ring of segments (rows 1 and 11, columns 1 and 14) is considerably higher than in the fiducial volume used in analysis (rows 2-10, columns 2-13). Surrounding the segments is the acrylic support structure and the acrylic containment tank of the inner detector. 
}
\label{fig-corr}
\end{figure}

Background simulations  of IBD-like events from cosmogenic background sources with the above shielding are shown in Fig.~\ref{fig-corr}.
Analysis topology cuts vetoed events with  extra  energy deposits not associated with  the  segments containing the positron and neutron signals. 
These cuts lose
effectiveness near the edge of the detector as information of background neutron scatters is lost.  
The expected rate of IBD backgrounds  in the outermost segments is 10-100 times that of the innermost segments. 
Requiring that the accepted IBD events  originate in an inner
"fiducial"  region (removing the outermost segments and ends of each segment close to the photomultipliers (PMTs) lowers the expected background rate
below the IBD signal rate. 
Thus the conventional passive shielding elements discussed above are augmented by a layer of active shielding  
that is very effective in identifying background events.
 
During reactor operation, the thermal neutron rate in the experimental room was measured to be 
$\sim$2$/{\rm cm}^2/{\rm s}$~\cite{Ashenfelter:2015tpm}. 
For PROSPECT, thermal neutrons can cause singles from $\gamma$-rays emitted from neutron captures on materials near the detector. 
This source of singles can be suppressed by a hermetic enclosure rich in ${}^{10}{\rm B}$ which has a large thermal neutron cross-section and minimal gamma emission. 
PROSPECT used this guidance for background suppression within the weight and height constraints of the HFIR site, described in Section~\ref{Design-constraints}, to design the shielding described in Section~\ref{sec-passive}. 

\subsection{Achieved parameters}
\label{sec-parameters}
\begin{figure*}
\centering
\includegraphics[clip=true, trim=0mm 50mm 0mm 40mm,width=0.99\textwidth]{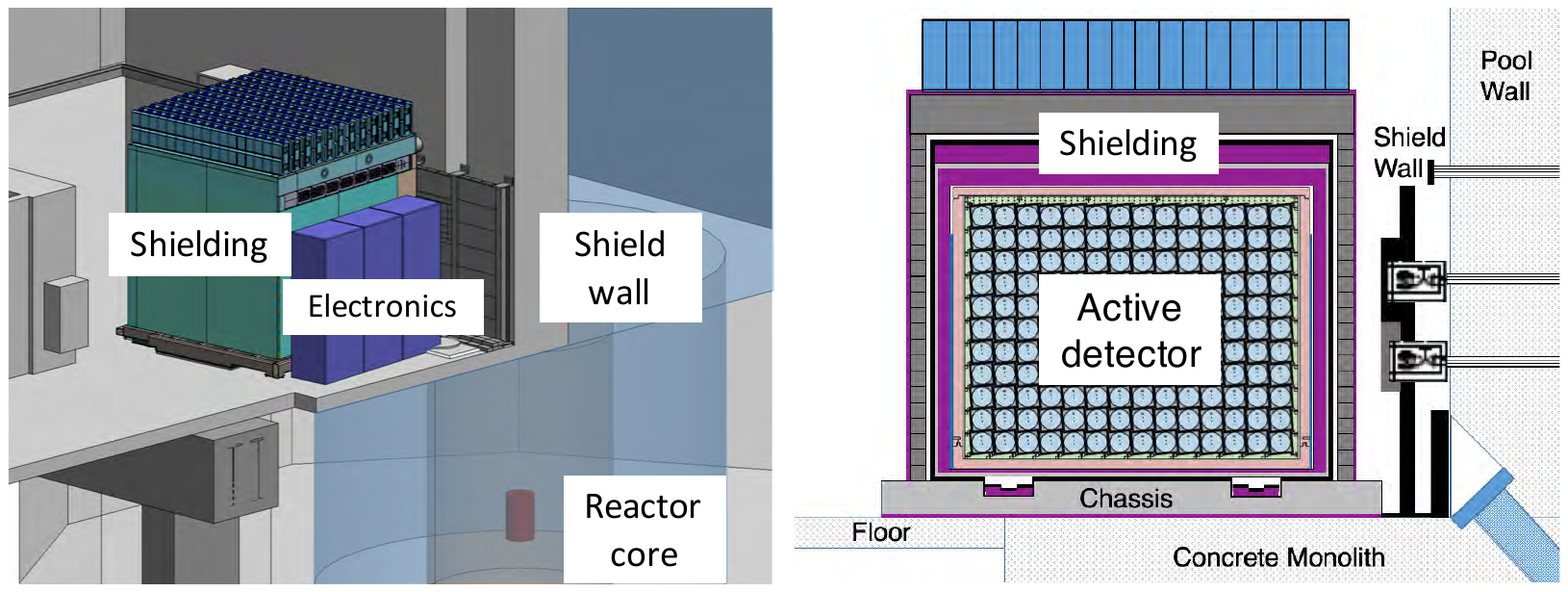}
\caption{
(left)  Layout of the PROSPECT experiment.  
The detector is installed in the HFIR Experiment Room next to the water pool and 5\,m above the HFIR reactor core (red). The floor below contains multiple neutron beam-lines  and  scattering experiments.
(Right) Schematic showing the active detector volume divided into 14 (long) by 11 (tall) separate segments and surrounded by nested containment vessels  and shielding layers.
Shield walls cover penetrations in the pool wall associated with high backgrounds.
}
\label{fig-layout}
\end{figure*}

The layout of the experiment at HFIR is shown in Fig.~\ref{fig-layout}.
Detector parameters are:
\begin{enumerate}
\item Active LiLS volume 1.176\,m wide $\times$ 2.045\,m long  $\times$ 1.607\,m tall, 3760\,liters, 3.68\,metric tons. 
\item Segmentation 14 (long) by 11 (tall). Square segment cross-section of 0.145~m by 0.145~m.
\item Reconstructed $z$-position resolution (along the length of the segment) 0.05\,m.
\item Center of the reactor core to center of the detector at the nearest position $7.93\pm 0.1$\,m. Detector movement to  baselines of 9.1 and 12.4~m possible (shown in Fig.~\ref{fig-baselines}).
\item Baseline coverage $\pm1$\,m for a single position.
\item Energy resolution of 4.5\,\% (RMS) at 1~MeV.
\item Fraction of non-LiLS mass in the target region 3.4\,\%.
\end{enumerate}

\section{ Experimental facility }
\label{sec-facility}

\subsection{Overview}
PROSPECT is installed in the HFIR Experiment Room at ground level, one floor above the HFIR
core and containment vessel as shown in  Fig.~\ref{fig-layout}.  
A one-meter-thick concrete wall  separates the  room from the reactor water pool. 
The nominal water level  in the pool is 3.1~m above the detector center.
Part of  the detector rests on a  solid, polygonal shaped, concrete monolith surrounding and supporting the reactor pool and structure. 
The rest of the detector is supported by a 0.15-m-thick steel reinforced  concrete  floor over 
a large room containing multiple thermal neutron scattering experiments and  cold neutron beam-lines. 
A  0.20-m-thick steel reinforced  concrete roof is 5.5~m above the detector center.

\subsection{Design constraints}
\label{Design-constraints}
\begin{figure*}
\centering
{\includegraphics[clip=true, trim=0mm 50mm 0mm 50mm, width=6.2in]{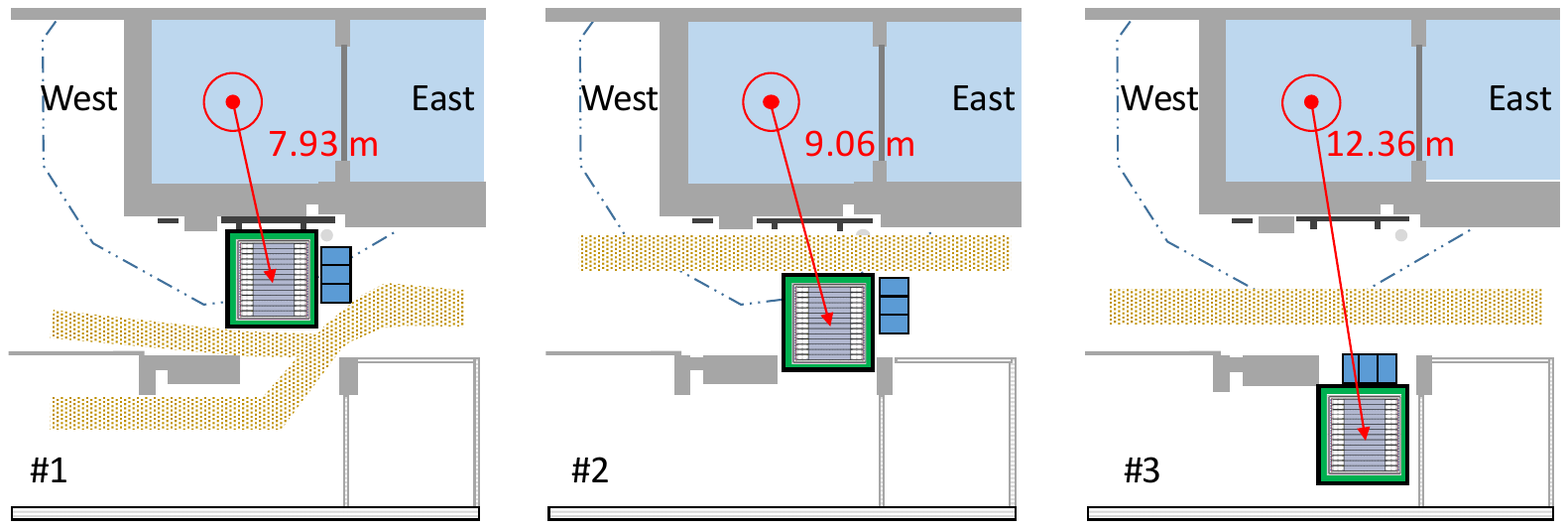}}
\caption{
 Plan view of PROSPECT detector locations in the HFIR Experiment Room. The detector is initially installed in Position 1 at an estimated baseline (final survey pending) of  ($7.93\pm 0.1$)~m from the center of the reactor core to the center of the active detector.
Moves to Position 2 (9.06~m) or Position 3 (12.36~m) are planned.  The chassis footprint (green) and inner detector are shown. 
Electronics racks (dark blue), reactor water pool (light blue) and  reactor vessel and core (red) are also shown.  
A dashed line shows the shape of the underlying concrete monolith. 
Required walkways and clearances that limit possible positions are also shown in beige.}
\label{fig-baselines}
\end{figure*}

Detector size, weight, and position were significantly constrained by safety considerations and the geometric limitations
of the experiment room. 
A maximum floor loading of 3670~kg/m$^2$ (750~lb/sq. ft) was imposed on the detector plus passive shielding.
The detector footprint was limited by the need to maintain adequate walkways past the detector for access to other HFIR facilities
and to allow the detector to be moved to alternate baselines. 
A simplified layout of  detector positions at HFIR is shown in Fig.~\ref{fig-baselines}.

The door into the experiment room limited the width of large items to be less than 2.95~m. 
Overhead piping and lighting limited the height as well.  In addition, doors to other experimental apparatus in the room  could not be occluded. To satisfy these criteria the detector plus passive shielding envelope
was required to be less than 2.95~m (wide) by 3.25~m  (long) by  3.25~m (tall) and to weigh less than 34,090~kg.

To maximize the size of the active detector within the above constraints, 
detector segments are installed parallel to the reactor wall as seen in Fig.~\ref{fig-baselines}. 
As a result  every detector segment  contains a small range of baselines and has an expected rate asymmetry from one end to the other. The effect is quite small as the expected flux asymmetry between the ends of the closest segment is 0.43\,\%.

\subsection{Baselines}
Three possible baseline positions are possible, in order to optimize the sterile neutrino search sensitivity.  
Figure~\ref{fig-baselines} shows the near(1) and proposed middle(2) and far(3) positions.
The detector is initially installed in position 1.
The average baseline can be increased from 7.93~m to 12.36~m  by moving from the near to far position. 
Only the  orientation of the electronic racks changes with position.  

\subsection{Fixed local shielding}
\label{sec-fixed}
\begin{figure}
   \centering
      \includegraphics[width=0.49\textwidth]{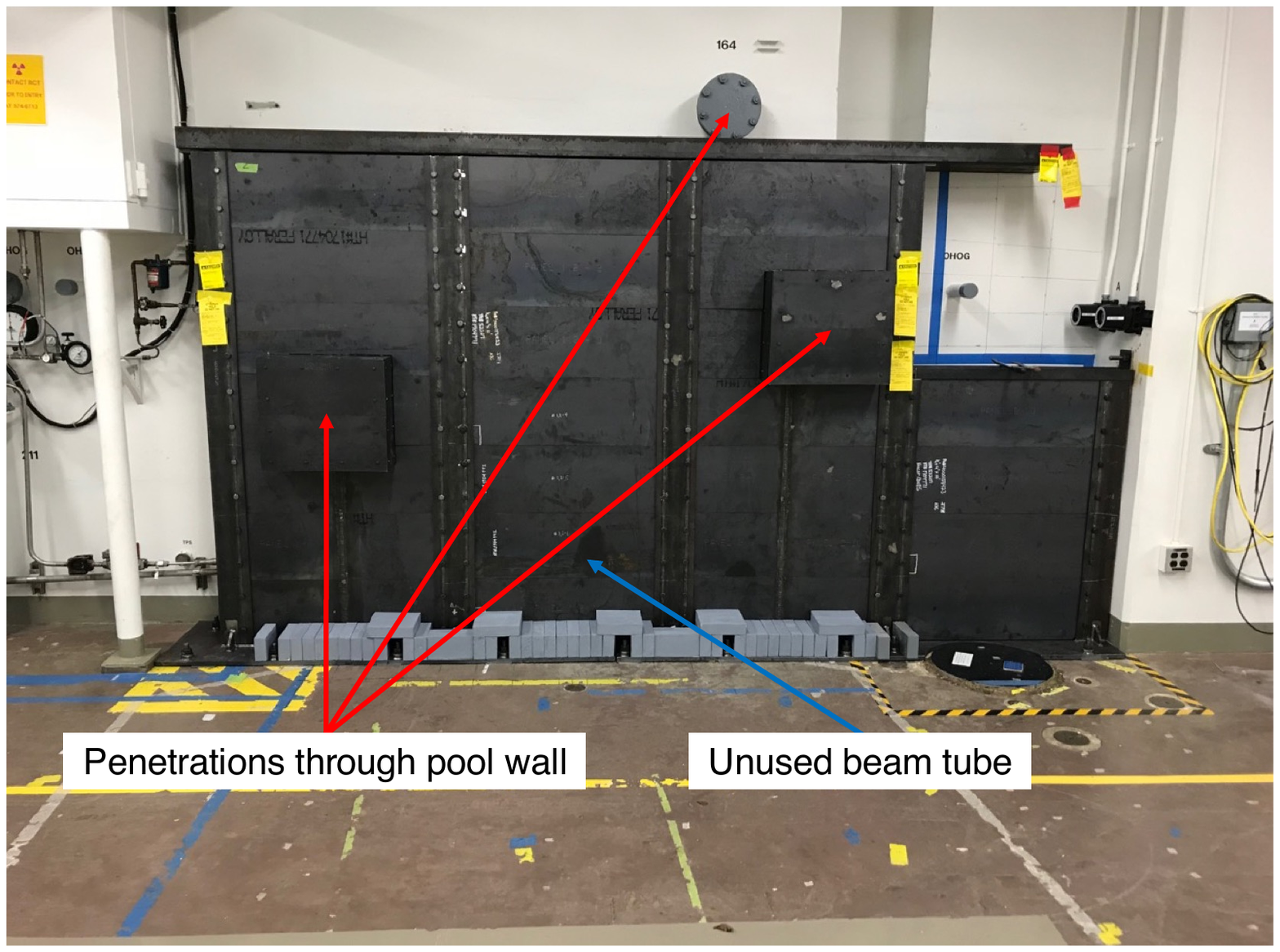}
   \caption{Photograph of the local shield wall. Red arrows mark the location of pipes penetrating to the reactor pool. A blue arrow marks the location of the unused EF-4 beam line that points directly to the reactor vessel.  The tall portion sections of the wall contain 100~mm of lead. 
    }
   \label{fig:Wall}
\end{figure}

The concrete wall between the reactor and detector is penetrated by several pipes and unused beam lines. Each is a potential background source during reactor operation. Scans with a NaI(Tl) crystal~\cite{Ashenfelter:2015tpm,Heffron:2017,Hackett:2017} identified  the most significant sources.  The largest $\gamma$-ray  source was the EF-4 beam line directly in front of the detector. Although plugged by a concrete-filled pipe, the EF4 region is a thin spot in the shielding. 
As mentioned in Section~\ref{sec-shielddesign}, a lead filled shielding wall (shown in Fig.~\ref{fig:Wall})
was installed close to the concrete pool wall to eliminate backgrounds from these sources. The central part of the wall is 3.0~m wide and 2.1~m tall.  Shorter flanking walls on each side completed the design. Protective cages were installed around two of the pipes penetrating the wall. The lead thickness in the central part of the wall was typically 0.10~m. The far left and right hand sections  were 0.05~m thick.  A stand alone mini-wall 0.10~m thick was added between the local shield wall and the EF4 opening to provide additional suppression of this source. Steel supports for the wall  were sturdy and robust and designed to withstand seismic loads as required by safety codes.

\section{Detector}  
\label{sec-AD} 

\subsection{Summary} 
\label{summary}
\begin{figure*}
\centering
\includegraphics[width=6.2in]{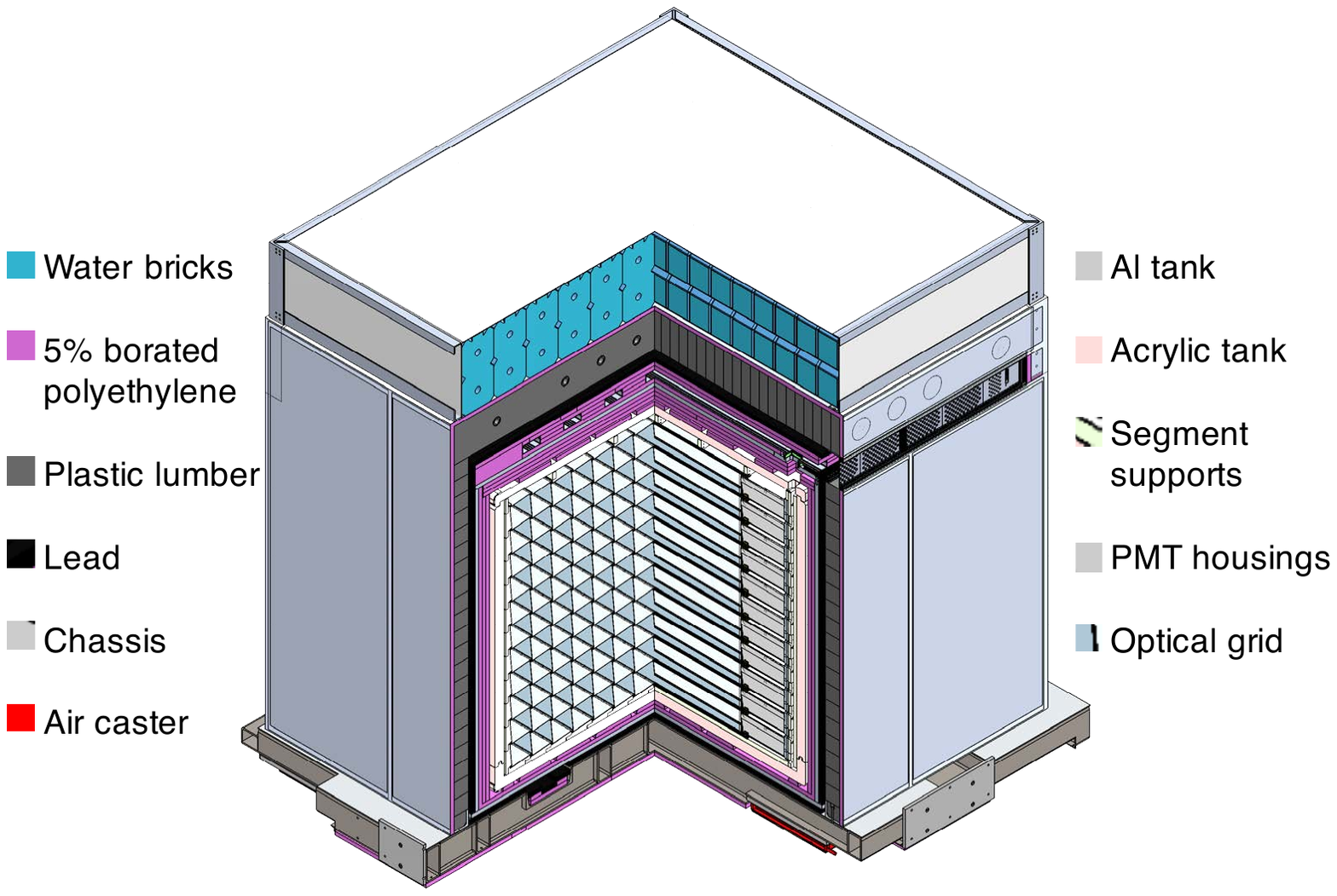}
\caption{
A cutaway view of the 3D detector and shielding assembly model. The inner detector, inside the acrylic tank (rose), is segmented into an eleven by fourteen grid by reflective optical separators. The active detector is defined as the LiLS filled portion of the optical grid viewed by PMT housings (beige) on either end. The housings and grid are supported by acrylic segment supports (light green). The acrylic tank is surrounded by borated polyethylene (purple) and a secondary aluminum  tank (light gray). More details are shown in Figs. 9-14.
}
\label{fig-side}
\end{figure*}

The PROSPECT detector  shown in Fig.~\ref{fig-side} consists of an inner detector filled with LiLS, inner and outer containment vessels (tanks), shielding and detector  movement
elements, and data acquisition (DAQ) and control electronics housed in three electronic racks.  
All components within the acrylic inner vessel were  tested for compatibility with the LiLS. 
The active LS volume is divided into 14 by 11 segments  by reflective optical separators held together at the edges by 3D printed hollow plastic rods.
Segments are parallel to the reactor pool wall on the north side of the detector.
Each segment is viewed on the east and west ends by PMTs enclosed  in acrylic housings.
The housings are several mm smaller in cross-section than the optical segments to allow LS or gas to flow into or out of each segment volume during the filling procedure.
The housings support the corner rods and define the segment geometry.
Selected  rods contain tubes for the insertion of radioactive sources into the active volume.
Other rods contain  optical diffusers midway along the segment length coupled to the optical calibration system. 
Acrylic segment supports tie the housings together and support the outermost optical separators and corner rods.
The detector was transported while dry to ORNL and filled onsite. The top layer of optical separators is covered by 
a few cm of LiLS.  
An expansion volume filled with  nitrogen  cover gas fills the remaining space inside the acrylic vessel providing room
for volume changes with temperature.

The inner detector has several unique design features:
\begin{itemize}
\item{A $^6$Li doped liquid scintillator that provides a very localized energy deposition from the neutron capture which is easily separated from $\gamma$-ray backgrounds of similar energy.
The high light yield and transparency produce an energy resolution of approximately 4.5\,\% at 1 MeV. }
\item{A reflective grid separates the active volume into 154 segments   of uniform volume.
Neighboring segments share optical separators made of a low-mass carbon fiber core covered by  laminated reflective and fluorinated ethylene propylene (FEP) film. }
\item{A tessellated segment structure that minimizes non-reflective surfaces in the optical volume while providing access for multiple optical or radioactive calibration sources. }
\item{Cross talk between segments of less than 1\,\%. The optical separators have an opaque carbon fiber core preventing transmission through the optical separator. The front windows of the PMT housings protrude $\approx1$\,cm into the optical grid, minimizing light transmission between segments. }
\item{PMTs inside the LiLS. The PMTs are mounted inside acrylic housings filled with mineral oil. Low cost conical reflectors in the MO improve the light collection efficiency
in the corners. Gaps between housings are filled with LiLS.  The mineral oil and LiLS provide a low background buffer on both ends of the segment structure. }
\end{itemize}

A series of nested, nearly hermetic shielding and structural layers surround the inner detector. 
From the inside to outside,  the active segments are surrounded on the sides by
the segment support structure, a 0.063~m thick acrylic tank wall, a mixed layer of 0.025~m water or borated polyethylene, 0.025~m to 0.075~m of borated polyethylene shielding, 
a 0.025~m thick outer aluminum tank wall, a 0.025~m layer of lead, 
0.10~m of structural polyethylene timbers, 0.025~m of borated polyethylene shielding, and an outer aluminum covering.
As seen in Fig.~\ref{fig-side} the order of materials from bottom to top is similar, but with less shielding below and more shielding above to combat cosmogenic backgrounds.

\subsection{Lithium loaded liquid scintillator} 
\label{sec-scintillator}

The conceptual design of the PROSPECT detector (AD) required a liquid scintillator (LS) with both very good PSD for background rejection of fast neutron and ambient $\gamma$-ray  background (i.e. better than the linear alkylbenzene used in Daya Bay or RENO experiments) and high light yield for energy resolution. The compactness of the AD as well as the length-scale of the segmentation strongly preferred doping with a neutron capture agent yielding only charged particles and thus a topologically compact capture signature. Furthermore, a low-toxic, non-flammable formulation was needed to support ease of deployment within the HFIR reactor building.  Based on several prototyping studies, a light yield better than 8000 optical photons per MeV was determined to meet energy resolution requirements.  Though there exist certain challenges related to chemistry, doping with $^6$Li yields an $\alpha$ and a $^3$H with a Q-value of 4.78~MeV ($0.55$~MeV$_{ee}$), 
providing an ideal compact mono-energetic signal.

To meet these requirements, the PROSPECT collaboration developed a novel lithium-doped liquid scintillator (LiLS) formulation based on a commercially available product.  Doping of up to 0.2\,\% $^6$Li by mass is supported by the addition of a surfactant to the base LS.  The surfactant in combination with an aqueous $^6$LiCl solution forms a thermodynamically stable microemulsion, ensuring material uniformity.  This approach also allows the addition of radionuclide solutions for calibration purposes as described in Section~\ref{sec:IntrinsicSources}. In practice the doping fraction is an optimization of cost and reduced capture time (background rejection) and the final LS was doped to 0.08\,\% $^6$Li.
The mass fraction of carbon and hydrogen content were determined  from combustion analysis as C( $84.34 \pm 0.11$ \%) and  H($9.69 \pm 0.21$ \%).

The LiLS was manufactured at the Brookhaven National Laboratory (BNL) from commercial chemicals. 
LiLS consists of a nonionic surfactant, 10~mol/L aqueous $^6$Li chloride, 2,5-diphenyloxazole (PPO) and 1,4-bis(2-methylstyryl)benzene (bis-MSB) in a commercial, di-isopropylnapthalene (DIN)-based scintillator (EJ-309\footnote{https://eljentechnology.com/products/liquid-scintillators/ej-301-ej-309. Certain trade names and company products are mentioned in the text or identified in illustrations in order to adequately specify the experimental procedure and equipment used. In no case does such identification imply recommendation or endorsement by the National Institute of Standards and Technology, nor does it imply that the products are necessarily the best available for the purpose.}).  The surfactant is an ether-based glycol. The $^6$LiCl was purified and supplied by the National Institute of Standards and Technology (NIST) from enriched lithium carbonate material produced at ORNL. The PPO and bis-MSB were obtained from Research Product International\footnote{https://www.rpicorp.com/}. 
The LiLS density is $0.9781\pm0.0008$\,g/cc.

PROSPECT plans to run for four years making long-term LS stability a priority.  To this end, the collaboration carried out comprehensive material compatibility and stability studies.  All materials considered for use in the inner detector and that were to be in contact with LiLS were soaked in samples of LiLS for extended periods.  Ultra-violet (UV)-vis emission and transmission spectra of the LiLS over the wavelength range 260~nm to 850~nm were periodically compared against reference LS samples.  
Typically, changes were seen as increased absorption in the 425~nm to 500~nm range.  Based on these tests the inner detector materials were restricted to specific tested lots of polylactic acid plastic (PLA), polytetrafluoroethylene (PTFE),  FEP, polyether ether ketone (PEEK), acrylic (clear, black, and white), Viton\textregistered\footnote{https:/www.chemours.com/Viton}, and Acrifix\textregistered~2R\footnote{https://www.acrifix.com/product/acrifix/} as an adhesive. 

Equally important is the long term stability of the $^6$Li doping. The thermodynamically stable microemulsion phase of the LiLS is achieved over a range of aqueous fractions. With higher or lower aqueous content, the LiLS is unstable. With respect to long-term stability, the high aqueous fraction phase is particularly worrisome as an emulsion prone to phase separation over time is formed. Dynamic light scattering and centrifugation experiments, similar to those described in~\cite{ISI:000413983300030}, confirmed that the LiLS formulation used in PROSPECT is stable against phase separation.  Also of concern is oxygen quenching due to interaction with air.  Oxygen quenching effects were studied as well as being observed in prototypes~\cite{Ashenfelter:2018cli}.  For these reasons a cover gas of boil-off nitrogen was maintained over the LiLS at all times. 

The PROSPECT LiLS was produced by first purifying raw components and then mixing in stages in a reaction vessel.  The LiCl was added as a final step.  Preparation and mixing were carried out as follows.  Solutions of 10~mol/L lithium chloride were prepared in 1~L batches from 95.37\,\% $^6$Li (by atom, as reported by the supplier) enriched lithium carbonate and analytical grade concentrated (37\,\% by mass) hydrochloric acid according to
\begin{equation}
\rm{Li}_2\rm{CO}_3 + 2\rm{HCl}\rightarrow 2\rm{LiCl} + \rm{H}_2\rm{O} + \rm{CO}_2. \label{eqn:LiClProd}
\end{equation}

LiCl solutions were filtered and passed through an anion exchange chromatography column\footnote{Bio-Rad AG 1-X4, 100 to 200 mesh  http://www.biorad.com}, which efficiently retained the dissolved iron impurity (presumably in the form of FeCl$_{4-}$) responsible for an initial yellow coloration.

Six individual lots of purified material were analyzed for optical transmittance, LiCl concentration, HCl concentration, and density. All lots showed transmittance over the wavelength range 260~nm to 547~nm that compared favorably to a commercially available solution of purified 8~mol/L LiCl. For the combined lots, the LiCl concentration was 9.98~mol/L and the HCl concentration was 0.088~mol/L. The density of the combined lots of LiCL solution was 1.206~kg/L. In total, 86~L (104~kg) of 10~mol/L LiCl solution were prepared.

The production of the LiLS commenced in January 2017. All the tubing, filtration system, liners, and mixing system were pre-cleaned with high purity ethanol, rinsed with 18.2~M$\Omega$cm pure water, and dried with nitrogen gas. All systems were then sealed in an inert environment until use. The scintillator mixing/synthesis system was a double-jacketed 90~L Chemglass\footnote{https://www.chemglass.com/} reactor with several injection ports made of Teflon\textregistered\footnote{https:/www.chemours.com} for chemical inoculation. 
All raw materials were introduced into the reactor at different mixing stages with different time parameters. 
After each synthesis, the $^6$Li-doped scintillator was discharged through a 2-micron glass filter in a 316-stainless-steel filtration house and stored in a 55-gallon drum. 
Each drum was equipped with a 5-micron perfluoroalkoxy alkanes (PFA) inner bag and a 5-micron outer polypropylene liner. The maximum storage capacity of each drum is limited to 180 liters (80\% full). 
A total of 5,040 liters were produced in 56 production batches and distributed in 28 drums by June 2017. 
These drums were kept in a nitrogen environment before shipment to the experimental site at ORNL. 
 The optical transmission spectra of the drums were consistent and no absorbance variations over 1\,\% were observed in the six month storage period. 
Mixing of the batches and filling of the AD are discussed in Section~\ref{sec-liquidprep}.

\subsection{Optical lattice}  

The 1.176~m wide  $\times$ 2.045~m long $\times$ 1.607~m tall antineutrino target is separated into a 14 by 11 grid of segments whose lengths run roughly perpendicular to a line formed by the core-detector baseline.  
Each segment is 1.176~m in length and has a 0.145~m\,$\times$\,0.145~m square cross-sectional area. 
This optical grid consists of low-mass, highly specularly reflective optical separators held in position by white 3D-printed support rods.  
These two primary optical grid components are further supported and constrained on both ends by PMT housings, and on the other four sides by acrylic segment supports. 

\begin{figure*}
\centering
\includegraphics[clip=true, trim=0mm 15mm 0mm 9mm,width=0.9\textwidth]{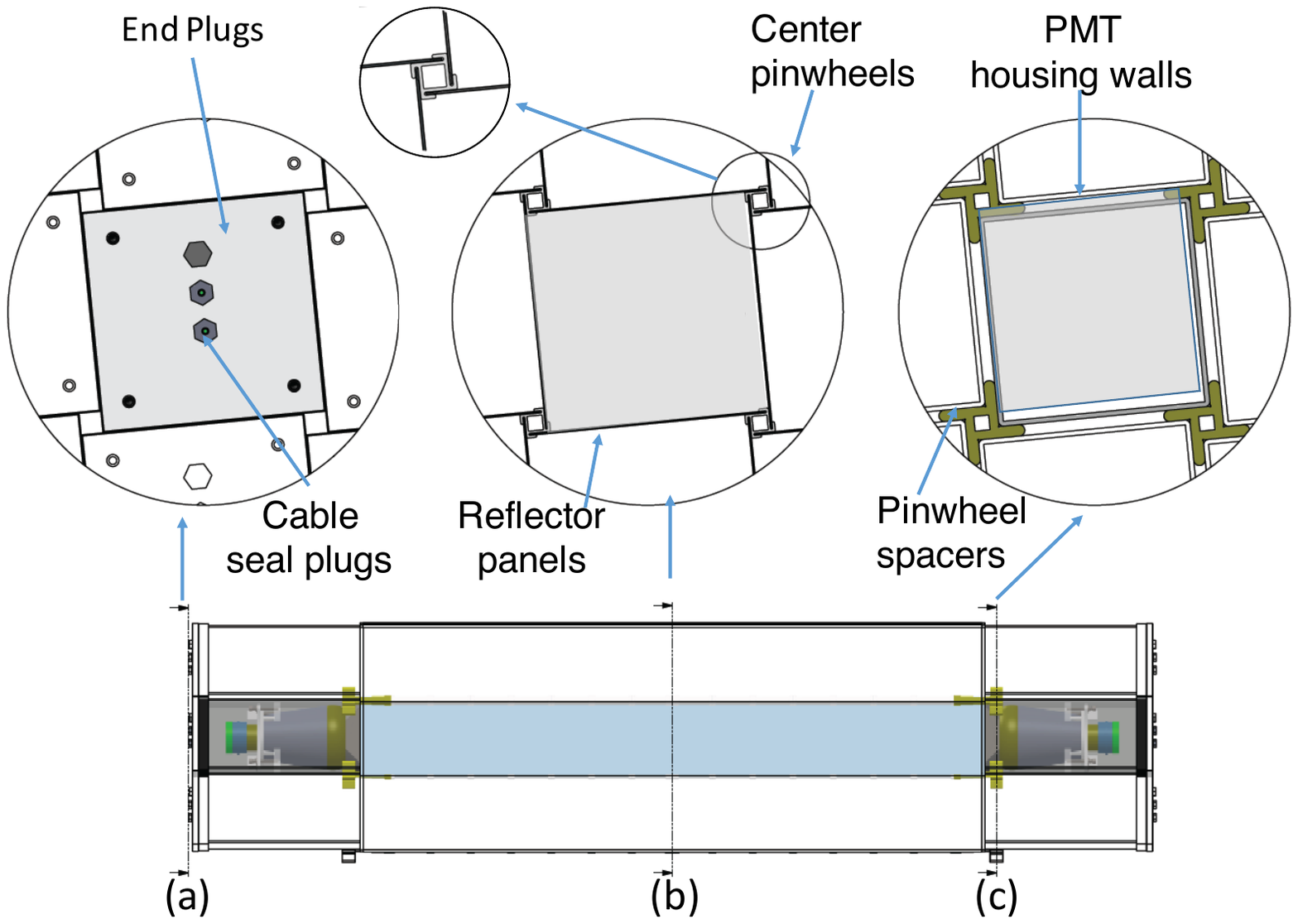}
\caption{
(Bottom) A single PROSPECT segment surrounded by neighboring segments.
PMT housings are inserted into the optical grid on each end. 
The opaque PMT housing is drawn transparent to reveal the PMT inside.  
Plane (a)  shows the PMT housing end plugs. 
PMT housings are supported by the end plugs and the pinwheel spacers at an angle of $5.5^{\circ}$ as shown in plane (c).
Plane (b)  shows the center pinwheels and optical separators,  
The complex shape of the pinwheels can be better seen  in Fig.~\ref{fig:pinwheel_types}.
}
\label{fig-universe}
\end{figure*}

Scintillation light produced by an antineutrino interaction is efficiently propagated down the length of a segment with minimal cross-talk by the specular  optical separators, which comprise $\sim$99\,\% of the total interior surface of each segment.  
In addition to supporting the optical separators, the support rods contain axes running along the entire length along each corner of each segment, allowing for calibration source deployment throughout the active detector volume.
The total mass of these two components of the segmentation system comprise less than 3\,\% of the total target mass, reducing the loss of IBD positron energy in non-scintillating regions. 
A drawing of a single detector segment's optical grid components are shown in Fig.~\ref{fig-universe}.

To achieve the physics goals of the experiment, the components of the PROSPECT optical grid must 
exhibit a high degree of dimensional uniformity to enable assembly of the detector and ensure uniformity of segment volumes 
and be chemically compatible with the liquid scintillator. 
Dimensional checks were made during assembly 
(Section~\ref{sec-assembly}) of the components ( optical separators and PMT housings) which determine the size of each segment.
The relative size variations (sigma) were all $<$ 0.1\% ensuring that the segment volumes were well within  1\% of each other.

Optical separators are composed of a carbon fiber backbone covered on both sides with adhesive-backed 3M DF2000MA\footnote{https://www.3m.com/} specularly reflecting film, an optically clear adhesive film, and a thin surface layer of FEP film.  
All layers are adhered to one another utilizing cold pressure lamination, and outer scintillator-compatible FEP film layers on each side are heat-sealed to one another to prevent scintillator contact with the optical separator interior.
The glossy twill carbon fiber sheet substrate provides structural support and removes the risk of optical segment-to-segment cross-talk.  
The DF2000MA reflecting film is both highly reflective ($>99$\,\% at normal incidence ) and highly specular ($>95$\,\% at normal incidence) for photons above 400~nm.   
Light transport at higher incident angles is further enabled by total internal reflection at the optical interface of the surface FEP layer ($\sim$1.33 index of refraction) and the PROSPECT scintillator ($\sim$1.56 index of refraction).  
Extensive dimensional, optical, mechanical, and leak-tightness quality assurance checks  were performed on all production optical separators prior to use. 

\begin{figure}
\includegraphics[clip=true, trim=50mm 65mm 40mm 20mm,width=0.45\textwidth]{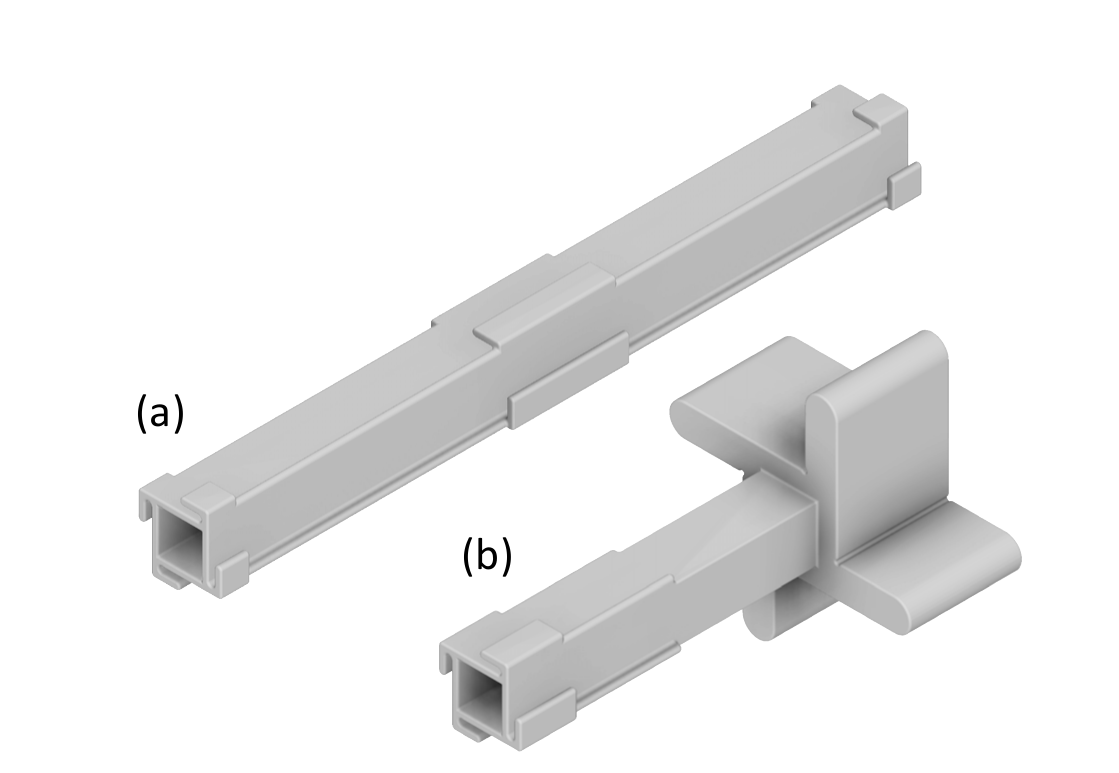}
\caption{Representative pinwheel types. (a) Central pinwheel - Three tabs per side hold the  optical separator in place. (b) End pinwheel - spacer arms separate the PMT housing bodies and support the pinwheel string.
\label{fig:pinwheel_types}}
\end{figure}

Pinwheel support rods were produced via filament-based 3D printing using a scintillator-compatible, white-dyed 100-micron polylactic acid filament.  
Support axes of $>$1.2~m total length are composed of shorter $\sim$150 mm rods of varying design strung onto a central Teflon tube or extruded acrylic rod, in the case of calibration and un-instrumented axes, respectively.  
Isometric drawings of  two pinwheel designs are shown in Fig.~\ref{fig:pinwheel_types}.
All sub-rods include multiple tabs which are used to grip each of four attached  optical separators.   
Sub-rods closest to the PMT housings contain additional thick profiles (Fig.~\ref{fig:pinwheel_types}b) that serve as the  
mechanical interface  between the optical grid and the PMT housings or acrylic supports on the outside of the detector.  
Other designs with two or three spacer arms were used at the corners and edges of the detector.
As with production optical separators, support rods underwent extensive optical and dimensional quality assurance checks (QA) prior to installation in the detector.  
Prior to QA, extensive preparation of 3D printed pieces was required to remove PLA flashing and support structures required for or produced during the 3D printing process.
Further details of the optical lattice construction are found in Section~\ref{det-assembly}.

\subsection{PMT modules } 

\begin{figure}
\centering
\includegraphics[clip=true, trim=0mm 10mm 0mm 10mm,width=3.1in]{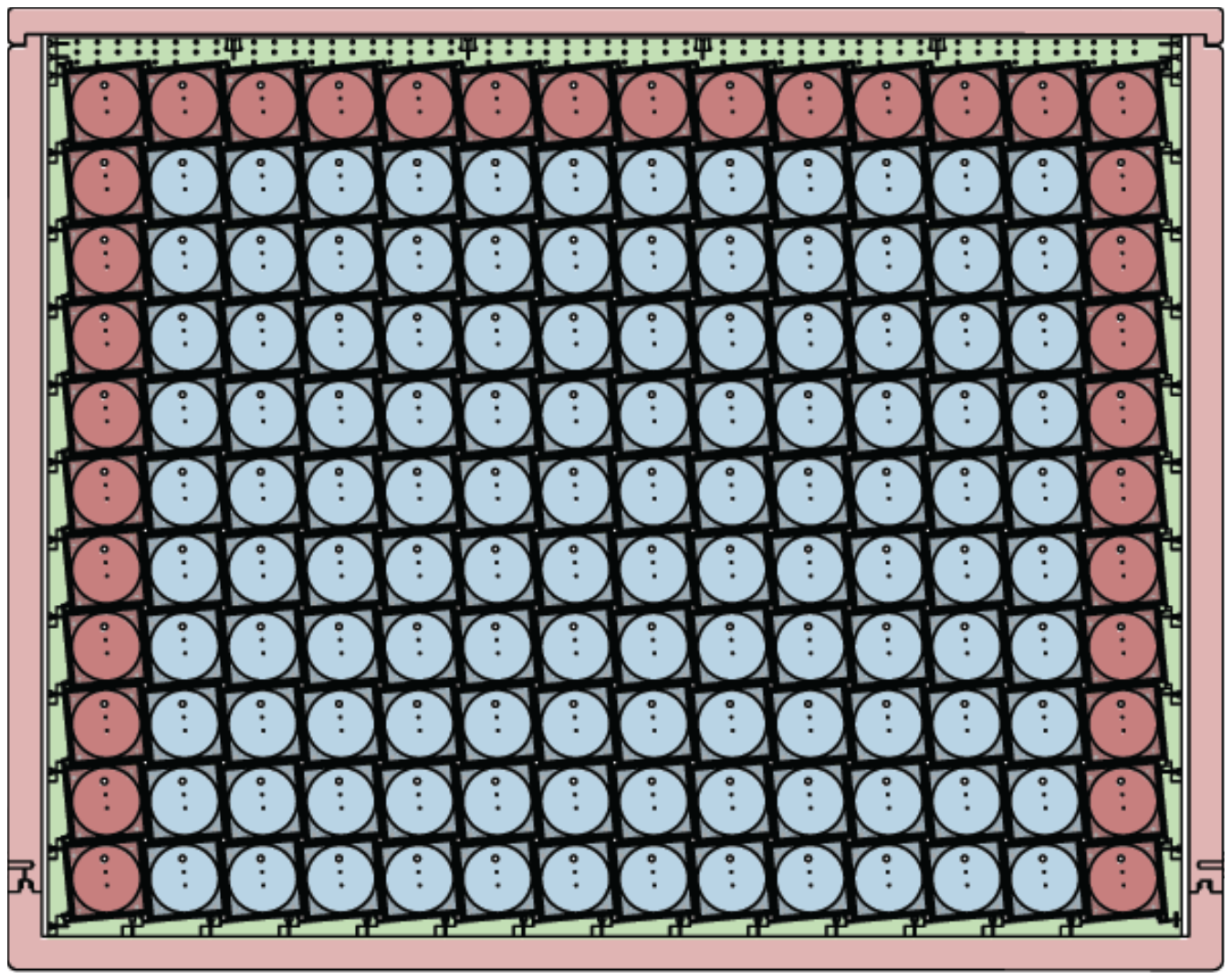}
\caption{
Cross-section of the active antineutrino detector showing the installation of 68 ET PMTs (red) in the outer columns and top row.  
The remaining detector segments are filled with
240 Hamamatsu PMTs (blue).
}
\label{fig-PMT_map}
\end{figure}

PMTs with similar characteristics from two manufacturers were chosen to expedite  PMT procurement. 
Detector segments  were made with one type or the other. 
240 Hamamatsu R6594 SEL  PMTs\footnote{https://www.hamamatsu.com/jp/en/product/optical-sensors/pmt/index.html} were used in the inner segments as shown in Fig.~\ref{fig-PMT_map}.
68 ADIT Electron Tubes 9372KB (ET) PMTs\footnote{http://www.et-enterprises.com} were used in the outer segments.
This mapping ensured that all of the PROSPECT segments in the fiducial region were of a uniform PMT type.

\begin{figure*}
\centering
\includegraphics[clip=true, trim=0mm 20mm 0mm 20mm, width=0.9\textwidth]{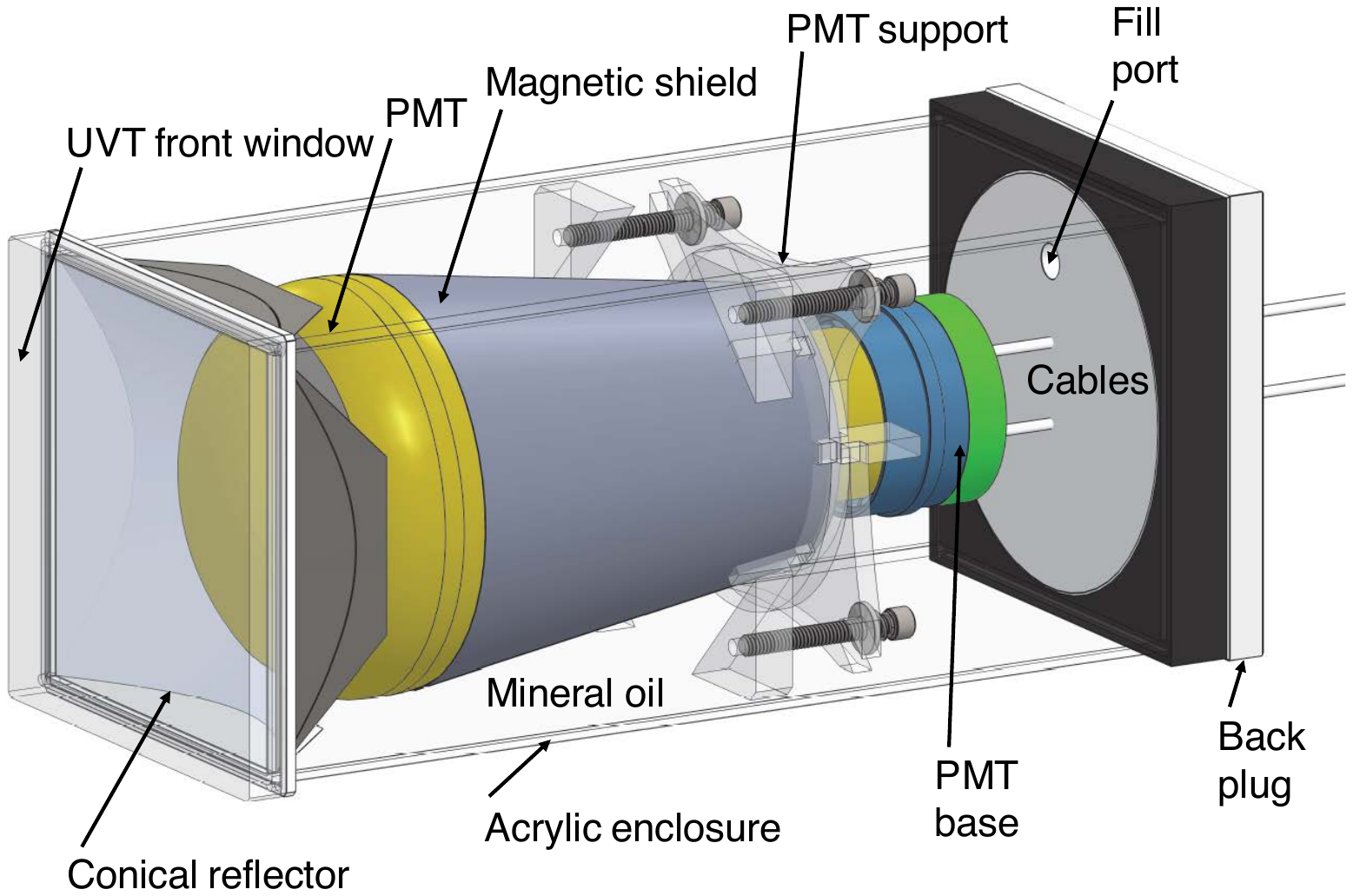}
\caption{
PMT housing module. }
\label{fig-PMT}
\end{figure*}

The major components of a PMT module are shown in Fig.~\ref{fig-PMT}.
The PMT housing is constructed from acrylic pieces  bonded together with Acrifix to make a roughly rectangular shape   350 mm long.
Slots are machined into the 144-mm-square  front window and back flange to accept the 3-mm-thick white acrylic side walls for bonding.
The 13-mm-thick  acrylic front window is constructed from ultra-violet transmitting acrylic (UVT).
The  19-mm-thick back flange is constructed from black acrylic and has a 130 mm diameter circular hole  to allow insertion of the PMT during assembly.
A 32-mm-thick clear back plug has a  cylindrical front section with an O-ring groove and a rear 145-mm-square section
and seals the housing module after all parts were installed.
Two cable seal plugs and a fill/test port connect to the module interior. 
Housings are supported by the back plug (Fig.~\ref{fig-cell_support}a) and by the pinwheel spacer arms at the front.
The rotational degree of freedom allowed by the back flange and plug configuration ensures that the front window and back plug are parallel.
The 132-mm-square cross-section of the sidewalls is purposely less than the front window and back plug   to 
provide tolerance against possible  construction variations.

A conical light guide is formed from a layer of  adhesive-backed  DF2000MA film and 1~mm thick acrylic. 
Rectangular reflector strips from the same material are adhered directly to the inside walls of the housing to complete the light guide.
The round PMT face is pressed into the light guide by an acrylic  plate at the rear of the housing. 
The different shapes of the Hamamatsu and ET PMT glass required different light guide shapes. 
A conical section of Hitachi Finemet\textregistered\footnote{https://www.hitachi-metals.co.jp/e/products/elec/tel/pdf/hl-fm4-k.pdf} surrounds the PMT to protect against stray magnetic fields.
Type specific PMT bases and sockets  push onto the PMT pins and connect to signal and high voltage cables which exit the rear plug.
The signal and high voltage (HV) cables are all made the same length (4.88~m) from RG188 cable and terminate in bulkhead connectors which are 
latter mounted on panels outside the aluminum tank.

After completion of all QA tests and PMT studies the housings are filled with an optical grade mineral oil. A 150~cc gas filled bag
inside the housing dampens any pressure variations due to thermal expansions. More construction details appear in Section~\ref{PMT-assembly}

\subsection{Segment supports}
\label{sec-supports}

\begin{figure}
\centering
\includegraphics[clip=true, trim=25mm 0mm 25mm 0mm,width=3.4in]{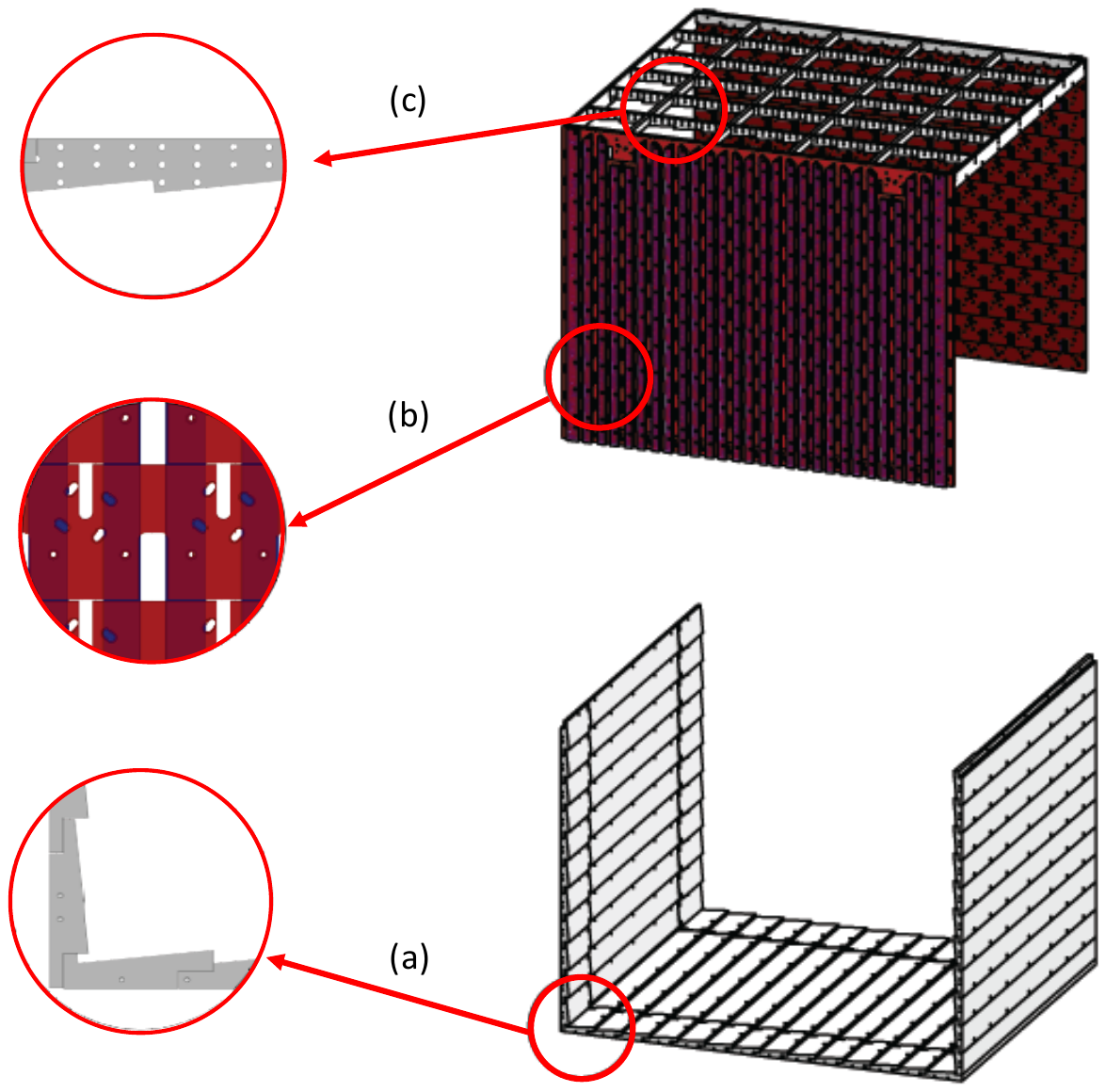}
\caption{Acrylic segment support structure.
(a) The wedge shaped planks of the segment support the two walls of PMT housings at the near and far faces.
The planks bolt together shiplap style and contain slots to position the pinwheel spacer arms correctly. 
The side walls constrain the outer rows of pinwheels and define the active detector volume.
(b) Horizontal planks are screwed into the backs of the PMT housings. 
Vertical planks stiffen the structure and form slots for the routing of cables and calibration tubes to the lid. 
(c) Baffles at the top tie the side and PMT walls together while holding the top reflector layer in place. 
Perforations in the baffles allow LiLS to cover the space above the top optical separator layer.}
\label{fig-cell_support}
\end{figure}

Machined acrylic segment supports underneath the bottom row of PMT housings hold the back plug of the PMT housings at the required $5.5^{\circ}$  tilt and 0.146~m (5.75 inch) pitch.
The wedge shaped acrylic planks bolt together ship-lap style and form the bottom and sides of the inner detector as shown in Fig.~\ref{fig-cell_support}a. 
The side supports hold the outermost layers of the optical grid in position and determine the size of the active volume.
Figure~\ref{fig-cell_support}b shows the horizontal and vertical planks that tie the backs of the PMT housings together. 
The structure is completed by machined acrylic baffles (Fig.~\ref{fig-cell_support}c)
on top which tie all sides together and hold the top reflectors in position.

\section{Calibration methods}
\label{sec-cal}

\begin{figure}
\centering
\includegraphics[clip=true, trim=0mm 0mm 0mm 0mm,width=3.1in]{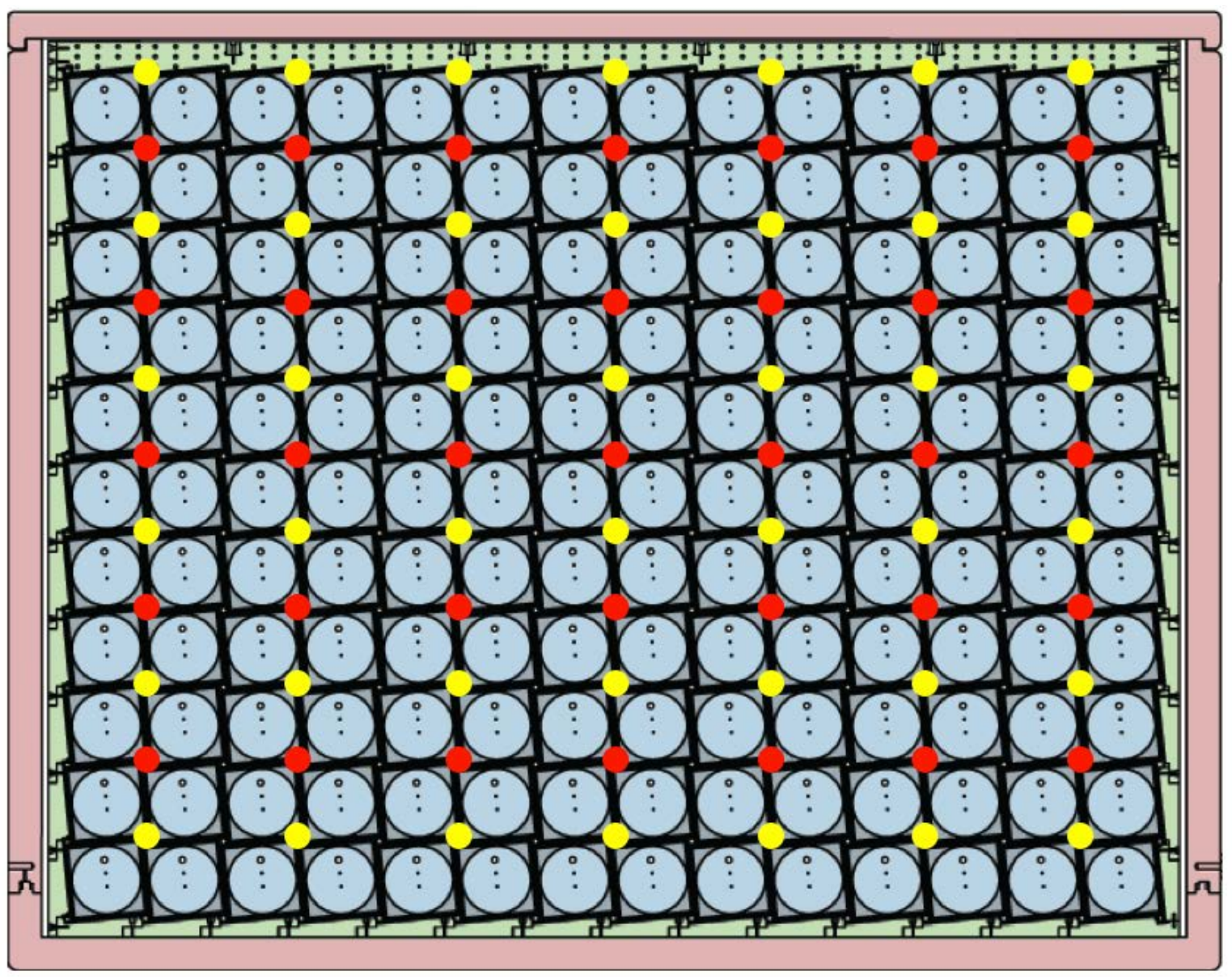}
\caption{Locations of the source tube (red) and optical insert (yellow) positions, in between the segments of the inner detector.
}
\label{fig:sourceplan}
\end{figure}

The timing and energy response of each PROSPECT segment is measured and tracked over time by a combination 
of optical reference signals, radioactive sources, and intrinsic radioactive backgrounds.
Optical diffusers located inside 42 center pinwheels can be pulsed over a range of intensities to measure timing offsets, 
determine single photo-electron responses and study PMT linearity. 
Radioactive sources can be positioned to any desired location 
along the length of 35 other locations by a source motor pushing or pulling a toothed drive belt attached to the source capsule. 
The locations of the optical and radioactive sources are shown in Fig.~\ref{fig:sourceplan}.
Analyses of time correlated signals in the PROSPECT data stream can cleanly identify neutron captures on $^6$Li ,
$^{214}{\rm Bi}\to^{214}{\rm Po}+\beta\to^{210}{\rm Pb}+\alpha$  or 
 $^{212}{\rm Bi}\to^{212}{\rm Po}+\beta\to^{208}{\rm Pb}+\alpha$   decays.
Additionally,  0.5~Bq of $^{227}$Ac  was  dissolved in the liquid scintillator to provide a source of
$^{227}{\rm Ac}\to^{219}{\rm Rn}+\alpha\to^{215}{\rm Po}+\alpha\to^{211}{\rm Pb}+\alpha$   decays.

\subsection{Optical calibration system}

\begin{figure*}
\centering
\includegraphics[clip=true, trim=0mm 40mm 0mm 20mm,width=6.4in]{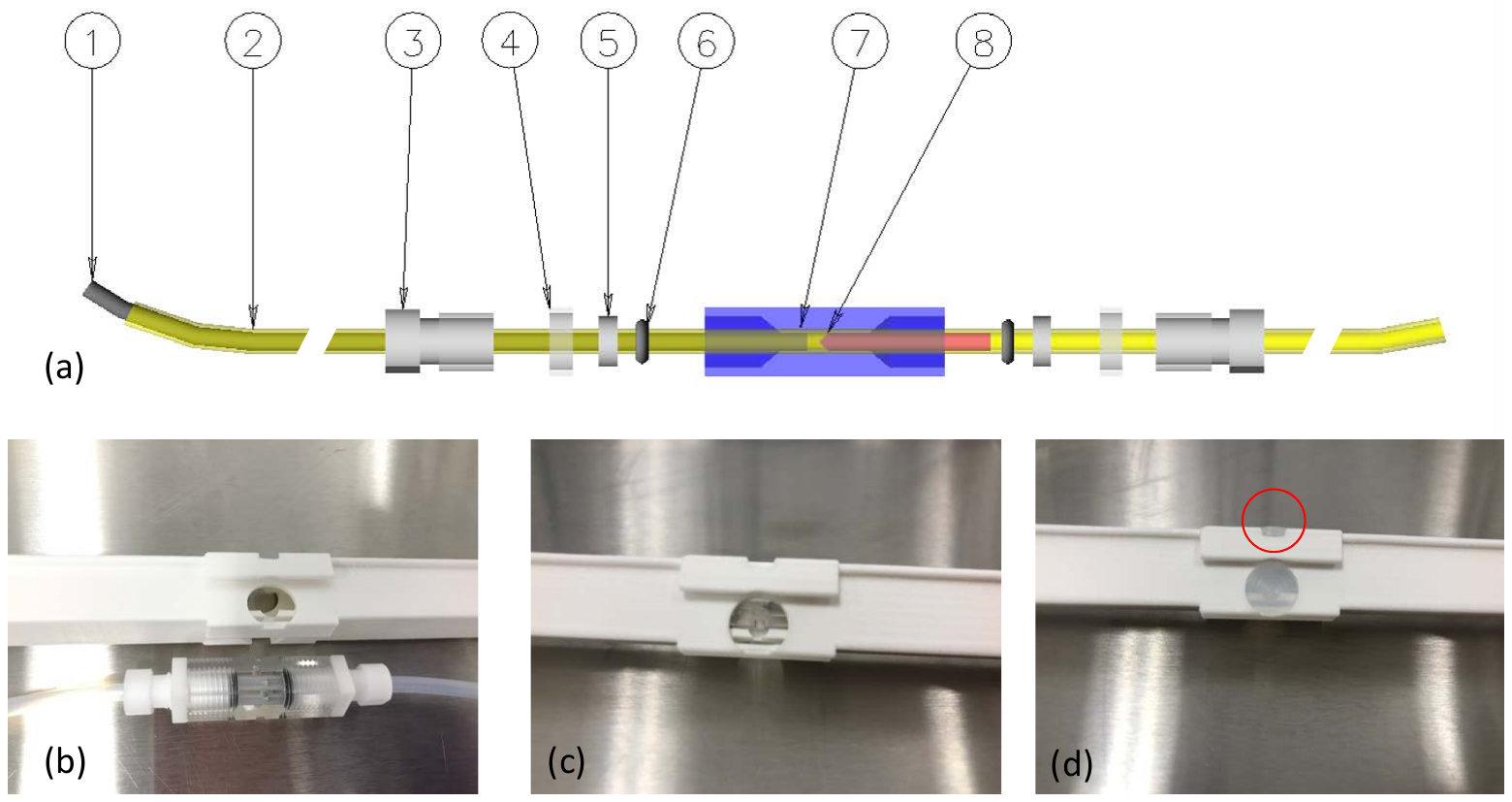}
\caption{(a) Components of the fiber optic assembly: (1) Fiber optic cable, (2) PTFE tube, (3) Compression nut, (4,5) spacer washers, (6) O-ring, (7) Square clear acrylic body, (8) Conical reflector. 
The fiber optic assembly, shown assembled in (b) is inserted into the square bore of the center pinwheel. (c) shows the assembly inserted in the pinwheel before being covered (d) by a diffusive Teflon disk. 
Most of the disk will be covered by a reflective optical separator (not shown), leaving only the small area shown circled in red in (d) inside the optical volume.
Pulsed light from fiber optic cable (1) is reflected into a radial direction by the conical reflector (8).  
The light passes through the acrylic body (7) and enters four Teflon diffusers embedded in the pinwheel rod before entering the center of the segment.  Each fiber optic assembly delivers light to four adjacent segments.  
\label{fig:OpticalCalib}}
\end{figure*}

Timing differences between segments, PMT west - PMT east balance within a segment and single photon equivalent (SPE) response of the PMTs are provided by light sources embedded in the pinwheel rods. Light from a pulsed laser is split multiple times and fed into 42 light guides.  The light guides are covered by PTFE tubing and fed to the center of the pinwheel rods.  Rods instrumented with a light fiber illuminate the center of four segments simultaneously through four Teflon diffusion disks in a four fold symmetric array embedded into the pinwheel rod common to those four segments. The arrangement is shown in Fig.~\ref{fig:OpticalCalib}.

The Optical Calibration System (OCS) consists of a laser pulser that delivers light into forty-two locations in the inner volume to service all 154 optical segments of the detector. The source of the optical calibration system is a 15~mW single mode fiber-pigtailed laser\footnote{Thorlabs LP450-SF15 https://www.thorlabs.com}  with a center wavelength of 450~nm. 
The laser is powered by a high performance laser diode driver \footnote{AVTECH model AVO-9A4-B-P0-N-DRXA-VXI-R5 https://avtech.com}. 
The driver supplies pulses up to 800~mA, with $<10$~ns width and 0.5~ns rise time, to drive the laser diode. 
The laser serves as the input to a custom single-mode fiber-optic splitter from Thorlabs, which splits the light into 48 output ports, 42 of which feed the optical diffusing units in the detector, leaving six spare output ports. 
The laser intensity is monitored with amplified photodiodes\footnote{Thorlabs PDA10A and PDA8 https://www.thorlabs.com} on two additional outputs of the splitter.
A 3.0~m long polyethylene optical fiber\footnote{Industrial Fiber Optics, IF 181L-3-0 https://www.i-fiberoptics.com/} runs from each of the output ports to a bulkhead on the outside of the detector package. From the inside of the bulkhead connection, another 5.5~m of the same fiber run through a set of icotek\footnote{http://www.icotek.com} fittings into the detector volume. Since the fibers are not scintillator compatible, they are encased in a 10 gauge Teflon sheath inside the inner detector volume. This cable and sheath then runs through the pinwheel rods to the longitudinal center, where each fiber terminates at an optical diffusing unit, a machined acrylic piece containing a reflective cone used to distribute the light radially. A Teflon diffusing cap is then used to both hold the acrylic optical diffusing unit in place inside the pinwheel and evenly distribute the light into the center of each of the four adjacent optical segments (See Fig. \ref{fig:OpticalCalib}).

By varying the laser driver current and pulse width the OCS light intensity can be varied from single photoelectrons per pulse to hundreds of photoelectrons per pulse.  In single photo-electron mode the OCS is used for gain calibrations of the 308 PMTs. At higher intensity the OCS is used to measure relative timing offsets between PMTs at 0.1~ns precision, to measure PMT non-linearity, and to monitor stability of the scintillator attenuation length. During normal operations the OCS is pulsed at between 10~Hz and 20~Hz, allowing for continuous monitoring of timing offsets and scintillator attenuation length. During dedicated OCS runs the rate can be increased up to $>1$~kHz.

\subsection{Radioactive source system}
\label{rad-cal}
\begin{figure}
\centering
\includegraphics[clip=true, trim=0mm 0mm 0mm 0mm,width=3.4in]{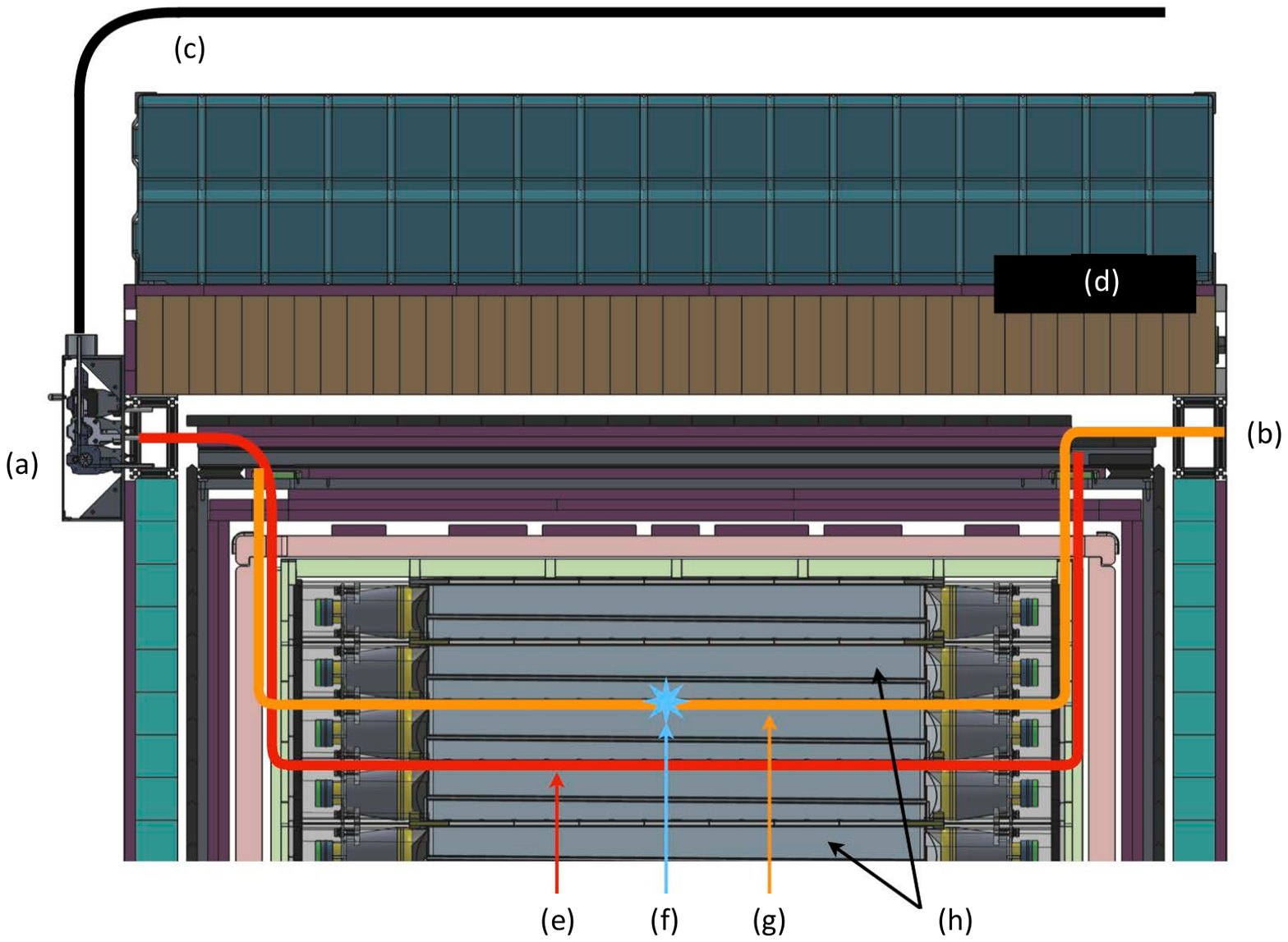}
\caption{End view of the detector showing the routing of a typical  source deployment tube ((e) red) and optical insert ((g) orange).
Also shown are (a) source drive motors, (b) optical fiber connector panel, (c) belt storage tube, (d) shielding, (f) light injection point and (h) detector segments.  
}
\label{fig:sourceplanB}
\end{figure}

The PROSPECT radioactive source calibration system is designed to move emitters of $\gamma$-rays, neutrons, and positrons through tubes routed into the active volume of the detector (as seen in Fig.~\ref{fig:sourceplanB}) to measure and calibrate the energy and position response of the detector as well as to study topological effects.  
There are thirty-five source tubes integrated with the optical array, spread out in a 5 by 7 grid.
PROSPECT currently deploys $^{137}$Cs, $^{60}$Co, $^{22}$Na, and $^{252}$Cf sources.  
The source map is shown in Fig.~\ref{fig:sourceplan}. 
A table detailing the sources and their uses is shown in Table~\ref{tab:sources}.
Each source can be repeatably positioned to within $\sim$1~mm with an absolute accuracy of $\sim$1~cm along the length of each source tube.

\begin{table}
\centering
\begin{tabular}{|c|c|c|c|}
\hline 
\textbf{Source} & \textbf{Decay} & \textbf{$\gamma$ energies (MeV)} & \textbf{Purpose} \\ 
\hline 
$^{137}$Cs & $\beta^-$ & 0.662 & $\gamma$-ray  \\
\hline
$^{22}$Na & $\beta^+$ & 0.511, 1.274 & positron energy \\
\hline
$^{60}$Co & $\beta^-$ & 1.173, 1.332 & $\gamma$-ray  \\ 
\hline 
$^{252}$Cf & $n$ (fission) & - & neutron response \\ 
\hline 

\end{tabular} 
\caption{Proposed $\gamma$-ray , positron, and neutron sources for calibration.}
\label{tab:sources}
\end{table}

\begin{figure}
\centering
\includegraphics[width=0.45\textwidth]{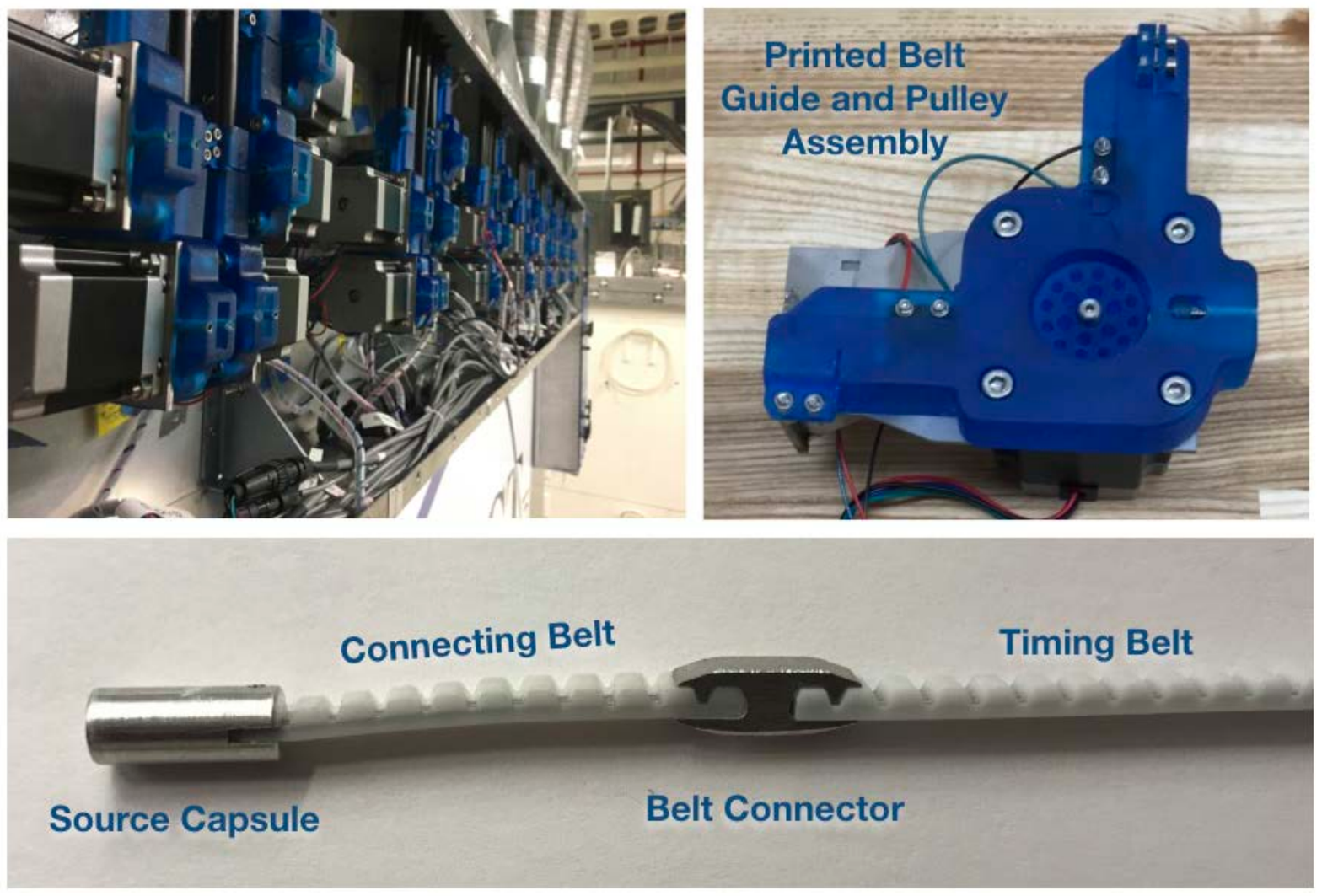}
\caption{Bottom: Source capsule attached to the drive belt. A short connecting belt is attached to the source and belt connector to make it easier to swap sources. Top Right: 3D printed belt guide and pulley.
Top Left: Source motors and belt assemblies.}
\label{fig:capsule}
\end{figure}

Each source is encapsulated into a small aluminum cylinder, sealed with a set-screw and epoxy (Fig.~\ref{fig:capsule}).
The capsule attaches to the belt with a stainless steel spring pin.
Each capsule is etched with a unique ID number that is recorded in the source control monitoring database.

Toothed drive belts (timing belts) are used to push the capsules into the detector along the length of the segments ``source tubes'' as well as to retract them. 
The timing belt width and stiffness must be correct to avoid buckling or excess friction in the tube.
A 3~mm wide, AT3 pitch, polyurethane belt reinforced with steel cords works well. 
The ``source tubes'' are annealed PTFE with a 0.0095 m OD and 0.0064 m ID. 

The timing belt is driven by a custom-made 3D printed pulley on a NEMA 23 stepper motor (Fig.~\ref{fig:capsule}).
The pulley is attached to the motor shaft to drive the belt, and a spring-loaded jockey keeps the timing belt held tightly to the timing belt pulley. 
A 3D printed belt guide keeps this assembly together and guides the belt from the source tube to the pulley, and out to a storage tube on top of the detector. 
It also contains two micro switches; one that stops the motor if the source capsule approaches the pulley, acting as a safety feature and as the home position of the source capsule, and another that prevents the belt from being deployed beyond the pulley.
The timing belt pulleys and motor housings were designed specifically for this system and 3D printed using a UV-cured resin.

\subsection{Intrinsic radioactive sources}
\label{sec:IntrinsicSources}

We make use of three radioactive sources present within the liquid scintillator itself. 
Two of these are intrinsic sources, collectively called ``BiPo'' decays, which arise from the fast coincidences of $\beta$-decays from $^{212}$Bi and $^{214}$Bi and the subsequent $\alpha$-decays of $^{212}$Po and $^{214}$Po. The bismuth isotopes arise from naturally occurring $^{232}$Th ($t_{1/2}=14$~Gyr) and $^{238}$U ($t_{1/2}=4.5$~Gyr), contaminants respectively.

A third source, $^{227}$Ac ($t_{1/2}=22
$~yr), was intentionally added to the LS to monitor the product of efficiency $\times$ volume for all detector  segments.
A chloride solution of $^{227}$Ac  was prepared from a commercial actinium source, and dissolved in the liquid scintillator at a concentration near 0.5~Bq, over the whole detector. These give rise to ``RnPo'' decays, namely the fast coincidence of $\alpha$-decays from $^{219}$Rn and $^{215}$Po ($t_{1/2}=1.78$~ms). Care was taken to ensure that the AcCl solution was dissolved uniformly into the scintillator before it was transferred to the detector.

 These three sources produce time correlated signals within the detector which are triggered and read into the normal DAQ data stream. The events are identified  for analysis by energy cuts, decay time distributions and pulse shape discrimination cuts which utilize the relatively long decay times of these processes (0.3-3 msec). Large event samples with minimal background contamination are accumulated by integrating over the detector exposure. 
 
\section{Containment vessels }  
A pair of nested inner (acrylic) and outer (aluminum) containment vessels (tanks) provide redundant protection against LiLS leaks. The space between the vessels is filled with borated polyethylene and water to reduce the stress on the acrylic tank walls and O-rings.

\subsection{Inner containment vessel }  

\begin{figure}
\centering
\includegraphics[clip=true, trim=165mm 89mm 0mm 10mm,width=3.4in]{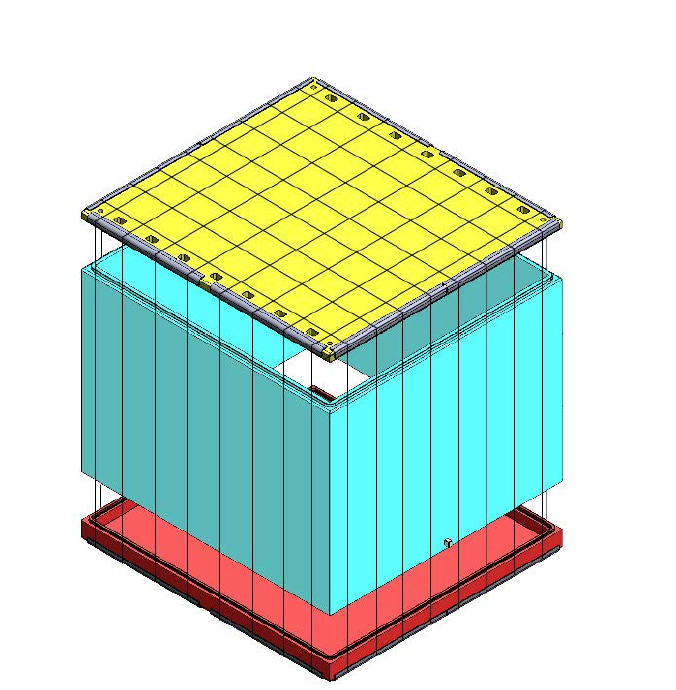}
\caption{
The acrylic containment tank consist of three pieces: a 64~mm thick base (red), four 64~mm thick walls bonded together (aqua), and a 51~mm thick lid (yellow). Sixteen cable loops compress the O-rings between the  wall and base.  Aluminum angles  and Teflon cushions (grey)  distribute the force evenly over the acrylic. 
}
\label{fig-acrylic}
\end{figure}

As noted in Section~\ref{sec-scintillator}, the known list of materials compatible with the $^6$Li doped liquid scintillator used in the PROSPECT detector is somewhat limited, i.e. acrylic, Teflon (PTFE, PFA and FEP), PVDF, PEEK, Viton.  
Furthermore, the proximity of the detector to a nuclear reactor adds the requirement of secondary containment.  
The practicality of access during assembly of the inner detector components imposed the need to lower the primary tank walls onto a base after assembly of the inner detector was completed.  
The inner primary containment vessel shown in 
Fig.~\ref{fig-acrylic}
is constructed from acrylic with a Viton seal between the base and vertical walls.  
A Teflon lined aluminum tank was considered, but the technology was uncertain and the presence of so much aluminum in unshielded proximity to the scintillator was undesirable.

The inner dimensions of the tank are 1.995~m (wide)  $\times$ 2.143~m (long) $\times$ 1.555~m (tall). The walls and base were specified to have a thickness of 0.0635~m to keep the longterm stress at or below 4.1~MPa (600~psi), thus maintaining dimensional stability for many years.   Fourteen rectangular holes (0.051~m $\times$ 0.076~m) provided passage for the numerous instrumentation cables.  A thin strip of Teflon along the top surface provided a cushion between the lid and the walls.

The bottom Viton seal presented several design challenges.  
A double seal was required to verify leak tightness after the final installation.  
A small passageway to the space between seals allows for leak checking in place without pressurizing the entire vessel.  
A tube extending to the outside of the detector allowed testing of the seal after the entire acrylic assembly was lowered into the aluminum tank and also after the entire detector was shipped from Yale to Oak Ridge. 
A second passageway with tube was added to allow for the possibility of purging the space between seals after the detector was filled with liquid. 

The original design of the seal  which had O-rings on either side of a wall tongue inserted into a groove on the base failed. It was impossible to control the lateral dimensions of this large acrylic object  well enough for a good seal. However, the flat horizontal surfaces at the bottom of the wall and top of the base were planar within a tight tolerance. A new seal design with an inner and outer O-ring vertically compressed between the wall and base was implemented.
Vertical compression was provided by the weight of the wall and a series of tensioned steel cables wrapped around the assembly.
More details are presented in Section~\ref{sec:Cables}.

The O-ring squeeze of the primary inner 3.2~mm diameter Viton cord was determined by a series of 2.4~mm thick PEEK spacers providing a nominal 20\,\% compression.  
This high value was chosen to allow a margin for the known deviations from flatness of the sealing surfaces.  
The inner Viton 75 cord was a custom fabrication, vulcanized and polished commercially.  
To minimize the total required compression force, the secondary outer seal was made from 6.35~mm diameter neoprene sponge cord.  
The outer O-ring seal is not exposed to LS, but only to the surrounding water.
A third back up seal was added in the form of 0.05~m wide marine tape applied to the 2.4~mm gap between walls and base around the entire perimeter of the detector.

\subsection{Secondary containment vessel}

An aluminum tank with internal dimensions 
of 2.205~m (wide)  $\times$ 2.255~m (long) $\times$ 1.982~m (tall)
was constructed to provide secondary containment for the scintillator, and to provide a protective support structure during shipping.  The lid was sealed to provide control of the gas environment around the detector.  This required the development of feedthroughs for 748 PMT cables, multiple gas and liquid lines, and  additional tubes for insertion of the calibration devices described in section~\ref{rad-cal}.

Material for the tank was 5083-H321 aluminum of 0.025~m thickness.  While this alloy is not the stiffest alloy available, it retains its properties after welding better than most other alloys.  
Commercial aluminum plates were not available in the sizes we needed so all walls were made by joining two plates with a friction stir weld. 
The walls are welded leak-tight to the base. 
The inside dimensions were chosen to provide generous clearance between the acrylic and aluminum tanks. That space was filled with sheets of borated polyethylene and demineralized water for absorption of thermal neutrons.  The lid was sealed to the walls using a flat neoprene sponge gasket.

\section{Detector movement and shielding}  

\subsection{Detector chassis}  

The multiple purposes served by the mechanical support structure, dubbed the ``chassis'', are to 
\begin{enumerate}
\item Enable detector installation.
\item Allow detector motion to multiple baselines.
\item Distribute the weight of the detector package to remain within the floor loading requirements.  
\item Enable tilting of the detector during scintillator filling  (Sec.~\ref{sec-detfill}).
\end{enumerate}

\begin{figure}
\centering
\includegraphics[clip=true, trim=20mm 40mm 20mm 40mm,width=3.4in]{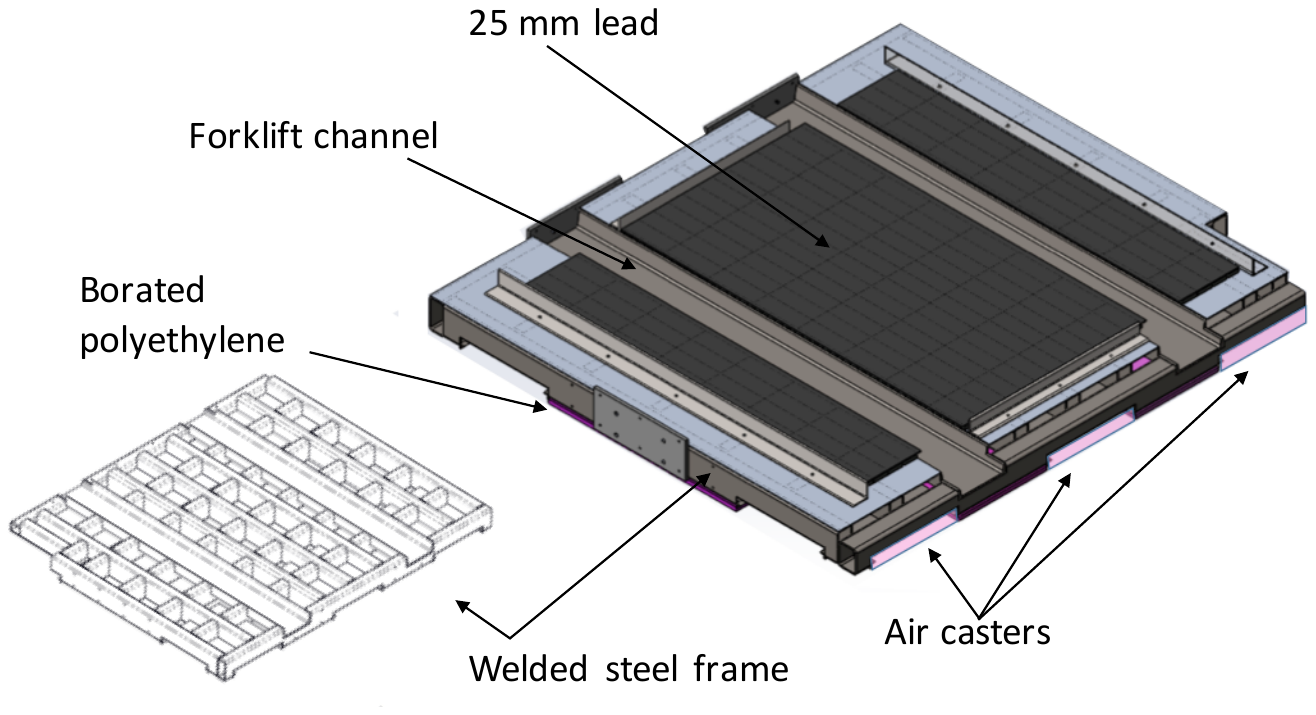}
\caption{
Detector support chassis. The welded 210~mm thick steel frame supports the detector during movement by the air caster system and distributes the weight of the detector over the maximum allowed floor area. Six air caster lifting pads slide into slots at the bottom of the detector. Two deep channels run across the frame at the top to allow a forklift to lower the detector onto the frame. A 25~mm borated polyethylene layer below and a 25~mm lead layer on top complete the passive shielding.}
\label{fig-chassis}
\end{figure}

The chassis, shown in Fig.~\ref{fig-chassis},  is a rectangular welded steel frame 2.946~m (wide)  $\times$ 3.242~m (long) $\times$ 0.21~m (tall) with a mass of 1786 kg. The frame has a  0.356~m $\times$ 0.691~m cut-out to avoid blocking door openings (Fig.~\ref{fig-baselines}), six slots on the sides to accept Aero-go\footnote{\sf https://www.aerogo.com} air casters that enable detector motion, and two C-channels on top to allow the detector to be loaded with a forklift. The air casters can raise the fully loaded chassis by $\sim$\,0.025~m to allow movement  to other baselines, and were used during the movement of the dry detector to Position 1 (Fig.~\ref{fig-baselines}) during installation (Sec.~\ref{sec-HFIRassembly}). 

The chassis was designed to deflect $<\!0.1\ {\rm mm}$ with all air casters in operation and $<\!0.3\ {\rm mm}$ if one of the six casters was non-operational. Borated (5\,\%) polyethylene sheets 0.025~m thick are attached to the top surface of all casters and the bottom surface of the chassis, save for the caster slots, to suppress backgrounds due to thermal neutrons. 

\subsection{Passive shielding}  
\label{sec-passive}
The passive shielding of the detector was designed based on background measurements and prototype operation~\cite{Ashenfelter:2015uxt} in the Experiment Room discussed in Sec.~\ref{sec-shielddesign}. 
Comparison of the prototype response to simulation showed that correlated ``IBD-like'' backgrounds were events with multiple neutron interactions in the active detector which either produced an in-time $\gamma$-ray  or had a neutron interaction that was mis-identified as a $\gamma$-ray  in addition to a captured thermal neutron. These events were primarily produced by high energy ($\sim$10~MeV to a few hundred MeV) cosmic neutrons. Spallation neutrons from interacting cosmic muons also contribute to the background but at a nearly negligible rate.

Hydrogenous material above the detector, followed by a 0.025~m lead layer and a 5\,\%-BPE layer, were determined to provide the best suppression of the high energy neutrons given the safety and geometric constraints as shown in Fig.~\ref{fig-side}. 
The aluminum containment vessel rests on 0.025~m thick lead bricks and the vessel supports walls of interlocking 0.025~m lead bricks. 
Approximately 0.127~m of BPE on top of the vessel support another 0.025~m thick layer of bricks. There are penetrations and openings in the BPE and lead on top to accommodate cables and services.  
Outside of the lead walls is a structure of 0.102~m $\times$ 0.102~m cross-section recycled high density polyethylene (HDPE) beams bolted together in a ``log cabin'' style.  
These walls support a roof of 0.064~m $\times$ 0.241~m cross-section HDPE beams.
To limit sagging, 
the roof beams are joined by eight steel pipes transverse to the beams and bolted at each end.  
The outer HDPE surfaces are covered with 0.025~m BPE to limit the effect of 2.2~MeV $\gamma$-rays produced by thermal neutron captures in the HDPE. 
The BPE is covered with thin (0.6~mm) aluminum sheet for fire safety.  
The passive shielding is completed on top by interlocking polyethylene ``WaterBricks''\footnote{\sf https://www.waterbrick.org} (0.15\,m $\times$ 0.23\,m $\times$ 0.46\,m) filled with tap water arranged on top of the roof and covered with a fiberglass blanket. 

\section{Detector monitoring and control}
\label{sec-monitoring}

Detector temperature is monitored in multiple locations using resistance temperature detectors (RTDs). Eleven RTDs  are mounted inside Teflon tubes in the LiLS volume, with another RTD sampling the temperature of the water between the acrylic and aluminum containment tanks. The RTDs are connected to readout  modules\footnote{Advantech ADAM 6015 http://advantech.com}, and read out every 60~s by the monitoring system.

The levels of the LiLS and water are measured by ultrasonic sensors\footnote{ToughSonic~14, TSPC-30S1-485, https://senix.com/wp-content/uploads/ToughSonic-14-Data-Sheet.pdf}  mounted at the top of the acrylic and aluminum tanks. The two LiLS sensors are mounted on opposite corners of the acrylic tank so as to be sensitive to the tilt of the detector during the filling operation. A single sensor measures the water height. The water sensor is coupled directly to a 1.57~m pipe that goes to the floor of the aluminum tank. The LiLS sensors are mounted horizontally in the restricted vertical space, coupling to 0.019~m (ID) by 1.78~m sample pipes via 90-degree acrylic reflectors. After calibrating for gas and pressure the sensors have a resolution better than 1~mm.

Additional sensors inside and outside the aluminum tank measure the humidity, pressure and temperature of the cover gas system.

\subsection{ High voltage system} 
\label{sec-pmthv}

Each PMT channel has an independent high voltage (HV) bias supply allowing the  gain of all tubes to be set to
 $5\times 10^5$.  
Sixteen channel ISEG HV modules\footnote{ISEG EH161030n https://iseg-hv.com/files/media/isegXdatasheetXEHSXenX21.pdf}   are housed in MPOD crates from Weiner~\footnote{www.wiener-d.com/sc/power-supplies/mpod--lvhv/mpod-crate.html}. 
A total of twenty ISEG modules are  in two  crates.  
HV control and logging is via custom software  over a local DAQ network. Current and voltage values are logged.  

\subsection{Nitrogen cover gas system}

To prevent oxygen from dissolving into the liquid scintillator and quenching the scintillation light, PROSPECT replaces the air in the volume above the liquid with pure nitrogen gas boil-off from a liquid nitrogen dewar.  The amount of nitrogen going into the detector is set by a mass flow controller with a range of zero to one standard liter per minute.  The nitrogen flow rate out of the detector is also monitored by a mass flow meter, followed by an oil filled bubbler.  The bubbler ensures that if the flow stops for some reason, outside air cannot flow back into the detector.   

The nitrogen pressure is monitored at various places in the flow path with both absolute and differential pressure transducers.  The amount of oxygen and water in the gas outlet is monitored using a pair of oxygen sensors and a combination pressure/temperature/humidity sensor. 

In addition to providing cover gas to the scintillator, the gas system can also be used to bubble dry nitrogen gas through the detector through a set of tubes located around the perimeter of the active volume.  It can also pressurize and monitor the space between the double O-ring seals on the acrylic containment tank.

\section{Data acquisition} 
\label{sec-daq}

The DAQ system for PROSPECT has been designed to balance several competing priorities. As described above, PSD analysis of LiLS signals from all 308 PMTs is critical to background rejection, therefore waveform digitization is a necessity. Furthermore, a wide dynamic range is required, spanning  the range from 0-14 MeV with good linearity and high resolution. This upper limit is defined by the desire to include the endpoint of cosmologically produced $^{12}$B for energy scale and linearity studies. Full waveform digitization of all PMT channels would result in a very large data stream at the 40 kHz data rates  when HFIR is operating. Consequently, an efficient triggering scheme that only transfers and records channels with data of interest was also a priority. 

The solution adopted for PROSPECT uses commercial Waveform Digitizer Modules (WFDs). 
The PMT anode signals are sent directly into WFD inputs without analog pre-processing, which is also a considerable simplification. All trigger decisions are derived from on-board digital processing of the resulting sample stream. 

The WFD model\footnote{CAEN-V1725 http://www.caen.it}  has a sample rate of $250$\,MHz and $14$\,bit depth per sample. Studies using prototype detector modules~\cite{Ashenfelter:2015aaa,Ashenfelter:2018cli} determined that these digitization parameters would meet the PSD and dynamic range requirements of PROSPECT. In particular, no significant PSD performance gain was found when testing 500\,MHz digitizers due to the long optical propagation lengths and resulting time dispersion within the PROSPECT segment geometry. While a higher sampling rate would have provided improved longitudinal position reconstruction, gains beyond the transverse segment size ($\sim$0.15~m) provide no significant physics or background rejection performance gains. On-board logic governs trigger and sample processing functionality. 
No on-board signal amplitude or PSD calculations are attempted, instead waveforms are recorded  for off-line analysis. 
This approach provides greater flexibility for optimization of the processing approach, at the expense of higher data rates.

\subsection{DAQ hardware}

A schematic of the DAQ hardware used by PROSPECT is shown in Fig.~\ref{fig:daqLayout}.
A total of twenty-one WFD modules are used to readout the 308  PMTs. These are operated in two VME crates\footnote{Weiner 6023 http://www.wiener-d.com/sc/powered-crates/vme} powering ten and eleven WFD modules respectively. 
All readout and control of the WFD modules is performed via two optical fiber link cards\footnote{CAEN A3818 Optical Controller PCI Express Cards http://www.caen.it} installed in individual DAQ control computers being used for this purpose. 
Each card supports four independent optical fiber links, with a single link supporting either two or three WFD modules. The acquisition processes running on the DAQ control PCs are coordinated by a run control computer. 

A single custom Logic Fan-In/Fan-Out module\footnote{757 NIM Logic Fan-In/Fan-Out http://www.phillipsscientific.com/pdf/757ds.pdf  }{PS-FIFO} is used for trigger signal distribution. This module is custom-ordered to have a single bank of 32 input and 32 output channels, i.e. any logic signal input is mirrored on the 32 output channels.

\subsection{DAQ triggering}

\begin{figure*}
\centering
\includegraphics[width=0.8\textwidth]{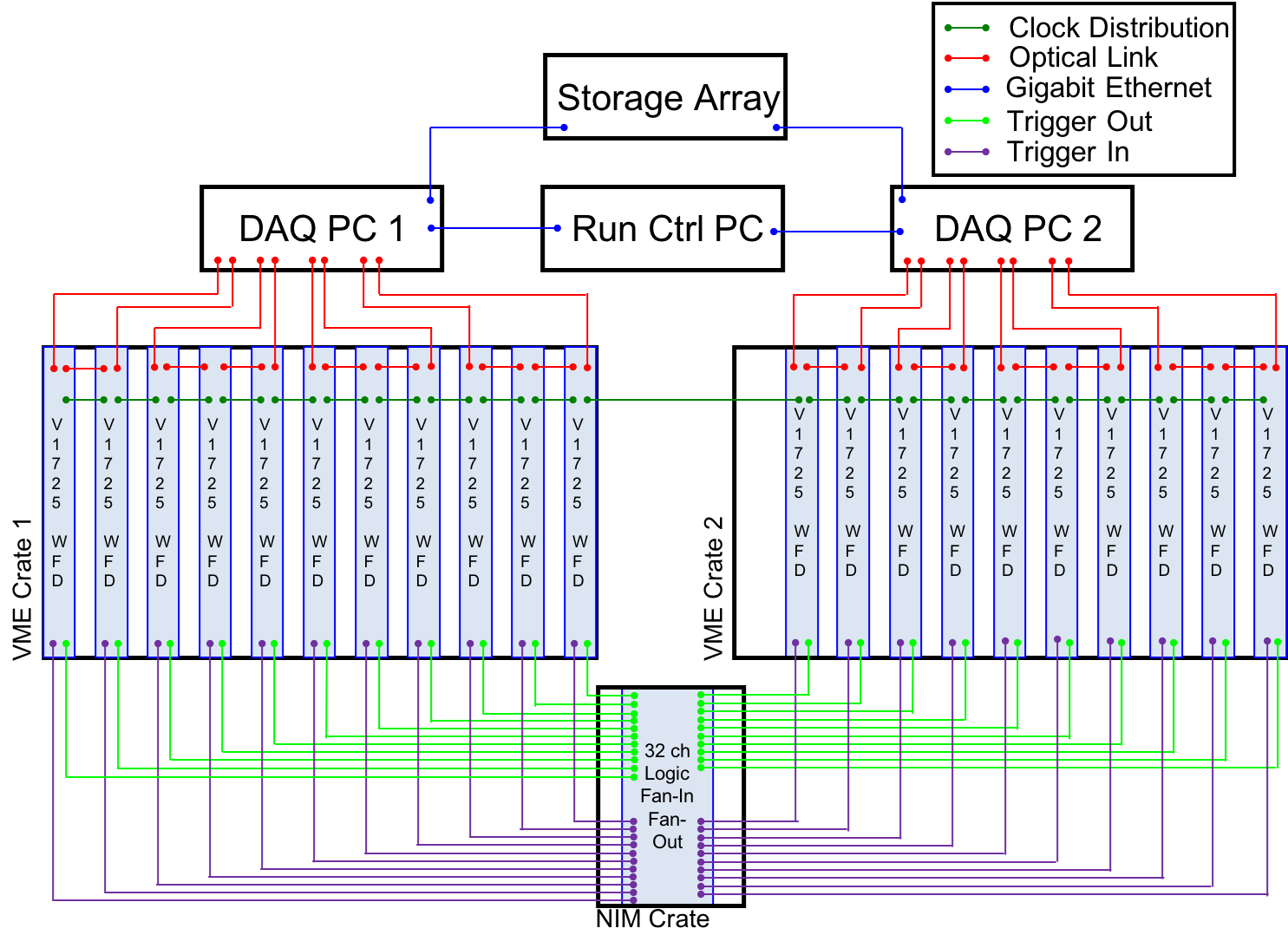}
\caption{Schematic diagram of the DAQ. } \label{fig:daqLayout}
\end{figure*}

The primary trigger functions are implemented in firmware on-board the WFD modules. 
Acquisition of  waveforms (148 samples long) by all WFD channels is triggered if both PMTs in any segment exceed a signal level of approximately five photoelectrons within a 64\,ns coincidence window.
As shown in Fig.~\ref{fig:daqLayout}, the acquisition of all channels on all WFD modules is achieved via a logic signal sent to every WFD module.
The waveform acquired for every PMT is examined via on-board firmware and compared to a secondary threshold. 
Acquired samples from an individual WFD channel are only recorded to disk in waveform regions that exceed a lower threshold signal level of approximately two photoelectrons, along with pre- and post-threshold regions of 24\, and 32\,samples, respectively.
We denote the trigger threshold as the ``segment'' threshold and the secondary threshold as the Zero Length Encoding (ZLE) threshold since it suppresses channels with zero or very small energy depositions.
Since the average segment multiplicity per trigger
is $\approx 3$, it is considerably more efficient to collect data only for those segments with energy depositions. 
However, it would also be inefficient to consider segments individually when making the trigger decision to acquire data - a prohibitive low individual segment threshold would have to be applied to collect all depositions of interest. 

This scheme is particularly important for the IBD positron measured in PROSPECT.
This will constitute a primary deposition, most likely limited to a single segment, by the slowing of the IBD positron, and  smaller depositions due to Compton scattering of 511~keV annihilation $\gamma$-rays. 
Having the ability to set a lower ZLE threshold enables efficient collection of energy deposited by annihilation $\gamma$-rays in segments near the primary interaction segment, while maintaining a manageable data rate.  

Raw waveforms are time stamped by the number of digitizer clock ticks from the start of the run using  the daisy-chained PLL-synchronized on board clocks. 
Timing offset calibrations between all channels are determined for 
each run using muon events for multi-cell coincidences. 
Any time stamp error would cause an  alignment jump in clock counts between boards ( never observed to date). Furthermore, if any board detects an unlock in the PLL signal, a signal is sent to the DAQ computer to cancel the run and log warnings.

Threshold values are set in terms of digitizer (ADC) counts  above baseline. 
Typical production settings for the segment and ZLE thresholds are 50\,ch and 20\,ch per PMT, corresponding to segment-level energy depositions of $\sim$100\,keV and $\sim$40\,keV, respectively.

\subsection{Data transfer and data rates}

Memory on-board the WFD modules is paged into two buffers. 
While one buffer is being filled with waveform data, the other is available for transfer to disk storage via the optical links. 
DAQ control software running on two independent computers continually polls the WFDs and transfer data when a buffer is filled. 
Typical trigger and data rates are given in Table\,\ref{tab:dataRates}. 

\begin{table*}[tbp]
  \begin{centering}
  \begin{tabular}{l|lll}
  Quantity/Run Condition& Reactor On & Reactor Off & Calibration \\ \hline
  Acquisition Event Rate (kHz) & 28 & 4 &  35 \\ \hline
  Segment Event Rate (kHz) & 115 & 35 &  190 \\ \hline
  Avg. Segment Multiplicity & 4.0 & 7.0 &  5.5 \\ \hline
  Max Opt. Link Rate (MB/s) & 3.0 & 1.0 &  7.2 \\ \hline
  Min Opt. Link Rate (MB/s) & 1.1 & 0.6 &  2.2 \\ \hline
  Data Volume per Day (GB) & 671 & 312 &  476 \\ \hline
  \end{tabular}
  \caption[dataRates]
  {Approximate data acquisition and transfer parameters for three typical operating conditions. The calibration case has five $^{137}$Cs sources deployed within the AD while the reactor is off.
  The average multiplicity is higher for the Reactor Off condition because muon and other cosmic events have  high multiplicity and these are are a greater fraction of events in this state. }
  \label{tab:dataRates}
  \end{centering}
\end{table*}

Data is transferred from the WFD modules to spinning disks on the two DAQ control computers. From there, it is immediately transferred to a multi-disk array for local storage. All acquisition related computers are connected via Gigabit Ethernet (Fig.~\ref{fig:daqLayout}).

\subsection{Clock distribution}

The V1725 WFD module can operate using either an internal or external clock. 
If a clock signal is received on the ``CLOCK IN'' input of a WFD module, it is mirrored on the ``CLOCK OUT'' output. 
One V1725 module is configured to act as the master clock for all modules, presenting a 62.5\,MHz differential clock signal to the ``CLOCK OUT'' output. 
Each successive module receives and mirrors this signal, so that the clock is distributed via a daisy chain from module to module. 
Between adjacent modules the daisy chain cables are approximately 0.05~m long. 
One longer cable ($\sim$1~m) is required to carry the clock signal between the two VME crates. 
The propagation delays inherent to this distribution scheme are measured and corrected for in data analysis. 


\section{Data processing and analysis framework} 
\label{sec-data}

Data is processed through multiple stages as described in this section. Processing time and resource estimates for each stage are given in Table~\ref{tab:dataProcess}.

\subsection{Raw data}

When the WFD memory buffer is full, raw waveform data is transferred via the optical link to the DAQ control PCs. That data is immediately written to disk in a compressed binary format, with one file being populated for each digitizer board per run. The run duration is typically one hour.

\subsection{Unpacked data}

An unpacking stage combines the raw data files from the multiple digitizer boards into a single file and converts the compressed binary format of the raw data.
The fundamental information, i.e. the digitizer waveforms, remains the same.
Thus, this step does not involve any physical or data analysis processing and only is a different format of the original data.
A channel map between the physical hardware channels and their ``logical'' functions ({e.g.} PMT positions in the detector) is included in the unpacked file.

\begin{table*}[tbp]
  \begin{centering}
  \begin{tabular}{l|lll}
  Processing Step/Run Condition& Reactor On & Reactor Off & Calibration \\ \hline
  Raw File Size (GB/run) & 29 & 13 &  22 \\ \hline
  Unpacked File Size (GB/run) & 30 & 13 &  23 \\ \hline
  Raw $\rightarrow$ Unpack processing time (CPU-min/file)& 98 & 44 &  77 \\ \hline
  DetPulse File Size (GB/run) & 8.2 & 3.7 &  4.9 \\ \hline
  Unpack $\rightarrow$ DetPulse processing time (CPU-min/file)& 58 & 26 &  37 \\ \hline
  PhysPulse File Size (GB/run) & 3.2 & 1.4 &  2.4 \\ \hline
  DetPulse $\rightarrow$ PhysPulse processing time (CPU-min/file)& 14 & 6.2 &  8.7 \\ \hline
  \end{tabular}
  \caption[dataRates]
  {Typical data file sizes and processing times for three typical operating conditions (Reactor On, Reactor Off, and Calibration). The file sizes given are for a typical run length of 1 hour, except for calibration, which is 10 mins. With typical availability of collaboration cluster computing resources, a year's worth of data can be processed in under four days. }
  \label{tab:dataProcess}
  \end{centering}
\end{table*}

\subsection{DetPulse data}
Unpacked data is processed through a custom software utility called PulseCruncher which converts digitized waveforms into a summary of the signal pulses in those waveforms, without applying any calibration.
PulseCruncher reads each digitized waveform and identifies signal pulses there.
The output of the PulseCruncher is a file containing DetPulse objects, each of which has the following attributes: event number from the WFD board trigger counter, PMT number, pulse area and height in ADC units, pulse arrival time at PMT, waveform baseline, pulse rise-time, and a PSD parameter. 

\subsection{PhysPulse data}
A calibration is applied in the next stage, converting uncalibrated DetPulses to calibrated PhysPulses.
The calibration is applied using a database storing the interpreted calibration results extracted from earlier data.
Applying the calibration combines information from both PMTs in a pulse's segment, so each PhysPulse is the combination of two DetPulses, including information about the segment as a whole and the signal in each of the two PMTs.
Each PhysPulse object contains the event number, segment number, pulse energy (MeV\textsubscript{ee}), pulse start time (in ns from run start), $\Delta t$ (time difference between the two combined PMT signals), estimated number of photoelectrons detected by each PMT, reconstructed position of the pulse along the segment axis, PSD parameter, and the identified particle type.

\section{Detector assembly at Yale}
\label{sec-assembly}

Most of the PROSPECT detector was assembled and tested at the 
Yale Wright Laboratory before shipment to ORNL. 
The unfilled (dry) detector included all active and passive components
inside the outer aluminum tank. 
Cables, gas, and liquid lines exited the aluminum lid via gas-tight feedthroughs.
Commissioning of the completed dry detector with cosmic rays 
and the light calibration system verified  the cabling and PMT mapping.
Cosmic ray signals in the PMT housing mineral oil  provided a sensitive baseline to compare detector performance before and after shipping.
Additionally, the outer plastic lumber pieces were test assembled at Yale and numbered for easy re-assembly onsite. 

\subsection{PMT module assembly}
\label{PMT-assembly}
PMT modules were assembled in a class 1000 clean room by 
teams of shifters from all collaborating institutions.
Internal parts were laser cut or machined externally, received and cleaned, then sub-assemblies and inner components were prepared for full module assembly.  
All components in contact with LiLS or mineral oil were rinsed in 10~M$\Omega$cm deionized  water (DI) before being soaked in a solution of ethanol or Alconox\textregistered\footnote{https:www.alconox.com/} (1\% by weight), depending on chemical compatibility, and then rinsed multiple times with  DI water until the collected rinse water measured 10~M$\Omega$cm.

\begin{figure*}
\centering
\includegraphics[clip=true, trim=0mm 45mm 0mm 40mm,width=6.0in]{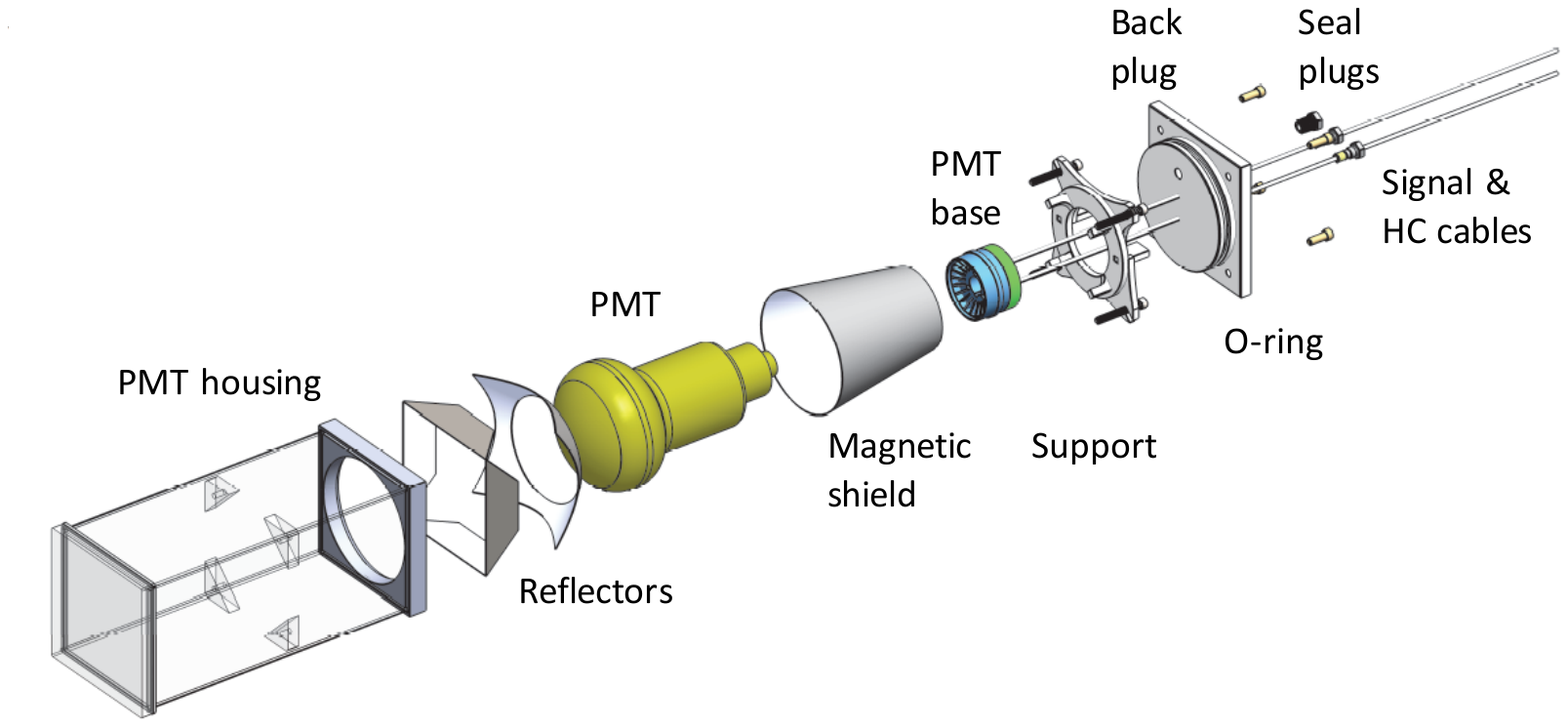}

\caption{
PMT assembly sequence. Starting with a cleaned, leak checked housing, reflectors are glued to the front side walls, the conical reflector is squeezed through the back opening and pushed against the front window. 
The PMT and magnetic shield are pushed against the conical reflector and secured in place with an acrylic support. 
A back plug assembly is made by threading the cables through the seal plugs and soldering to the PMT base. 
The base is pushed onto the PMT pins, seal plugs are tightened around the cables and temporary screws secure the plug to the back of the housing.}
\label{fig:Exploded}
\end{figure*}

The assembly sequence is shown in Fig.~\ref{fig:Exploded}.
After QA and cleaning of the acrylic housing, adhesive backed reflective film 
was applied on the inside walls near the front window in areas  not covered by the reflector cone, which was inserted next. 
In parallel, the internal support structure was cemented together with Weldon 16\textregistered\footnote{https:www.Weldon.com/}. 
The back plate of the module was pre-assembled by threading signal and HV cables through 
the PEEK plugs and acrylic end plug before the cables were soldered to the PMT base. 
Finemet magnetic shielding was slipped over the bulb of the tube, followed by the PMT support. 
The base was attached to the back of the PMT and the assembly lowered into the housing. 
An expansion bladder, made of ~150~cc plastic bubble wrap, was trapped between the Finemet and internal supports. 
The internal supports arms   were tightened to the sides of the housing until the bulb of the tube was snugly pressed against the reflector cone. 
The back plate (with Krytox\footnote{https://www.chemours.com} greased O-ring)  was inserted into the opening of the housing and retained by temporary nylon screws.  

A leak check was performed by pressurizing the module with 5.5~kPa (55 mbar) of nitrogen while submerged under water. 
Good modules were placed in a dark box for a current monitored burn-in at operating voltage (-1500~V) for 48 hours. 
The modules were then filled with mineral oil and  re-tested in the dark box to determine optical properties. 
Every module was cleaned as previously described  and thoroughly rinsed with DI water. 
PMT housings underwent a final 12 hour dark box test and resistance check prior to installation in the detector.

\subsection{Detector assembly}
\label{det-assembly}
Assembly of the inner detector on the acrylic tank base began at the Yale Wright Laboratory in early November 2017 inside a
soft-walled class 10000 cleanroom. The custom cleanroom had high ceilings to accommodate the detector and assembly scaffolding
and could split into two parts for overhead crane access.
A painted steel base on four Hilman\footnote{http://www.hilmanrollers.com} rollers held 
the assembly  at an ergonomic height, provided a level surface with flatness $<0.13$~mm and supported a
rigid frame  surrounding the assembly area.  
A rectangular frame  attached to  vertical posts could be mounted at adjustable heights
to provide a  reference for survey of the inner detector components as the detector was assembled row by row. 
The acrylic base was supported by an array of polyethylene 
blocks  to allow tensioning cables (Section~\ref{sec:Cables}) and lifting straps (Section~\ref{sec:Lifting}) to be threaded under
the completed assembly while still providing nearly uniform support to the acrylic baseplate. 

\begin{figure}
   \centering
      \includegraphics[clip=true, trim=0mm 0mm 0mm 0mm,width=3.4in]{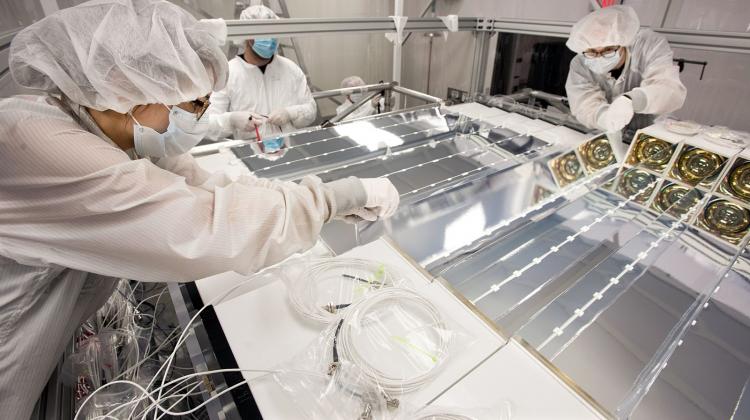}
   \caption{
Detector assembly midway through the top row. A vertical reflector optical separator is inserted into the pinwheel arms (white tabs) and between housings.
The white PMT housing bodies and clear front windows are visible on the near side while the far side shows the PMT faces and reflective cones.
The top reflector optical separators were installed after all PMT housings and vertical reflectors of that row were installed.}
   \label{fig:Assembly}
\end{figure}

The bottom layer of acrylic supports was installed, centered on the acrylic tank base and surveyed to initiate the detector assembly.
The lowest layer of reflector optical separators and pinwheel rods was installed, held in position by slots in the supports.
Vertical reflector optical separators and  PMT modules were installed in sequence, dividing the segments in that row, as seen in Fig.~\ref{fig:Assembly}.
The backs of the housings were held in place by horizontal acrylic planks that tied a given row to the layer of housings below.
Each row was completed by installing the upper horizontal reflector optical separators. 
The housing and pinwheel rod positions were surveyed. 
Teflon shims were added to the top of the pinwheel spacer arms or end plugs to minimize
any accumulated height variation produced during assembly. 
This process was repeated row by row. Each layer was supported by the layer underneath it.
The top support ribs were attached over the detector array, providing a vertical constraint to the reflector grid and tying the vertical walls of the segment supports together. 
Vertical acrylic bars were then mounted on the horizontal planks connecting the PMT housings to provide additional vertical constraint. 

The outer support structure was shimmed tightly against the acrylic base to prevent movement during shipping (Section~\ref{sec:shipping}).
O-rings for the face seal between the acrylic  tank side walls and the acrylic base were held  in position by additional shims  
and covered by a generous lubrication of Krytox grease. 
The clean room was opened,  the acrylic side walls were lifted over the completed assembly and then lowered  on to the O-rings. 
Temporary blocking was then installed to support the acrylic tank lid $\sim$0.60~m over the assembly to allow routing of the 
signal, HV cables, gas, bubbler and fill lines  through holes in the acrylic tank lid.
The lid was then lowered onto the side walls cushioned by a 0.381~mm 
Teflon layer, preventing acrylic to acrylic contact. 

\subsection{Tensioning cables}
\label{sec:Cables}

Sixteen stainless steel cables were looped over the lid and under the bottom of the acrylic tank to compress the wall onto the O-rings at the base
of the acrylic tank as seen in Fig.~\ref{fig-acrylic}. Tensioned to 1300N each by turnbuckles, these cables compress the O-rings by 20\,\% ensuring a positive seal. 
To  prevent direct contact between the wire rope and the acrylic tank, 2.5 mm-thick aluminum angles cushioned by 0.00635~m 
plastic strips were placed along the edges of the acrylic tank. 
The turnbuckles were placed on the top of the assembly to allow adjustments of the wire tension as needed.
A test port between the  double O-rings  was tested at 7~kPa 
to verify the seal before and after the acrylic tank was lifted.

\subsection{Final assembly}
\label{sec:Lifting}

\begin{figure}
\centering
\includegraphics[clip=true, trim=165mm 15mm 0mm 130mm,width=3.4in]{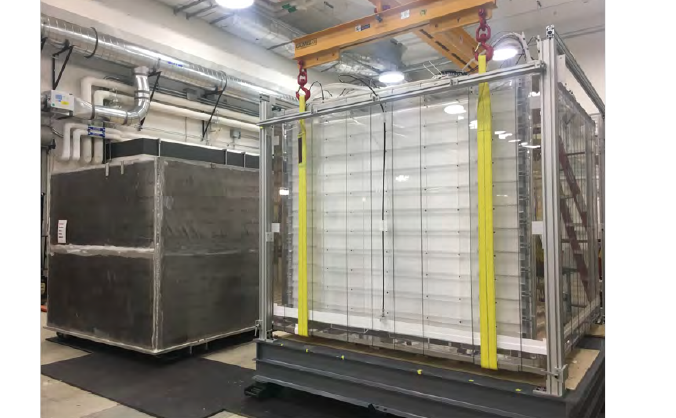}
\caption{ 
The inner detector on the right is ready for insertion into the outer aluminum tank shown on the left.
}
\label{fig-tanks}
\end{figure}

The aluminum tank was prepared  with a  BPE liner in the high bay of the Wright Lab. 
The completed inner detector assembly was wheeled from the cleanroom to a position next to the aluminum tank (Fig.~\ref{fig-tanks}).

Pre-stretched lifting straps were threaded underneath  the detector and attached to the shackles of a custom H-beam lifting fixture. 
The entire inner detector  assembly was lifted $\sim$2.5 m  and the aluminum tank  positioned underneath. 
The Hilman rollers  provided finer  positional control than horizontal movements of the crane and allowed fine tuning of the relative position 
as the crane lowered the inner assembly into place. 
The outer aluminum tank and inner acrylic tank were concentric within 1 cm.
The inner assembly was then shimmed in place using lengths of BPE.
The aluminum tank lid was positioned on blocking over the detector. 
Cables, calibration tubes, gas, fill, and sensor lines were all routed through their respective holes in the lid and the lid was lowered onto the 
aluminum tank walls and bolted in place. 
Icotek cable entry systems were mounted around each group of  cables and tubing. 
A special potting mixture of silicone caulk and graphite was poured over the icotek fittings to ensure the detector was light and gas tight. 
Signal and HV cables were laid in protective aluminum raceways fixed on the lid and routed to bulkhead plates. 
A brief dry commissioning of the electrical connections was performed prior to packing the detector for shipment to ORNL/HFIR, 
during which the detector was purged with argon and nitrogen.

\section{Detector installation into HFIR}

The main components of the PROSPECT detector were constructed or assembled off-site and shipped to ORNL for installation.
When possible, test assemblies of the shielding were made off-site to test fit and assembly techniques.
LiLS was shipped from BNL in Teflon-lined barrels to ORNL and pumped into an ISO Tank storage container~\cite{Band:2013zka}. 
The detector chassis was prepared with lead shielding and the air caster system before insertion into the HFIR experimental room.
The dry detector was placed onto the chassis and moved into its final location and then filled with LiLS.
Layers of lead, polyethylene, borated polyethylene and water containers were added to complete the detector shielding. 

\subsection{Shipment to ORNL}
\label{sec:shipping}

After dry commissioning of the assembled detector at Yale the aluminum tank containing the detector  was packed into a wooden shipping crate. The detector was cushioned by 0.1~m (4") foam (density 16 kg/m$^3$, 6 lbs/cu ft) underneath and by a ring of 0.05~m  (2") foam around the sides.
The crate was loaded into an enclosed air ride trailer and driven directly to ORNL. 
The detector was unloaded and stored under nitrogen cover gas in a HFIR maintenance facility. 

Shipment of the assembled detector was considered to be the  highest risk operation of the assembly and installation procedures.
To alleviate concerns about how well the detector would survive the shocks and vibrations of the road trip, prototypes
of the inner detector grid and a 3 by 3 array of PMT housings were subjected to  hours-long standardized vibration tests that mimicked 
the expected ride in an air ride trailer. No structural damage was observed. 
In particular, the fit of the optic segment components was quite snug and no abrasion of the thin Teflon coatings on the optical separators was observed.
Dry commissioning tests at ORNL  were very similar to the final tests at Yale, indicating no significant change in the internal detector elements.

\subsection{Liquid preparation}
\label{sec-liquidprep}

The LiLS filled drums were shipped to ORNL inside temperature controlled trucks in three batches. 
Bags that were continuously flushed with boil-off nitrogen were placed over each drum lid to limit oxygen intrusion while stored at ORNL.
A 20-ton Teflon lined shipping container (ISO tank)  previously used in the Dayabay experiment~\cite{An:2015qga,Band:2013zka}  was refurbished and cleaned at Yale.  
Several alcohol rinses of the tank interior were made in addition to a final rinse of EJ309.  
The tank was shipped to ORNL and fully purged with nitrogen. 

\begin{figure}
\centering
\includegraphics[width=3.2in]{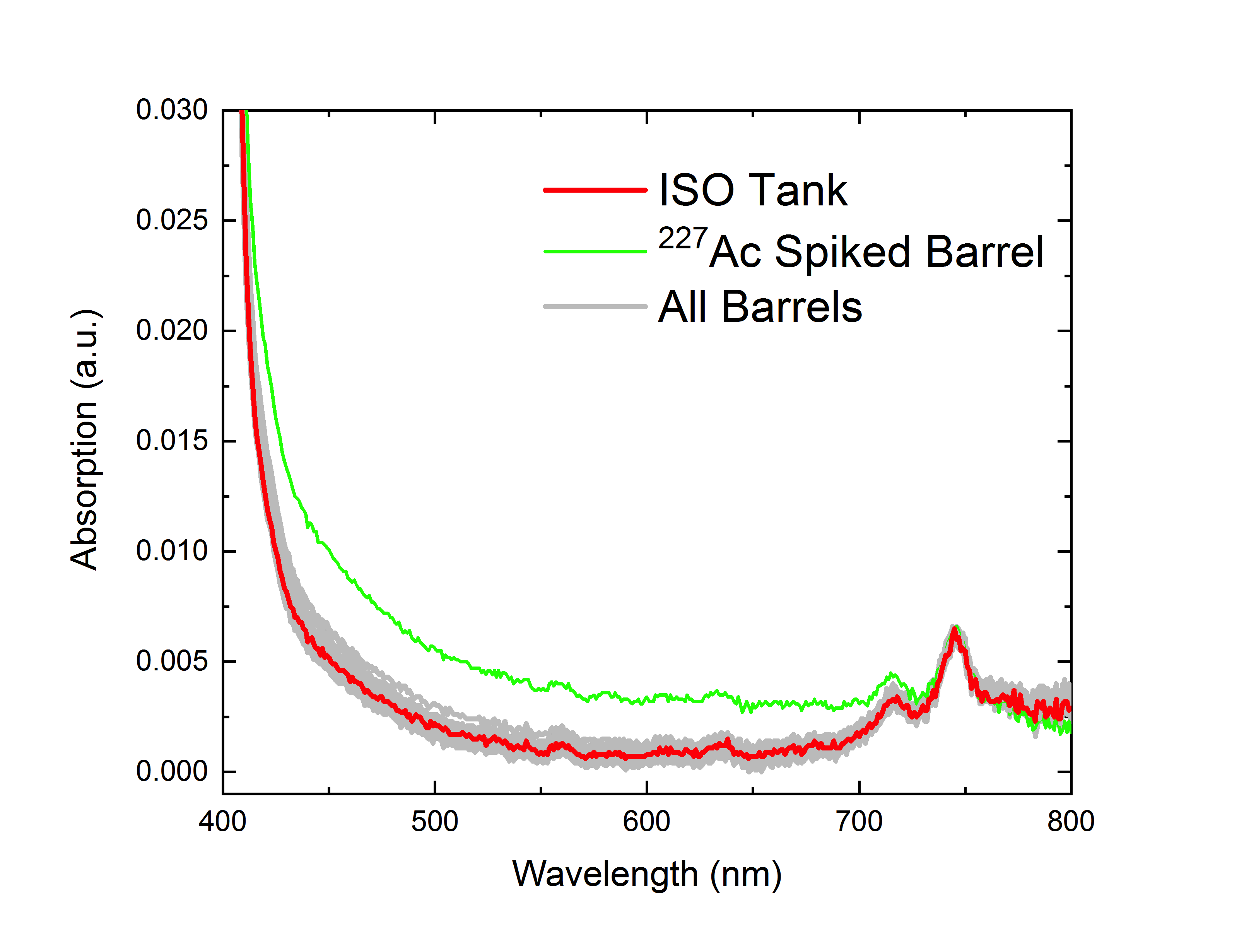}
\caption{
UV-Vis absorption spectra of the 28 drum samples (multiple colors) and the mixed ISO tank sample (red). 
Only the barrel spiked with actinium (light green) lies significantly outside the narrow range of spectra.}
\label{fig-UVis}
\end{figure}

A  pallet jack scale\footnote{Vestil PM-2748-SCL-LP https://vestil.motionsavers.com}   was used to weigh each pallet of four drums before and after pumping the LS contents from the drums into the ISO tank.  
The peristaltic pump utilized Teflon and Viton transfer lines to prevent contamination of the liquids. 
Care was taken to minimize the exposure to  air while opening each barrel and inserting the pump-out lines.
At two liters-per-minute, more than three days were needed to empty the barrels into the ISO tank.  
The barrel containing actinium was the fourth barrel emptied. 
Samples were taken from each drum and measured by a
UV-Vis spectrometer\footnote{Shimadzu UV-2700 https://www.shimadzu.com/}.
The UV absorption spectra of these samples are shown in Fig.~\ref{fig-UVis}.
The actinium barrel was the only barrel to show significant deviation from the average spectrum. 
All spectra were consistent with earlier measurements at BNL.
Nitrogen was bubbled through the liquid in the ISO tank for ten days to promote mixing of the different barrels. 
A sample from the mixed ISO tank is consistent with the expected average of all barrels.  A  total of 4841 kg  of LiLS was pumped into the ISO tank.

\subsection{Detector insertion into   HFIR}
\label{sec-HFIRassembly}

\begin{figure}[ht]
   \centering
      \includegraphics[clip=true, trim=0mm 0mm 0mm 0mm,width=3.4in]{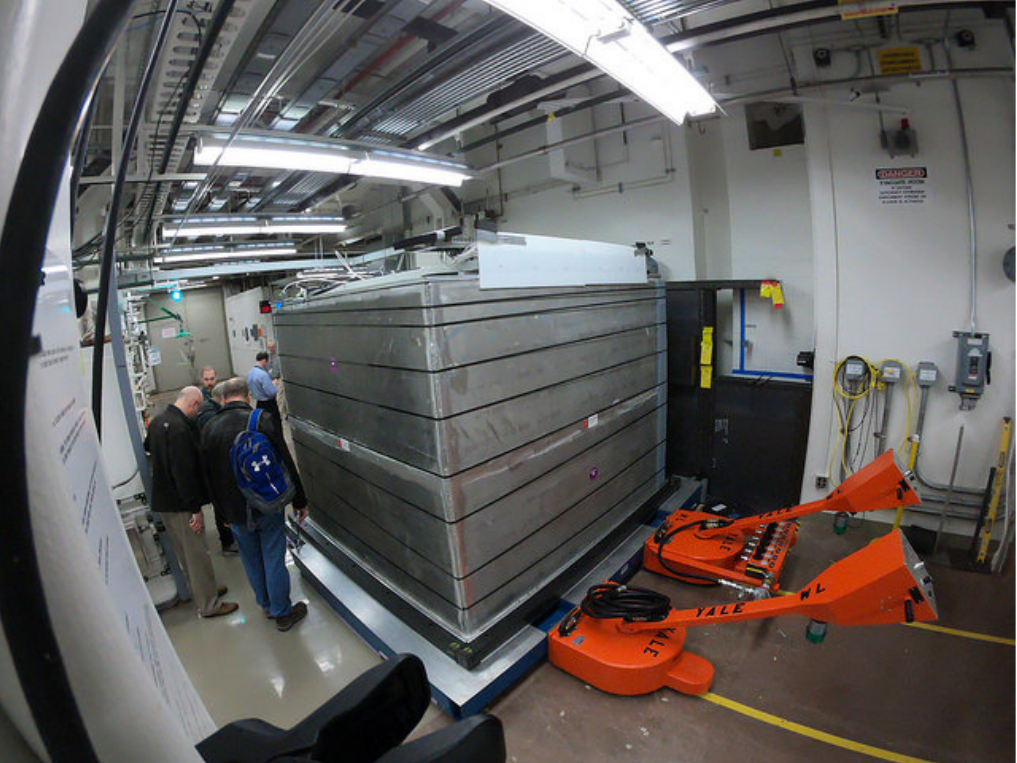}
   \caption{Fisheye view of the detector and chassis after being moved into Position 1 by the air casters and air drive motors (orange).}
   \label{fig:Move}
\end{figure}

The aluminum tank containing the PROSPECT detector elements was lifted by a large forklift, inserted through the
outer HFIR experimental room doors, and centered on previously installed chassis. The air caster system was then used to move the chassis
a few meters for installation of the north-side lead.  
The air casters were then used to move the detector/chassis assembly into Position 1 (see
Fig.~\ref{fig:Move}).

\subsection{Detector filling}
\label{sec-detfill}
The LiLS was stored for several weeks before the ISO tank was moved onto a truck bed and parked outside the outer door of the HFIR experimental
room.  
The tank was covered with a plastic tent to protect against the elements.  
A 19 mm Teflon pump-out line was routed through the door to the peristaltic pump previously used and to a detector fill line which went to the bottom of the acrylic tank.  
Although provisions were made to pass the pump-out line through a heat exchanger to equalize the LiLS and detector temperatures, 
no action was needed as the ISO tank and detector temperatures were within a few degrees of each other. 
Boil-off nitrogen from two dewars provided continuous cover gas flow into both the detector and ISO tank during the filling operation.

\begin{figure}
\centering
\includegraphics[clip=true, trim=10mm 10mm 0mm 20mm,width=3.1in]{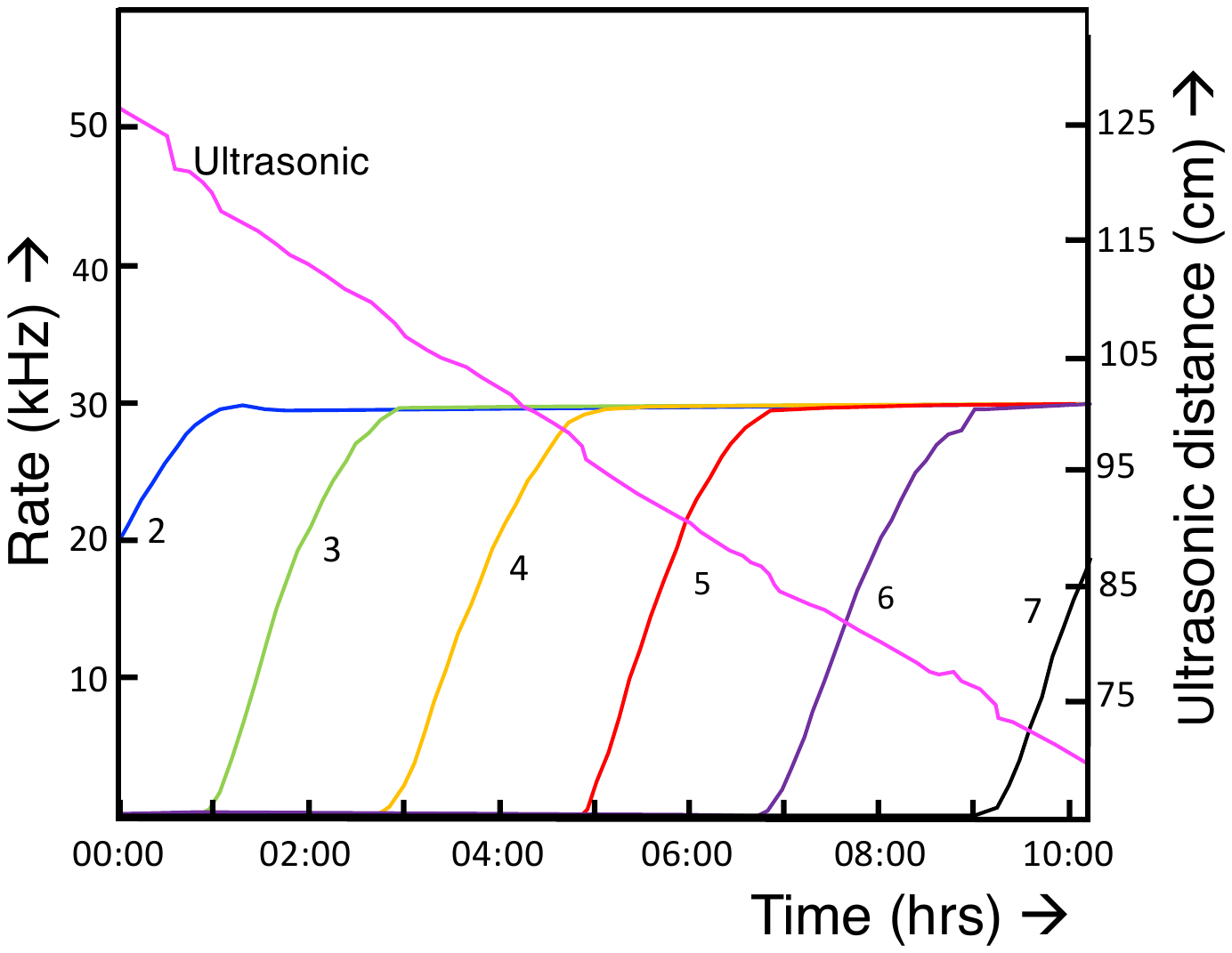}
\caption{
Ultrasonic sensor reading of the LiLS height and the trigger rate from detector segments in column 6 (labeled by row number) as a function of time partway through detector filling. 
The trigger rate (left axis) rises as soon as LiLS enters a given segment and saturates when that segment is completely filled.
The ultrasonic sensor measures the distance between the LiLS surface and the top-mounted sensor (right axis). Changes in slope near row transitions are visible.
}
\label{fig-Fill}
\end{figure}

The detector was tilted along its long axis by $0.7^{\circ}$ to prevent bubbles from being trapped in the optical grid structure. 
After purging the transfer lines, LiLS samples were taken for later study. 
The liquid was pumped at $\sim$3 liters per min. The height in the acrylic tank was measured by ultrasonic liquid level sensors and monitored by the DAQ system. 
The number of light pulses recorded by the PMTs varied strongly with the amount of liquid in a given segment and provided a 
clear indication when the LiLS started filling a given row of segments as seen in Fig.~\ref{fig-Fill}.
Changes in slope of the liquid level were also visible when the liquid level rose above segment boundaries. 

When the liquid level approached the top of the top segments, pumping was stopped and the PMTs were turned off to make a 
visual inspection of the liquid level through 2 acrylic windows on the detector lid. 
Liquid was then pumped to cover the upper segment completely. 
The detector was restored to level and $\approx 1$cm of LiLS was added. 
Water was pumped into the space between the acrylic tank and aluminum tank in several stages during the LiLS filling process.

The remaining LiLS in the ISO tank was pumped into three storage barrels and weighed. 
The difference between the weight of liquid pumped into the ISO tank and the storage barrels represented the weight of LiLS (4340 kg)
pumped into the PROSPECT detector after correcting for the various liquid samples. 
Similarly, the weight of the water pumped into the detector (403 kg) was determined from the weight of the drums before and after filling.

\subsection{Final assembly}
\label{sec-Finalassembly}

After the filling operation and subsequent commissioning checks
a lead layer of 0.025~m $\times$ 0.10~m $\times$ 0.30~m interlocking brick was stacked around the perimeter of the aluminum tank and secured by plastic strapping.
Rows of 0.10~m $\times$ 0.10~m recycled polyethylene lumber were stacked on each other log cabin style and secured together by lag screws.
The wall served as additional restraint for the lead bricks and supported the roof structure. Along the east and west faces transition boxes
were installed at the top of the walls to allow routing and connections of source and gas tubes (west side) and signal, HV, and monitoring  cables (east side).

Roof beams also of recycled polyethylene lumber were secured on top of the log cabin walls. 
A 0.025~m thick layer of borated polyethylene was added to cover the walls and top of the assembly. 
All plastic surfaces were then covered by thin aluminum sheets.
A 11 $\times$ 18 array of water filled containers  added to the roof completed the shielding assembly. 

HV, signal, and monitoring cables were routed from bulkhead connectors on panels in the east transition box  to three racks next to the detector.
These movable racks could be rolled 1.5~m from the detector for cabling access or secured to the detector for earthquake safety. 
Sources and source motors were then installed to complete
the PROSPECT detector installation.

\section{Performance}
\label{sec-performance}

PROSPECT began taking data in March 2018. 
Initial performance results are presented here, based on data taken during one partial Reactor On cycle and part of a Reactor Off cycle. 

\subsection{Response over longitudinal position}
\begin{figure}
\centering
\includegraphics[clip=true, trim=25mm 38mm 140mm 10mm,width=3.1in]{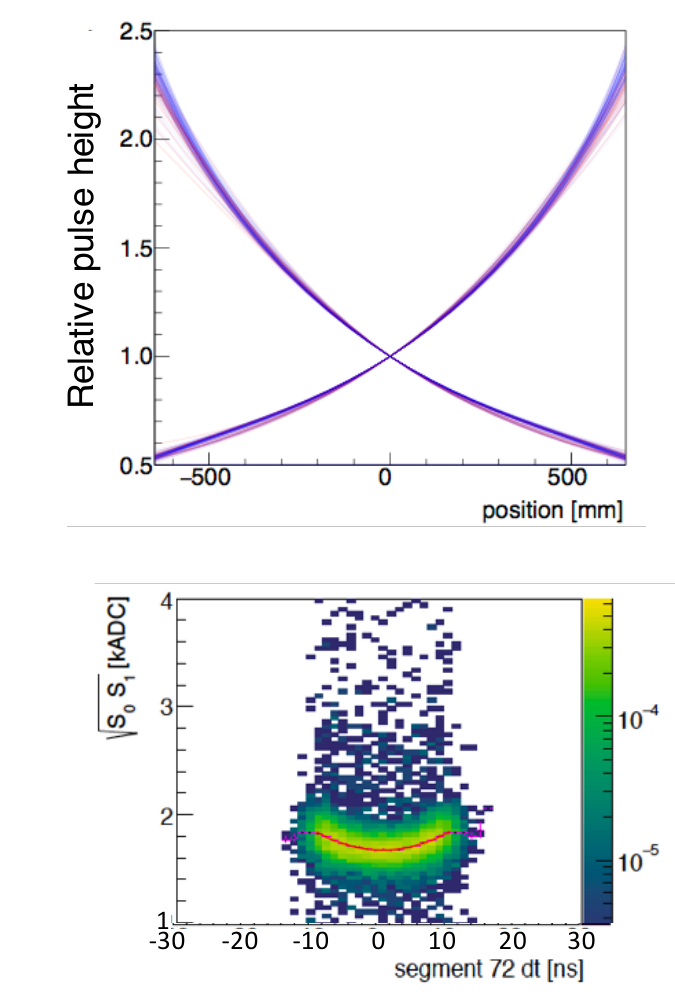}
\caption{
Shown on top are the relative pulse height distributions  for each of two PMTS in all 154 detector segments, scaled to be 1.0 at the detector center, as a function of longitudinal position (determined from timing) along the segment.
Hamamatsu (ET) PMTs are shown in blue (red).
All curves are approximately exponential. The bottom plot shows the geometric mean of the two PMT pulse heights (in 1000 ADC counts) for one arbitrarily chosen segment, demonstrating that the $z$-dependence is not purely exponential, but clearly correctable. The red line shows our parameterization.}
\label{fig:PHvsZ}
\end{figure}

Pulse heights (S0,S1) in the two PMTs on either end of a 
segment are combined to measure the energy deposited in that segment.
Figure~\ref{fig:PHvsZ} (top)
shows the relative pulse height of $^6$Li captures versus longitudinal ($z$) position along the length of a segment for all 154 segments. The z-dependence is approximately exponential. If the z-dependences were purely exponential then
an energy determination proportional to the geometric mean (S0S1) of the pulse heights would be independent of position.
The bottom of Fig.~\ref{fig:PHvsZ} 
scatterplots the geometric mean of the PMT signals for a sample of $^6$Li captures versus position.
The observed geometric means have a small remaining position dependence. The energy reconstruction algorithm uses the red line fit to this position dependence and the geometric mean of the PMT pulse heights to calculate the segment energy.

\subsection{Pulse shape discrimination}

\begin{figure}
\centering
\includegraphics[clip=true, trim=0mm 2mm 150mm 0mm,width=3.4in]{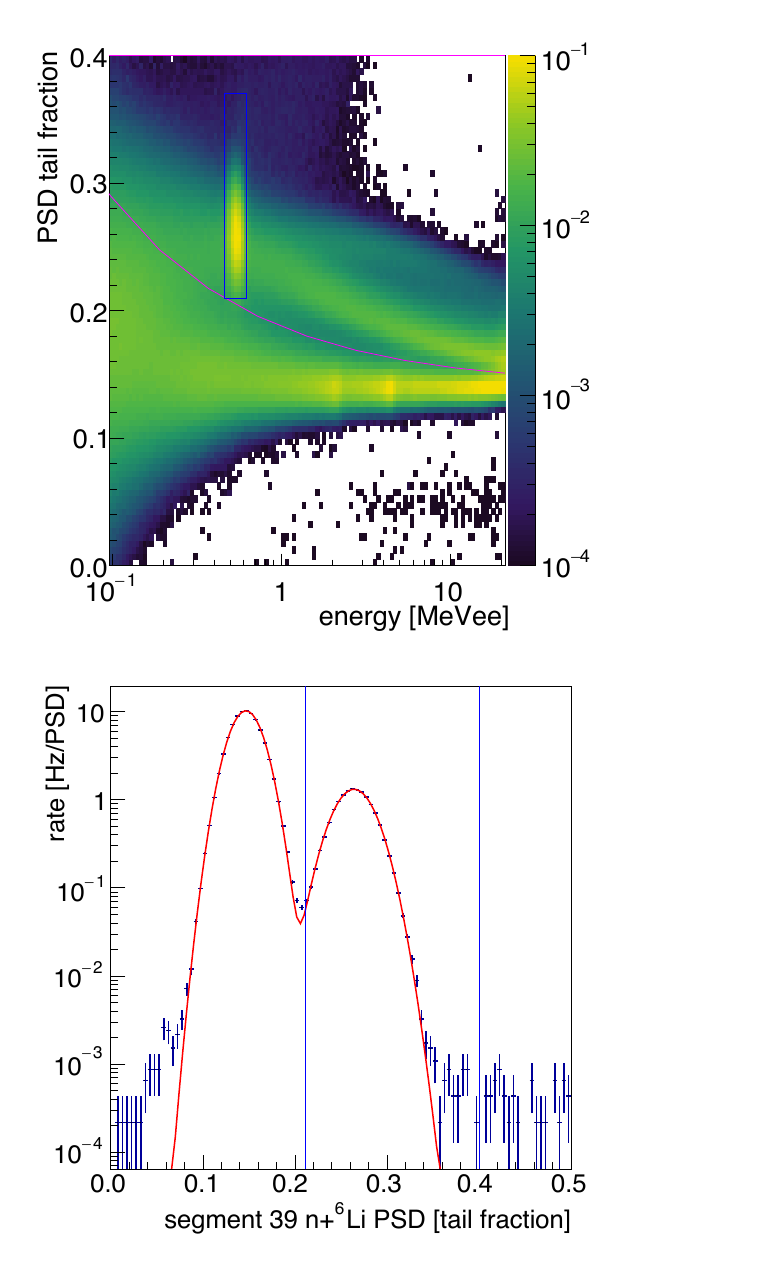}
\caption{Demonstration of PSD performance. To better highlight different event types, this plot displays prompt energy depositions correlated with a subsequent neutron capture on $^6$Li.  The top scatterplot shows the distribution of events according to the fraction of the pulse area in the tail, versus  (logarithm of the) energy. In the present analysis, the acceptance cut for $^6$Li is represented by the blue rectangle and the pink curve shows the upper cut for identifying electron-like signals as a function of energy.
The separation based on PSD is clear, with the lower histogram  showing the projection onto the PSD axis with the  blue lines showing the acceptance cut for $^6$Li.}
\label{fig:PerfPSD}
\end{figure}

Pulse Shape Discrimination is a critically useful tool for PROSPECT distinguishing the products of the reaction $n+^6{\rm Li}$ from electrons, photons, and other minimum ionizing background signals. 
The PSD tail fraction is the fraction of ADC pulse height in the tail window  (44ns to 100ns) divided by the full ADC integration window (-12ns to 100ns) where the times  are relative to the 50\% height of the leading edge of the pulse. 
Figure~\ref{fig:PerfPSD}
shows how this approach performs in PROSPECT, displaying a scatter plot of single pulses as a fraction of the total pulse area in the tail versus energy on a logarithmic scale. 
The horizontal band extending up to high energies with tail fraction near 0.1 is due to the many electron-like and minimum ionizing backgrounds. A clear collection of events with energy near 0.55 MeV and tail fraction near 0.25, are neutron capture events on $^6$Li. The two types of signals are well separated.

Interestingly, Fig.~\ref{fig:PerfPSD} also shows a long band extending to high energies, but with tail fraction near 0.25 at low energy, and decreasing as the energy increases. These are due to recoil protons from $np$ collisions of energetic cosmic ray neutrons. At the highest energies, the tail fraction decreases with decreasing ionization density.

\subsection{Electron/$\gamma$-ray backgrounds}

The IBD signal for an antineutrino interaction in PROSPECT, requires a prompt electron-like signal followed by a delayed neutron capture signal, that is, both classes of signals shown in Fig.~\ref{fig:PerfPSD}. 
Consequently, backgrounds to these signals are important to understand, and to minimize.

\begin{figure}
\centering
\includegraphics[width=2.75in]{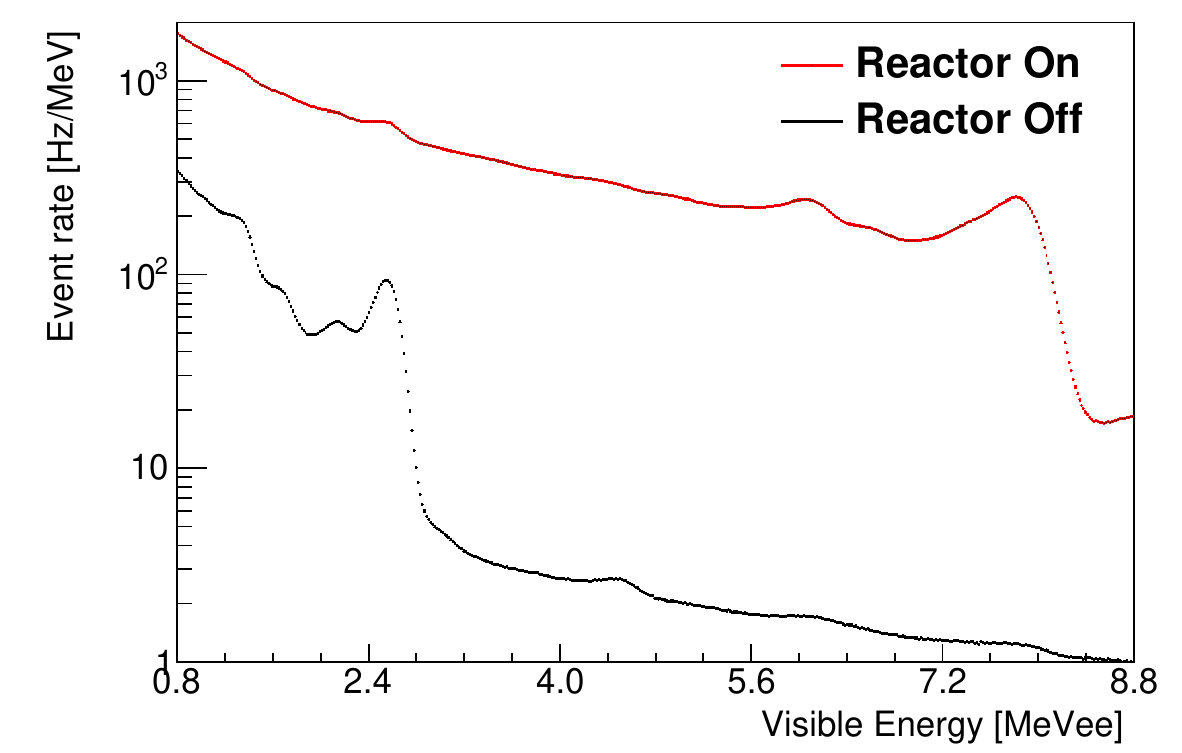}
\caption{
Energy distribution of electron-like signals in the PROSPECT detector, for Reactor On and Reactor Off samples. Radioactive background $\gamma$-ray  signals from $^{40}$K (1.4~MeV) and $^{208}$Tl (2.6~MeV) are evident. Higher energy structures are likely 5.9, 6.0, and 7.6\,MeV $\gamma$-rays  from neutron capture on $^{56}$Fe in the concrete rebar. The integrated electron-like singles rate is $\approx$5.2~kHz when the reactor is on, and $\approx$500~Hz when it is off.}
\label{fig:BkgdElike}
\end{figure}

\begin{figure}
\centering
\includegraphics[width=3.1in]{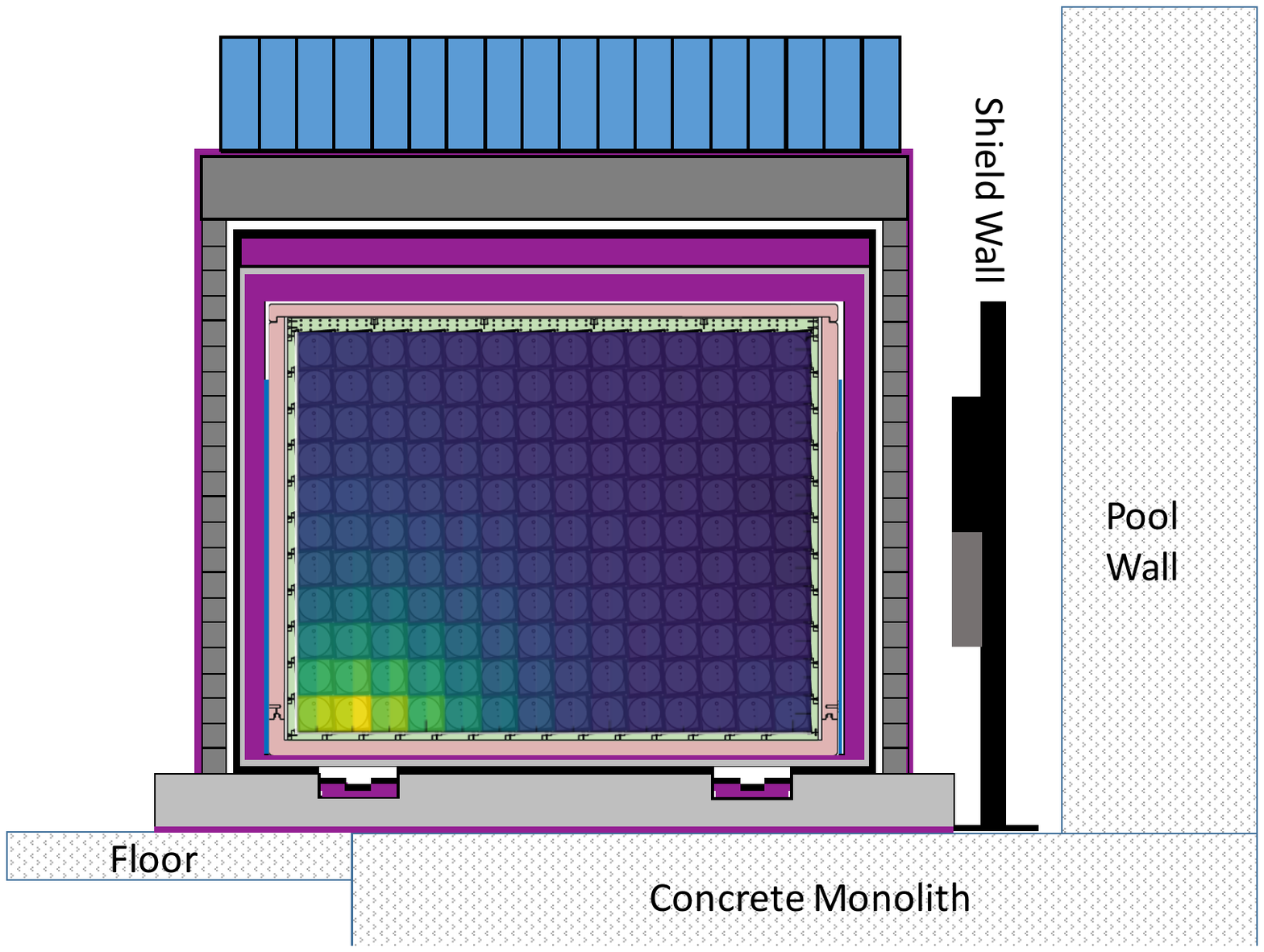}
\caption{
The rate per PMT of ($E\geq0.1$~MeV) 
as a function of segment and photomultiplier tube, in early PROSPECT data, with the Reactor On and with all shielding installed. Each square segment is subdivided to show the two PMT rates for each segment. The color scheme indicates rates from 200~Hz (dark blue) to 800~Hz (yellow).}
\label{fig:bkgd}
\end{figure}

The energy spectra of electron/gamma-like signals, for both Reactor On and Reactor Off, are shown in Fig.~\ref{fig:BkgdElike}.
The rate during  reactor operation is  much larger, as expected. Fig.~\ref{fig:bkgd}
displays the rate in each segment, for events with visible energy $E\geq0.1$~MeV, during an initial Reactor On period, after all of the shielding had been installed.
Demonstrating the effectiveness of the local shield wall, segments at the end of the detector toward the reactor are uniformly quiet, with rates  $\leq200$~Hz.  
Rates in segments at the opposite end of the detector are higher, closer to 800~Hz.  This region of the detector not only extends past the shielding monolith below and thus sees a significantly thinner floor, but is also above a break in the lead shielding due to the forklift channel. 
The shielding in the channel area will be modified to mitigate the effect due to the forklift channel.

\subsection{Neutron capture energy resolution}

\begin{figure}
\centering
\includegraphics[clip=true, trim=0mm 5mm 150mm 0mm,width=3.0in]{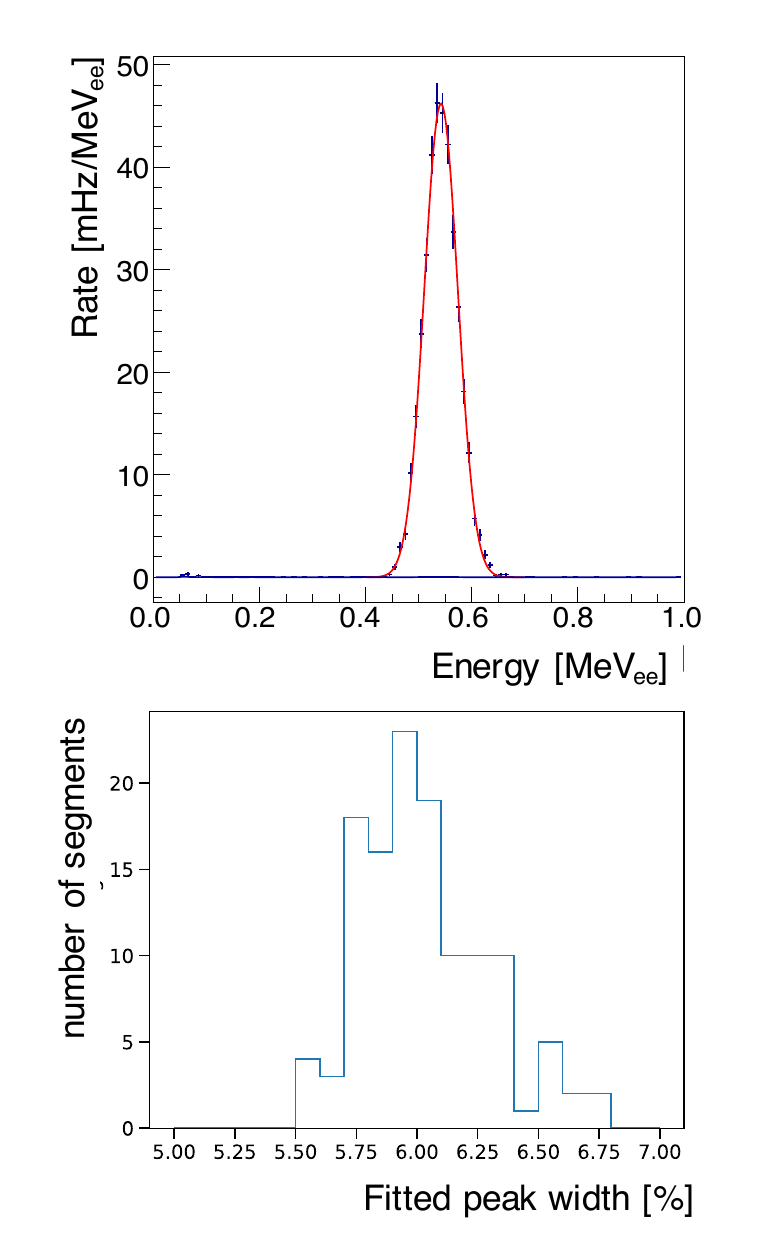}
\caption{
(Top) The measured energy distribution (in electron-equivalent MeV) of neutron capture events on $^6$Li is shown for a typical detector segment. Only events whose energy deposition is confined to that single segment are plotted. A Gaussian fit measures the segment energy resolution. (Bottom) The width of the Gaussian fit for all segments are histogrammed to show the segment to segment variation in energy resolution.}
\label{fig:nCaptureEnergy}
\end{figure}

The signal for delayed neutron captures after the PSD selection shown in Fig.~\ref{fig:PerfPSD} is robust. Figure~\ref{fig:nCaptureEnergy}
histograms the capture energy  distribution observed in  an arbitrarily selected single segment. 
Entries are selected by identifying a neutron capture in delayed coincidence with a fast neutron recoil. 
The bottom figure plots the standard deviation of the observed peaks in each of the 154 segments, as determined by a fit of the energy for capture events in a single run.

\subsection{Reactor associated events}

An IBD event consists of a prompt positron signal, followed by a delayed neutron capture signal. These two signals are selected by a preliminary analysis based on their energy and pulse shape. Backgrounds to IBD occur because of true prompt/delayed coincident processes; for example $n+^{12}{\rm C}\to n^\prime+^{12}{\rm C}^*$ where the 4.4~MeV photon from $^{12}{\rm C}^*$ de-excitation provides the prompt and the inelastically scattered neutron thermalizes and captures. Of course, backgrounds to IBD can also come from random accidental coincidences of prompt and delayed type signals.

\begin{figure}
\centering
\includegraphics[clip=true, trim=38mm 25mm 38mm 50mm,width=3.1in]{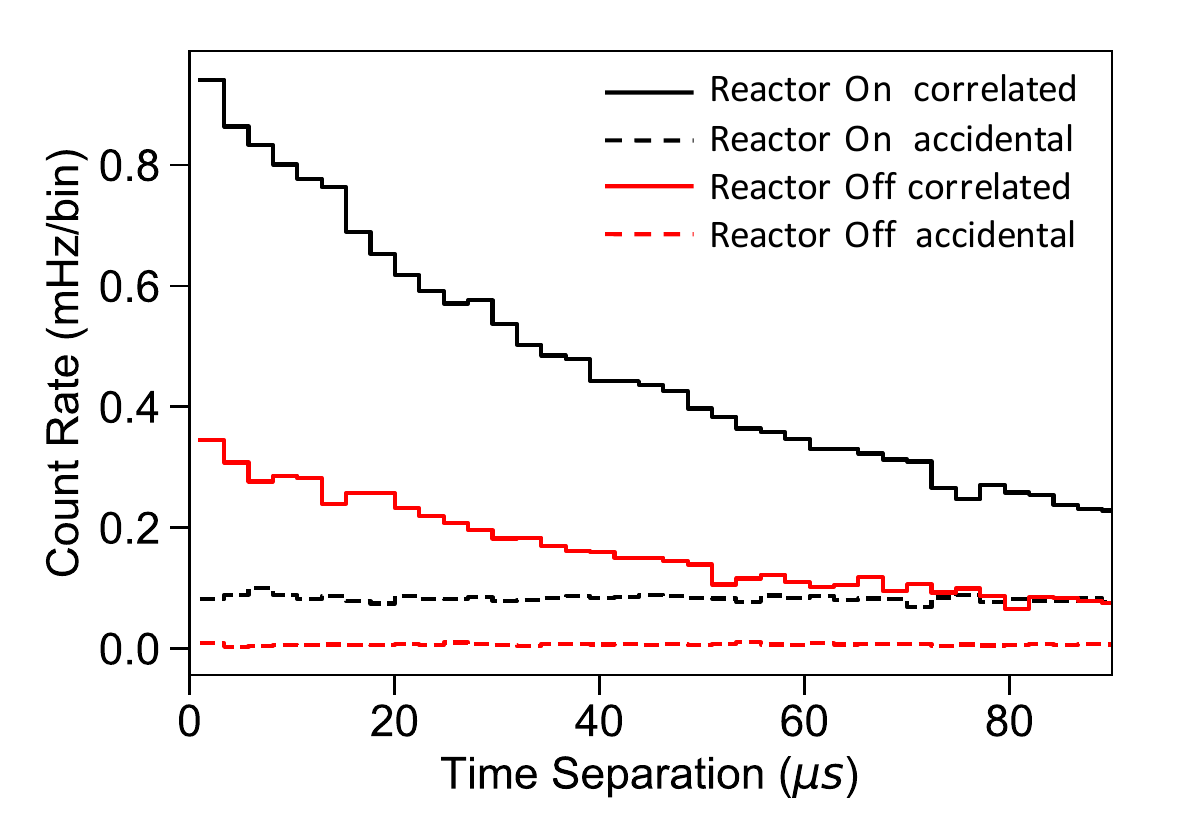}
\caption{
Histograms of the rate (per 2~$\mu$s bin) of the time distribution between ``prompt'' and ``delayed'' events. In ``correlated'' events the ``prompt'' precedes the ``delayed''
signal. ``accidentals'' have the wrong time ordering (i.e.\ the 
``delayed'' signal is earlier than the ``prompt'' signal).
The accidentals integrate over a 10~ms window for increased statistical precision.}
\label{fig:PerfCorrAcc}
\end{figure}

Figure~\ref{fig:PerfCorrAcc}
shows the prompt-delay time distribution for IBD candidates with the Reactor On and Off. An approximately $40~\mu$s time constant for ``correlated'' events is evident. Correlated events are present in both the Reactor On and Reactor Off samples, but the rate is higher by about a factor of two with the Reactor On. The accidental rate is flat, and very close to zero for the Reactor Off.

\begin{figure}
\centering
\includegraphics[width=3.1in]{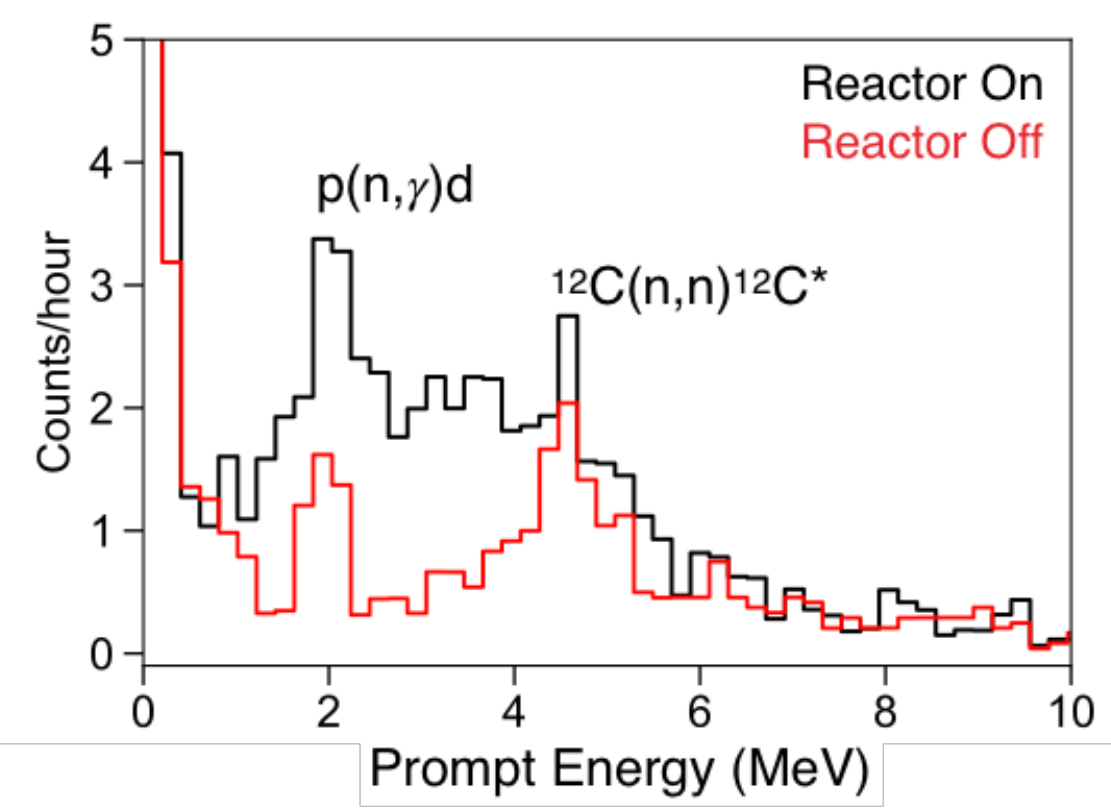}
\caption{
The prompt energy spectra for correlated events with the Reactor On and Reactor Off, for the first 24 hours of data in each case. Both spectra show prominent prompt energy peaks near 2.2~MeV and 4.4~MeV, but the spectra difference between the two dat sets has the expected general shape of a reactor antineutrino spectrum.}
\label{fig:24Hours}
\end{figure}

The prompt energy spectra for correlated events, after subtracting the accidental background, are shown in Fig.~\ref{fig:24Hours}
for roughly 24 hours of data with Reactor On and Off. The Reactor Off data are dominated by two peaks, near 2.2~MeV and 4.4~MeV. We interpret these as cosmic ray neutron capture on protons and inelastic neutron scattering from $^{12}$C, respectively, where the delayed neutron capture most likely comes from another neutron in the same cosmic ray air shower. The difference between the Reactor On and Reactor Off spectra has a shape consistent with the product of the reactor antineutrino spectrum and the IBD cross section. Further analysis development may reduce the prominence of the Reactor Off peaks.

\section{Conclusion}
\label{sec-conclusion}
We have constructed, installed and operated, a multi-ton, highly segmented, movable antineutrino detector at the 
High Flux Isotope Reactor at ORNL. 
PROSPECT operates well on the surface of the Earth with $< 1$~m of overburden within 7~m  of a research reactor. 
A custom $^6$Li-doped liquid scintillator provides both excellent light yield and discrimination between particle types through pulse shape discrimination.
An energy resolution of  better than 4.5\% at 1 MeV has been achieved.  
Signals from the neutron capture on $^6$Li are very localized and using PSD, distinct from the most common $\gamma$-ray  backgrounds.
A robust antineutrino signal was observed in less than one day of data with preliminary analyses. 
Time-correlated backgrounds from cosmogenic
neutron showers are well measured during Reactor Off data.  
A  signal to correlated background ratio of better than one-to-one has been demonstrated ~\cite{Ashenfelter:2018osc}.
The unique reflective grid design provides space for  both optical and radioactive sources at multiple locations
in the active detector volume to track detector performance. 
Energy calibrations are stable with time.
Initial results of a sterile neutrino search are being published and a measurement of the antineutrino energy spectrum
from  $^{235}$U  is in progress.

\section*{Acknowledgements }

This material is based upon work supported by the following sources: US Department of Energy (DOE) Office of Science, Office of High Energy Physics under Award No. DE-SC0016357 and DE-SC0017660 to Yale University, under Award No. DE-SC0017815 to Drexel University, under Award No. DE-SC0008347 to Illinois Institute of Technology, under Award No. DE-SC0016060 to Temple University, under Contract No. DE-SC0012704 to Brookhaven National Laboratory, and under Work Proposal Number  SCW1504 to Lawrence Livermore National Laboratory. This work was performed under the auspices of the U.S. Department of Energy by Lawrence Livermore National Laboratory under Contract DE-AC52-07NA27344 and by Oak Ridge National Laboratory under Contract DE-AC05-00OR22725. This work was also supported by the Natural Sciences and Engineering Research Council of Canada (NSERC) Discovery program under grant \#RGPIN-418579 and Province of Ontario.

Additional funding was provided by the Heising-Simons Foundation under Award No. \#2016-117 to Yale University. J.G. is supported through the NSF Graduate Research Fellowship Program and A.C. performed work under appointment to the Nuclear Nonproliferation International Safeguards Fellowship Program sponsored by the National Nuclear Security Administration's Office of International Nuclear Safeguards (NA-241).

We further acknowledge support from Yale University, the Illinois Institute of Technology, Temple University, Brookhaven National Laboratory, the Lawrence Livermore National Laboratory LDRD program, the National Institute of Standards and Technology, and Oak Ridge National Laboratory. We gratefully acknowledge the support and hospitality of the High Flux Isotope Reactor and Oak Ridge National Laboratory, managed by UT-Battelle for the U.S. Department of Energy.


\section*{References}
\bibliographystyle{model1a-num-names}
\bibliography{NIMcites}

\begin{thebibliography}{28}
\expandafter\ifx\csname natexlab\endcsname\relax\def\natexlab#1{#1}\fi
\providecommand{\bibinfo}[2]{#2}
\ifx\xfnm\relax \def\xfnm[#1]{\unskip,\space#1}\fi
\bibitem[{Mueller et~al.(2011)Mueller, Lhuillier, Fallot, Letourneau, Cormon
  et~al.}]{Mueller:2011nm}
\bibinfo{author}{T.~Mueller}, \bibinfo{author}{D.~Lhuillier},
  \bibinfo{author}{M.~Fallot}, \bibinfo{author}{A.~Letourneau},
  \bibinfo{author}{S.~Cormon}, et~al., \bibinfo{journal}{Phys.Rev.}
  \bibinfo{volume}{C83} (\bibinfo{year}{2011}) \bibinfo{pages}{054615}.
\bibitem[{Huber(2011)}]{Huber:2011wv}
\bibinfo{author}{P.~Huber}, \bibinfo{journal}{Phys.Rev.} \bibinfo{volume}{C84}
  (\bibinfo{year}{2011}) \bibinfo{pages}{024617}. \bibinfo{note}{Erratum-ibid.
  C85 (2012) 029901}.
\bibitem[{Mention et~al.(2011)Mention, Fechner, Lasserre, Mueller, Lhuillier
  et~al.}]{Mention:2011rk}
\bibinfo{author}{G.~Mention}, \bibinfo{author}{M.~Fechner},
  \bibinfo{author}{T.~Lasserre}, \bibinfo{author}{T.~Mueller},
  \bibinfo{author}{D.~Lhuillier}, et~al., \bibinfo{journal}{Phys.Rev.}
  \bibinfo{volume}{D83} (\bibinfo{year}{2011}) \bibinfo{pages}{073006}.
\bibitem[{Abazajian et~al.(2012)}]{Abazajian:2012ys}
\bibinfo{author}{K.~N. Abazajian}, et~al., \bibinfo{title}{{Light Sterile
  Neutrinos: A White Paper}}, \bibinfo{type}{Technical Report}
  \bibinfo{number}{FERMILAB-PUB-12-881-PPD}, {FermiLab}, \bibinfo{year}{2012}.
\bibitem[{Kopp et~al.(2013)Kopp, Machado, Maltoni, and Schwetz}]{Kopp:2013vaa}
\bibinfo{author}{J.~Kopp}, \bibinfo{author}{P.~A.~N. Machado},
  \bibinfo{author}{M.~Maltoni}, \bibinfo{author}{T.~Schwetz},
  \bibinfo{journal}{JHEP} \bibinfo{volume}{1305} (\bibinfo{year}{2013})
  \bibinfo{pages}{050}.
\bibitem[{An et~al.(2016)}]{An:2015nua}
\bibinfo{author}{F.~P. An}, et~al., \bibinfo{journal}{Phys. Rev. Lett.}
  \bibinfo{volume}{116} (\bibinfo{year}{2016}) \bibinfo{pages}{061801}.
  \bibinfo{note}{[Erratum: Phys. Rev. Lett.118,no.9,099902(2017)]}.
\bibitem[{Abe et~al.(2014)}]{Abe:2014bwa}
\bibinfo{author}{Y.~Abe}, et~al., \bibinfo{journal}{JHEP}
  \bibinfo{volume}{1410} (\bibinfo{year}{2014}) \bibinfo{pages}{086}.
  \bibinfo{note}{[Erratum: JHEP1502,074(2015)]}.
\bibitem[{Seo et~al.(2018)}]{Seo:2016uom}
\bibinfo{author}{S.~H. Seo}, et~al., \bibinfo{journal}{Phys. Rev.}
  \bibinfo{volume}{D98} (\bibinfo{year}{2018}) \bibinfo{pages}{012002}.
\bibitem[{Gandhi et~al.(2015)Gandhi, Kayser, Masud, and
  Prakash}]{Gandhi:2015xza}
\bibinfo{author}{R.~Gandhi}, \bibinfo{author}{B.~Kayser},
  \bibinfo{author}{M.~Masud}, \bibinfo{author}{S.~Prakash},
  \bibinfo{journal}{JHEP} \bibinfo{volume}{11} (\bibinfo{year}{2015})
  \bibinfo{pages}{039}.
\bibitem[{Ashenfelter et~al.(2016)}]{Ashenfelter:2015uxt}
\bibinfo{author}{J.~Ashenfelter}, et~al., \bibinfo{journal}{J. Phys.}
  \bibinfo{volume}{G43} (\bibinfo{year}{2016}) \bibinfo{pages}{113001}.
\bibitem[{Ashenfelter et~al.(2018)}]{Ashenfelter:2018osc}
\bibinfo{author}{J.~Ashenfelter}, et~al., \bibinfo{journal}{Phys. Rev. Lett.}
  \bibinfo{volume}{121} (\bibinfo{year}{2018}) \bibinfo{pages}{251802}.
\bibitem[{Capozzi et~al.(2015)Capozzi, Lisi, and Marrone}]{Capozzi:2015bpa}
\bibinfo{author}{F.~Capozzi}, \bibinfo{author}{E.~Lisi},
  \bibinfo{author}{A.~Marrone}, \bibinfo{journal}{Phys. Rev.}
  \bibinfo{volume}{D92} (\bibinfo{year}{2015}) \bibinfo{pages}{093011}.
\bibitem[{{Swartout, J.A. and Boch, A.L. and Cole, T.E. and Cheverton, R.D. and
  Adamson, G.M. and Winters, C.E.}(1964)}]{HFIR}
\bibinfo{author}{{Swartout, J.A. and Boch, A.L. and Cole, T.E. and Cheverton,
  R.D. and Adamson, G.M. and Winters, C.E.}}, \bibinfo{title}{{The Oak Ridge
  High Flux Isotope Reactor}}, \bibinfo{type}{Technical Report}
  \bibinfo{number}{4042265}, {Oak Ridge National Laboratory},
  \bibinfo{year}{1964}.
\bibitem[{Way and Wigner(1948)}]{Way:1948zz}
\bibinfo{author}{K.~Way}, \bibinfo{author}{E.~P. Wigner},
  \bibinfo{journal}{Phys. Rev.} \bibinfo{volume}{73} (\bibinfo{year}{1948})
  \bibinfo{pages}{1318--1330}.
\bibitem[{Avignone and Greenwood(1980)}]{Avignone:1980qg}
\bibinfo{author}{F.~T. Avignone, III}, \bibinfo{author}{C.~D. Greenwood},
  \bibinfo{journal}{Phys. Rev.} \bibinfo{volume}{C22} (\bibinfo{year}{1980})
  \bibinfo{pages}{594--605}.
\bibitem[{Klapdor and Metzinger(1982)}]{Klapdor:1982sf}
\bibinfo{author}{H.~V. Klapdor}, \bibinfo{author}{J.~Metzinger},
  \bibinfo{journal}{Phys. Lett.} \bibinfo{volume}{112B} (\bibinfo{year}{1982})
  \bibinfo{pages}{22--26}.
\bibitem[{Vogel(2007)}]{Vogel:2007du}
\bibinfo{author}{P.~Vogel}, \bibinfo{journal}{Phys. Rev.} \bibinfo{volume}{C76}
  (\bibinfo{year}{2007}) \bibinfo{pages}{025504}.
\bibitem[{{Chandler, D., Betzler, B., Hirtz, G., Ilas, G., and Sunny,
  E.}(2016)}]{HFIR_model}
\bibinfo{author}{{Chandler, D., Betzler, B., Hirtz, G., Ilas, G., and Sunny,
  E.}}, \bibinfo{title}{{Modeling and Depletion Simulations for a High Flux
  Isotope Reactor Cycle with a Representative Experiment Loading}},
  \bibinfo{type}{Technical Report} \bibinfo{number}{ORNL/TM-2016/23}, {Oak
  Ridge National Laboratory}, \bibinfo{year}{2016}.
\bibitem[{{C.J. Werner, et al.}(2018)}]{MCNP}
\bibinfo{author}{{C.J. Werner, et al.}}, \bibinfo{title}{{MCNP6.2 Release
  Notes}}, \bibinfo{type}{Technical Report} \bibinfo{number}{LA-UR-18-20808}, {
  Los Alamos National Laboratory}, \bibinfo{year}{2018}.
\bibitem[{Heeger et~al.(2013)Heeger, Littlejohn, Mumm, and
  Tobin}]{Heeger:2012tc}
\bibinfo{author}{K.~M. Heeger}, \bibinfo{author}{B.~R. Littlejohn},
  \bibinfo{author}{H.~P. Mumm}, \bibinfo{author}{M.~N. Tobin},
  \bibinfo{journal}{Phys. Rev.} \bibinfo{volume}{D87} (\bibinfo{year}{2013})
  \bibinfo{pages}{073008}.
\bibitem[{Ashenfelter et~al.(2016)}]{Ashenfelter:2015tpm}
\bibinfo{author}{J.~Ashenfelter}, et~al., \bibinfo{journal}{Nucl. Instrum.
  Meth.} \bibinfo{volume}{A806} (\bibinfo{year}{2016})
  \bibinfo{pages}{401--419}.
\bibitem[{Heffron(2017)}]{Heffron:2017}
\bibinfo{author}{B.~A. Heffron}, \bibinfo{title}{{Characterization of Reactor
  Background Radiation at HFIR for the PROSPECT Experiment}}, Master's thesis,
  {University of Tennessee}, \bibinfo{year}{2017}.
\bibitem[{Hackett(2017)}]{Hackett:2017}
\bibinfo{author}{B.~T. Hackett}, \bibinfo{title}{{DANG and the Background
  Characterisation of HFIR for PROSPECT}}, Master's thesis, {University of
  Surrey}, \bibinfo{year}{2017}.
\bibitem[{Bergeron et~al.(2017)Bergeron, Mumm, and Tyra}]{ISI:000413983300030}
\bibinfo{author}{D.~E. Bergeron}, \bibinfo{author}{H.~P. Mumm},
  \bibinfo{author}{M.~A. Tyra}, \bibinfo{journal}{{J. Radioanal. Nucl. Chem.}}
  \bibinfo{volume}{{314}} (\bibinfo{year}{{2017}}) \bibinfo{pages}{{767--771}}.
  \bibinfo{note}{{23rd International Conference on Advances in Liquid
  Scintillation Spectrometry (LSC), Copenhagen, Denmark, APR 30-MAY 05, 2017}}.
\bibitem[{Ashenfelter et~al.(2018)}]{Ashenfelter:2018cli}
\bibinfo{author}{J.~Ashenfelter}, et~al., \bibinfo{journal}{JINST}
  \bibinfo{volume}{13} (\bibinfo{year}{2018}) \bibinfo{pages}{P06023}.
\bibitem[{Ashenfelter et~al.(2015)}]{Ashenfelter:2015aaa}
\bibinfo{author}{J.~Ashenfelter}, et~al., \bibinfo{journal}{JINST}
  \bibinfo{volume}{10} (\bibinfo{year}{2015}) \bibinfo{pages}{P11004}.
\bibitem[{Band et~al.(2013)}]{Band:2013zka}
\bibinfo{author}{H.~R. Band}, et~al., \bibinfo{journal}{JINST}
  \bibinfo{volume}{8} (\bibinfo{year}{2013}) \bibinfo{pages}{P09015}.
\bibitem[{An et~al.(2016)}]{An:2015qga}
\bibinfo{author}{F.~P. An}, et~al., \bibinfo{journal}{Nucl. Instrum. Meth.}
  \bibinfo{volume}{A811} (\bibinfo{year}{2016}) \bibinfo{pages}{133--161}.

\end{thebibliography}

\end{document}